\documentclass[aps,pre,preprint,groupedaddress]{revtex4-2}

\usepackage{amsmath}
\usepackage{amssymb}
\usepackage{bm}
\usepackage{upgreek}
\usepackage{empheq}
\usepackage{multirow}

\begin{document}

\title{A Markov chain approximation of switched Fokker-Planck equations for a model of on-off intermittency in the postural control during quiet standing}


\author{Yasuyuki Suzuki}
\email[]{suzuki@bpe.es.osaka-u.ac.jp}
\affiliation{Graduate School of Engineering Science, Osaka University, Osaka 5608531, Japan}

\author{Keigo Togame}
\affiliation{Graduate School of Engineering Science, Osaka University, Osaka 5608531, Japan}

\author{Akihiro Nakamura}
\affiliation{Graduate School of Engineering Science, Osaka University, Osaka 5608531, Japan}

\author{Taishin Nomura}
\email[]{taishin@bpe.es.osaka-u.ac.jp}
\affiliation{Graduate School of Engineering Science, Osaka University, Osaka 5608531, Japan}


\date{\today}

\begin{abstract}
The intermittent on-off switching of feedback control is considered as a major mechanism of postural stabilization during human quiet standing, which can be modeled by switched-type hybrid stochastic delay differential equations with unstable subsystems. Dynamics of the model can also be described by the corresponding switched-type Fokker-Planck (FP) equations. Here, we develop a comprehensive numerical recipe to simulate switched-type FP equations in the case that the probability current is conserved at the switching boundary, as is the case for noise-free models exhibiting $C^{0}$-continuity for solutions at the boundary, but in a way extendable to cases with discontinuous jump. Specifically, the FP equations are approximated by a finite state Markov chain model using the finite element method. Then, dynamics of the Markov chain model, including time evolution of probability density function (PDF), stationary PDF, and power spectrum of postural sway are analyzed. We further investigate how the stationary PDF alters as values of important parameters of the model change. Dynamics of the Markov chain model are compared with Monte Carlo-based dynamics of the model, by which the developed numerical recipe is validated. The obtained Markov chain model forms a basis of our future investigations of the intermittent postural control as a Markov decision process.
\end{abstract}

\maketitle

\section{Introduction}
It has been a common view that homeostasis of living organisms is not static and unvarying \cite{Billman_2020}. Instead, it is a dynamic process that can change an internal state as required to survive external challenges, which is often accompanied by complex fluctuations of the internal state \cite{Bernhardt_2020}. Such a homeostatic state can be considered as the stochastic manifestation of nonlinear oscillations, including chaotic dynamics \cite{Fang_2019}. Underlying mechanisms of such oscillatory dynamics are often associated with time delay and switching-like behaviors in a feedback control loop in the presence of random noise. Cardiovascular control provides such a classical example of homeostasis at the macroscopic whole-body level. For example, Mayer waves in blood pressure may be caused by a feedback delay time of the baroreflex \cite{Seydnejad_2001}. Moreover, a stochastic switched feedback control is a leading hypothesis for the cause of $1/f$ fluctuation in the heart rate variability \cite{Ivanov_1998}. Similar classes of dynamics have been observed also in human motor behaviors \cite{Milton_1989}. Importantly, pathophysiology of those physiological dynamics may be expressed by altered complexity in fluctuations via transitions or bifurcations from one attractor to another as parameters of the control systems change, which has been associated with dynamical diseases \cite{Glass_2015,Belair_2021}.

In this study, we consider a model of postural stabilization of bodily mechanical plants against the gravitational toppling force, which is achieved by the neural feedback control of movement during human quiet stance, in light of the fact that the underlying mechanisms for postural maintenance could be analogous to the homeostasis. Upright posture during quiet stance exhibits complex fluctuations, referred to as the {\it postural sway}. Characterizations of postural sway and similar movement fluctuations during manually balanced stick at the fingertip have been challenged by a number of studies \cite{Maurer_2004,Kiemel_2011,Cabrera_2002,Collins_1994}. One of the important unsolved questions is the mechanistic cause of postural sway that exhibits a type of scaling behavior with $f^{-3/2}$ in the low frequency band at $f\in [0.01\sim 1.0]$ Hz, typically observed in healthy young people \cite{Collins_1994,Yamamoto_2015,Santos_2015}. Moreover, such a characteristic behavior can be lost in elderly and patients with neurological diseases \cite{Suzuki_2020}. The latest researches suggest that intermittent switching between ON and OFF in the active postural feedback controller for stabilizing a bodily inverted pendulum, referred to as the intermittent control during quiet stance \cite{Eurich_1996,Bottaro_2008,Asai_2009,Nomura_2013,Nomura_2020}, is a major mechanism of postural stabilization as well as postural fluctuation during human quiet standing, perhaps as a natural consequence of resolving a trade-off between minimizations of energy expenditure and erroneous postural deviation from the upright position in the presence of feedback time delay \cite{Suzuki_2020,Asai_2009,Michimoto_2016}.

The intermittent control model analyzed in this study is a novel but well-established hypothesis in the field of human postural control \cite{Richmond_2021,Xiang_2018,Tigrini_2022}. It is a hybrid dynamical system that switches between two unstable subsystems in a state-dependent manner, driven by additive white noise \cite{Suzuki_2020,Asai_2009,Nomura_2020}. The model exploits two types of instability for stabilizing upright posture. One is associated with the off-subsystem (open-loop control system), which exhibits unstable dynamics of purely mechanical inverted pendulum-like human body pinned around the ankle joint in the absence of the active feedback control, due to the fact that intrinsic, i.e., passive ankle stiffness is not sufficient for stabilizing the upright posture \cite{Loram_2002}, whereby the upright posture is characterized by a saddle-type unstable equilibrium with stable and unstable manifolds in the phase space. The other is associated with the on-subsystem (closed-loop control system), which is modeled by a time-delay proportional (P) and derivative (D) feedback control system, if the delayed PD controller were adopted persistently without switching, particularly with small values of the $P$ and $D$ gains that lead to a delay-induced unstable oscillation around the upright equilibrium. In the intermittent control model, an appropriate mechanism of state-dependent switching between those two unstable dynamics makes the overall dynamics stable, where the switching function is implemented in such a way that the feedback controller is switched OFF when the state vector of the inverted pendulum is near the stable manifold of the saddle, and it is switched ON otherwise \cite{Asai_2009,Nomura_2020}. The important role played by the stable manifold in the intermittent control model is conceptually similar to that in the OGY chaos control that also exploits a stable manifold of a saddle point to make chaotic dynamics periodic \cite{Ott_1990}. A difference between them is a way to exploit the stable manifold. A chaotic meandering behavior and a delay-induced unstable oscillation are exploited by the system to make the state point close to the stable manifold in the OGY and the intermittent control model, respectively. A remarkable feature of the intermittent control model is that it can achieve flexibility and stability of upright posture simultaneously, typically as in healthy young people, where the model owes flexibility of the ankle joint to the null-impedance ($P=D=0$) of the off-subsystem as well as to the small impedance (small $P$ and $D$ values) of the on-subsystem. To the contrary, postural instability in patients with Parkinson's disease may be caused by inflexible rigidity of the ankle joint via a loss of intermittent ON-OFF switching and large values of $P$ and $D$ gains that compensates the loss of intermittency \cite{Suzuki_2020}.

In this study, we perform a detailed numerical analysis of the intermittent control model as a switched stochastic delay system with unstable subsystems (the on- and off-subsystems), driven by additive Gaussian white noise. The on-subsystem is described by a stochastic delay differential equation (S-DDE), for which there are several analytical approaches developed so far, such as the small delay approximation for the corresponding Fokker-Planck (FP) equation that describes time evolutions of a probability density function (PDF) representing probabilities for a state point to be located at given positions in the state space \cite{Longtin_1999}. The off-subsystem is simply governed by a stochastic ordinary differential equation (S-ODE) with no time-delay, for which dynamics of PDF are also governed by its FP equation. Therefore, dynamics of the intermittent control model can be analyzed by using switched-type hybrid FP equations for the subsystems. Because fewer studies focused on the switched-type hybrid FP equations \cite{Bect_2006,Bect_2008,Kumar_2007,Wang_2020}, particularly with regard to S-DDEs, we put our major concern here to develop a comprehensive numerical recipe to represent and simulate hybrid FP equations, and apply it to the analysis of the intermittent control model. Because it is difficult to consider issues on time-delay and switched-type hybrid systems simultaneously, we approximate a deterministic part of the S-DDE for the on-subsystem by an ODE with no delay (but a delay time is included in coefficients of the resultant approximate ODE), based on the small delay approximation \cite{Stepan_2000}, i.e., Taylor expansion of a retarded state, rather than using a more rigorous stochastic correspondence for S-DDEs \cite{Longtin_1999}. In this way, the on-subsystem is approximated by the S-ODE. Thus, we have two S-ODEs with no retarded states: one for the on-subsystem and the other for the off-subsystem, and corresponding FP equations that form a much simpler switched-type hybrid FP equations. Moreover, because the intermittent control model without noise exhibits $C^{0}$-continuity for solutions at the switching boundary, stochastic dynamics of the model can be analyzed under the assumption on the conservation of probability current at the boundary, which also makes our analysis of the switched-type hybrid FP equations simple. Because a solution of each of two FP equations is a Markov process, dynamics of the hybrid FP equations can be approximated by a finite state Markov chain, if we consider a limited area of the state space and the FP equations are discretized in time and space. Here, such a discretization is performed by using the finite element method (FEM), and then stochastic switching dynamics of the intermittent control model, including time evolutions of PDFs, stationary PDFs, and power spectral density functions (PSD) of model-simulated postural sway are analyzed in detail using the Markov chain approximation of the hybrid FP equations. We also investigate how the stationary PDF alters as values of important parameters of the model change, as considered in \cite{Suzuki_2020} and \cite{Nomura_2013}. Moreover, we compare dynamics of the Markov chain model with those of approximate S-ODE+S-ODE hybrid model and the original S-DDE+S-ODE hybrid model based on Monte Carlo simulations, by which the developed numerical recipe and the Markov chain model are validated.

This paper is organized as follows. In Section \ref{sec:TheIntermittentControlModel}, we introduce the intermittent control model briefly. In Section \ref{sec:FEMBasedAnalysis}, we reformulate the intermittent control model as the hybrid S-ODEs, and as the hybrid FP equations. Then, we develop a comprehensive numerical recipe to approximate the hybrid FP equations by the finite Markov chain model based on the FEM. We perform detailed analysis of the Markov chain model and the original intermittent control model in Section \ref{sec:ComparisonWithMonteCarloSimulations}. Finally, we provide concluding remarks and future perspectives.

\section{The intermittent control model}
\label{sec:TheIntermittentControlModel}
\begin{figure}[htbp]
  \includegraphics[keepaspectratio=true,height=80mm]{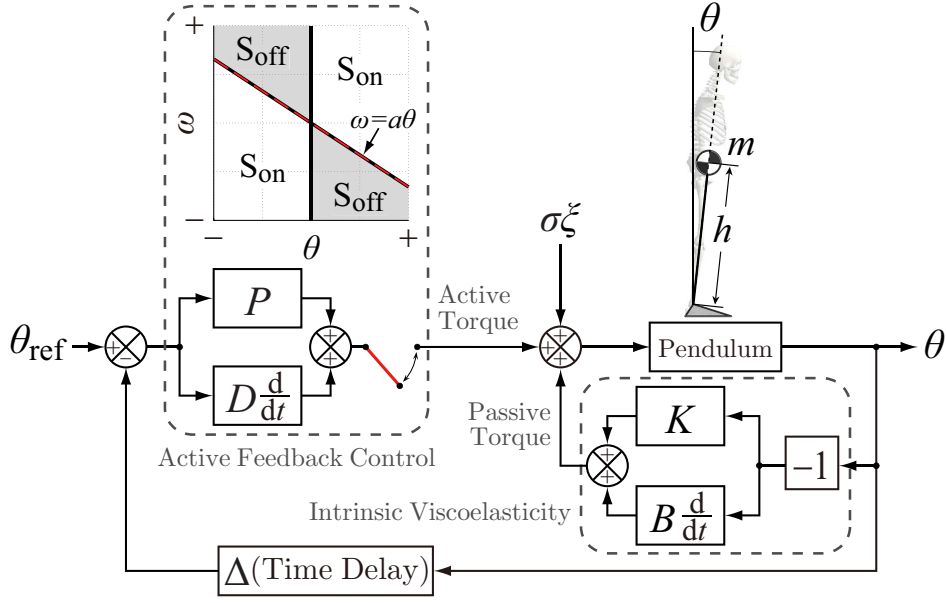}
  \caption{Block diagram of the intermittent control model for stabilizing human quiet standing}
  \label{fig:Model.eps}
\end{figure}

Human upright posture is modeled by a single inverted pendulum that rotates around a pinned joint, representing the ankle joint, in the sagittal plane (Fig.~\ref{fig:Model.eps}), as in many other previous studies \cite{Maurer_2004,Asai_2009,Winter_1998}. The equation of motion of the model can be linearized around the upright position for a small tilt angle $\theta(t)$ at time $t$, which is represented as
\begin{equation}
  I \ddot{\theta}(t) = m g h \theta(t) + \tau(t) + \sigma \xi(t),
\label{eq:MotionEquation}
\end{equation}
where $m$ is the mass of the pendulum, $h$ is the distance between the ankle joint and the center of mass (CoM) of the pendulum, $I$ is the moment of inertia around the ankle joint, and $g$ is the gravitational acceleration. $\xi(t)$ represents the Gaussian white noise with zero mean and unity standard deviation. $\sigma$ represents the noise intensity. $\tau(t)$ represents the ankle joint torque, which is formulated as
\begin{equation}
  \tau(t) = - K \theta(t) - B \omega(t) + \tau_{\text{act}}(t),
\label{eq:AnkleTorque}
\end{equation}
where $\omega(t)\equiv\dot{\theta}(t)$ is the angular velocity, $K$ and $B$ represent the coefficients of passive elasticity and viscosity for the rotational motion of the pendulum around the ankle joint. The term $\tau_{\text{act}}(t)$ represents the active feedback control torque, which is generated by active muscle contractions according to a sequence of time-varying neural commands from the central nervous system. It has been shown that $K<mgh$ \cite{Loram_2002}, by which the upright posture cannot be stabilized by the passive impedance of the ankle joint alone \cite{Morasso_2002}. Thus, the active feedback control torque $\tau_{\text{act}}(t)$ is indispensable for stabilizing upright posture.

\begin{table}
  \caption{Variables and parameters for the Intermittent control model}
  \label{tab:ModelParameters}
    \begin{tabular}{clc}
      \hline
      Symbol  &  Description  &  Value/Unit  \\
      \hline
      $\theta$  &  Tilting angle from vertical line  &  --- rad  \\
      $\omega$  &  Angular velocity of the pendulum  &  --- rad/s  \\
      $m$  &  Mass of pendulum  &  $60$  kg  \\
      $h$  &  Distance from ankle joint to Center of Mass  &  $1.0$ m  \\
      $I$  &  Moment of Inertia around ankle joint  &  $m h^{2}$  kg m$^{2}$  \\
      $K$  &  Passive elastic coefficient of ankle joint  &  $0.8 mgh$ Nm/rad  \\
      $B$  &  Passive viscosity coefficient of ankle joint  &  $4.0$ Nms/rad  \\
      $P$  &  Proportional gain of active feedback for ankle joint  &  $0.25mgh$ Nm/rad  \\
      $D$  &  Derivative gain of active feedback for ankle joint  &  $10.0$ Nms/rad  \\
      $a$  &  Slope of the boundary line  &  $-0.4$  \\
      $\Delta$  &  Delay in the active feedback loop  &  $0.2$ s  \\
      $\sigma$  &  Noise intensity  &  $0.2$  Nm \\
      \hline
    \end{tabular}
\end{table}

Traditionally, $\tau_{\text{act}}(t)$ has been modeled simply by a proportional and derivative feedback (PD) controller with a feedback time-delay $\Delta$ s \cite{Maurer_2004}, which is formulated as
\begin{equation}
  \tau_{\text{act}}(t) = - P \theta(t-\Delta) - D \omega(t-\Delta),
\label{eq:ActiveTorque_Cont}
\end{equation}
where $P$ and $D$ represent the proportional and derivative gains, respectively. The postural control model with this delay PD feedback controller is referred to here as the {\it continuous control model}, in contrast to the intermittent control model.

\begin{figure}[htbp]
  \includegraphics[keepaspectratio=true,height=85mm]{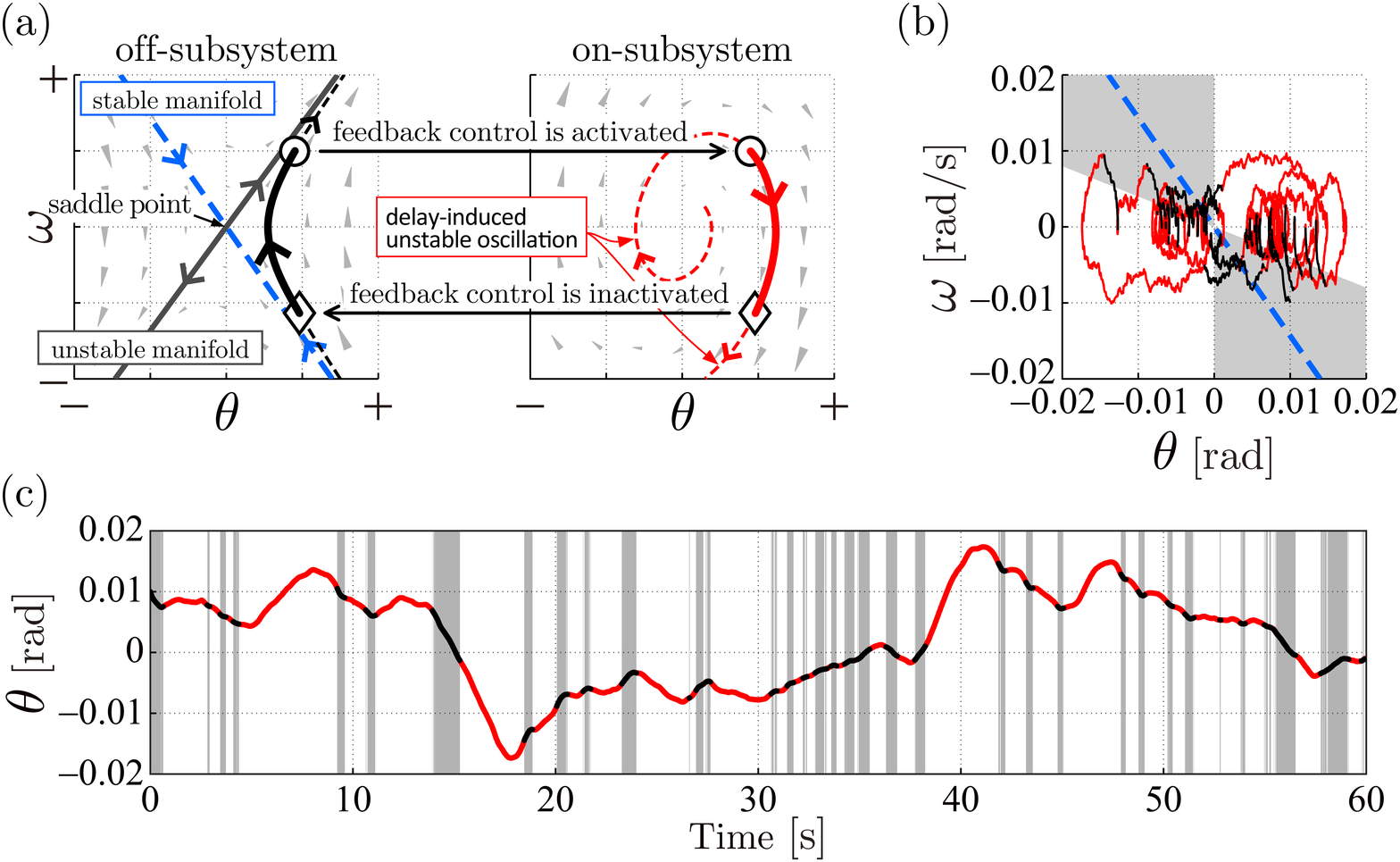}
  \caption{Schematic illustrations and sample dynamics of the intermittent control model. (a) A schematic diagram for explaining state-dependent switching behavior between the on- and off-subsystems. (b) A sample trajectory of the intermittent control model. Gray and white regions represent the off-region $\text{S}_{\text{off}}$ and the on-region $\text{S}_{\text{on}}$, respectively. Blue dashed line is the stable manifold of the off-subsystem. (c) A model-simulated postural sway, corresponding to the trajectory in (b). In (b) and (c), the trajectory and waveform are colored by black and red curve segments, when the state point is located in the off and the on regions, respectively.}
  \label{fig:IntermittentControl.eps}
\end{figure}

In the intermittent control model, the delay PD feedback controller is switched OFF in a state-dependent manner (Figs.~\ref{fig:Model.eps} and \ref{fig:IntermittentControl.eps}), which is formulated as
\begin{equation}
  \tau_{\text{act}}(t) = \left\{\begin{array}{cl}
    - P \theta(t-\Delta) - D \omega(t-\Delta),  &  \text{if } (\theta(t-\Delta), \omega(t-\Delta))^{\text{T}} \in \text{S}_{\text{on}},  \\
    0,  &  \text{otherwise if } (\theta(t-\Delta), \omega(t-\Delta))^{\text{T}} \in \text{S}_{\text{off}},
  \end{array}\right.
\label{eq:ActiveTorque_Int}
\end{equation}
where $\text{S}_{\text{on}}$ and $\text{S}_{\text{off}}$ represent the ON and OFF regions in the $\theta$-$\omega$ plane. Specifically, in the intermittent control model analyzed in this study, the $\theta$-$\omega$ plane is separated by two lines of switching boundaries defined as $\theta=0$ and $\omega=a\theta$ with a slope of $a$ (Fig.~\ref{fig:Model.eps}).

Note that the PD feedback control model used for the on-subsystem is exactly the same as the traditional continuous control model \cite{Maurer_2004,Winter_1998}, if it is used persistently without switching mechanisms. Indeed, $\tau_{\text{act}}$ in Eq.~(\ref{eq:ActiveTorque_Int}) for the intermittent control model becomes identical to $\tau_{\text{act}}$ in Eq.~(\ref{eq:ActiveTorque_Cont}) for the continuous control model as the slope parameter $a\to-\infty$. Note also that typical $P$ and $D$ gains of the on-subsystem used for the intermittent control model are much smaller than those used for the continuous control model. The upright equilibrium of the PD feedback control model as the on-subsystem of the intermittent control model with such small $P$-$D$ gains is unstable, whereas that of the traditional continuous control model with large $P$-$D$ gains is stable. See Table \ref{tab:ModelParameters} for a typical parameter values employed for the intermittent control model.

Figure \ref{fig:IntermittentControl.eps} illustrates typical steady-state dynamics of the intermittent control model, using schematic diagram (Fig.~\ref{fig:IntermittentControl.eps}(a)), a simulated sample trajectory in the $\theta$-$\omega$ phase plane (Fig.~\ref{fig:IntermittentControl.eps}(b)), and the corresponding postural sway time series (Fig.~\ref{fig:IntermittentControl.eps}(c)). One can observe that the sway pattern exhibits a slow oscillatory component with a relatively large amplitude despite the small noise intensity. The phase portrait of the model illustrates features of the intermittent control model, in which the state point in the off-region $\text{S}_{\text{off}}$ moves slowly along the stable manifold, approaching the saddle point transiently, followed by a fall away from the saddle point along the unstable manifold until the state point goes out from the off-region $\text{S}_{\text{off}}$. When the PD controller is activated in $\text{S}_{\text{on}}$, the delay-induced unstable oscillatory dynamics moves the state point closer to the stable manifold in the off-region, at which the PD controller is switched off (see Figs.~\ref{fig:IntermittentControl.eps}(a) and (b)). Since this scenario works for both right and left halves of the phase plane, the noisy dynamics exhibit transitions between left and right planes, in addition to the on-off transitions, generating a butterfly-shaped trajectory (Fig.~\ref{fig:IntermittentControl.eps}(b)). The PSD for this fluctuation shows the power-law-like behavior ($f^{-\beta}$) at the low-frequency regime with the scaling exponent about $\beta=3/2$ as in human postural sway \cite{Nomura_2013}.

By the small delay approximation \cite{Stepan_2000}, the on-subsystem of the intermittent control model as the DDE can be approximated by the following ordinary differential equation (ODE) with no retarded state variables:
\begin{equation}
  \left(I - D \Delta\right) \ddot{\theta}(t) + \left(B + D - P \Delta\right) \dot{\theta}(t) + \left(K + P - mgh\right) \theta(t) = \sigma \xi(t),
\label{eq:2nd_order_ODE}
\end{equation}
using Taylor expansions of the retarded state variables of $\theta(t-\Delta)\simeq\theta(t)-\Delta\omega(t)$ and $\omega(t-\Delta)\simeq\omega(t)-\Delta\ddot{\theta}(t)$. Then, the state space representation of the intermittent control model is expressed by the following hybrid system that switches between two first order ODE systems, i.e., between the ODE-approximated on-subsystem and the off-subsystem.
\begin{equation}
  \frac{\mathrm{d}}{\mathrm{d}t} \left(\begin{array}{c}
    \theta(t)  \\
    \omega(t)
  \end{array}\right) = \left\{\begin{array}{cl}
    \mathbf{A}_{\text{on}} \left(\begin{array}{c}
      \theta(t)  \\
      \omega(t)
    \end{array}\right) + \bm{\Sigma}_{\text{on}} \left(\begin{array}{c}
      0  \\
      \xi(t)
    \end{array}\right),  &  \text{if } \left(\begin{array}{c}
      \theta(t)  \\
      \omega(t)
    \end{array}\right) \in \text{S}_{\text{on}},  \\
    \mathbf{A}_{\text{off}} \left(\begin{array}{c}
      \theta(t)  \\
      \omega(t)
    \end{array}\right) + \bm{\Sigma}_{\text{off}} \left(\begin{array}{c}
      0  \\
      \xi(t)
    \end{array}\right),  &  \text{otherwise if } \left(\begin{array}{c}
      \theta(t)  \\
      \omega(t)
    \end{array}\right) \in \text{S}_{\text{off}},
  \end{array}\right.
\label{eq:StateSpaceRep_int}
\end{equation}
where $\mathbf{A}_{\text{on}}$, $\bm{\Sigma}_{\text{on}}$, $\mathbf{A}_{\text{off}}$, and $\bm{\Sigma}_{\text{off}}$ are defined as follows:
\begin{align}
  \mathbf{A}_{\text{on}} &= \left(\!\begin{array}{cc}
    0  &  1  \\
    {\displaystyle\frac{mgh-K-P}{I-D\Delta}}  &  {\displaystyle\frac{-B-D+P\Delta}{I-D\Delta}}
  \end{array}\!\right),
\label{eq:A_on}\\
  \mathbf{A}_{\text{off}} &= \left(\!\begin{array}{cc}
    0  &  1  \\
    {\displaystyle\frac{mgh-K}{I}}  &  {\displaystyle\frac{-B}{I}}
  \end{array}\!\right),
\label{eq:A_off}\\
\bm{\Sigma}_{\text{on}} &= \left(\!\begin{array}{cc}
    0  &  0  \\
    0  &  {\displaystyle\frac{\sigma}{I-D\Delta}}
  \end{array}\!\right),
\label{eq:Sigma_on}\\
\bm{\Sigma}_{\text{off}} &= \left(\!\begin{array}{cc}
    0  &  0  \\
    0  &  {\displaystyle\frac{\sigma}{I}}.
  \end{array}\!\right).
\label{eq:Sigma_off}
\end{align}
Note that, due to the ODE-approximation procedure, the system matrix $\mathbf{A}_{\text{on}}$ and the matrix $\bm{\Sigma}_{\text{on}}$ representing the noise intensity of the on-subsystem are now functions of the feedback gain $D$ and the delay time $\Delta$.

\section{FEM-based analysis of the ODE-approximated model}
\label{sec:FEMBasedAnalysis}
\subsection{Problem setting}
We formally rewrite Eq.~(\ref{eq:StateSpaceRep_int}) into the following set of two S-ODEs:
\begin{align}
  \mathrm{d}\mathbf{X} &= \mathbf{f}_{\text{A}}(\mathbf{X})\mathrm{d}t + \bm{\Sigma}_{\text{A}} \mathrm{d}\mathbf{W},  \quad \text{if } \mathbf{x} \in \Lambda_{\text{A}},
\label{eq:SDE_2D_hybrid_A}\\
  \mathrm{d}\mathbf{X} &= \mathbf{f}_{\text{B}}(\mathbf{X})\mathrm{d}t + \bm{\Sigma}_{\text{B}} \mathrm{d}\mathbf{W},  \quad \text{otherwise if } \mathbf{x} \in \Lambda_{\text{B}},
\label{eq:SDE_2D_hybrid_B}
\end{align}
where $\mathbf{X}=(\Theta,\Omega)^{\text{T}}$ is the two-dimensional random variable vector that takes real values $\mathbf{x}=(\theta,\omega)^{\text{T}}$. Moreover, $\mathbf{f}_{\text{A}}(\mathbf{X})\equiv\mathbf{A}_{\text{on}}\mathbf{X}$ and $\mathbf{f}_{\text{B}}(\mathbf{X})\equiv\mathbf{A}_{\text{off}}\mathbf{X}$ represent the deterministic vector fields of the on- and off-subsystems, respectively, according to Eqs.~(\ref{eq:A_on}) and (\ref{eq:A_off}). The noise term $\mathrm{d}\mathbf{W}$ is an infinitesimal increment of the standard Wiener process $\mathbf{W}$, whose formal derivative $\mathrm{d}\mathbf{W}/\mathrm{d}t=(0,\xi)^{\text{T}}$ is associated with the Gaussian white noise defined in Eq.~(\ref{eq:MotionEquation}). $\bm{\Sigma}_{\text{A}}\equiv\bm{\Sigma}_{\text{on}}$ and $\bm{\Sigma}_{\text{B}}\equiv\bm{\Sigma}_{\text{off}}$ are the intensity matrices of additive noise. Thus, $\bm{\Sigma}_{k}\mathrm{d}\mathbf{W}$ represents the motor noise that drives the system for $k\in\{\text{A},\text{B}\}$. The regions in the $\theta$-$\omega$ plane for selecting the on- and the off-subsystems are denoted as $\Lambda_{\text{A}}\equiv\text{S}_{\text{on}}$ and $\Lambda_{\text{B}}\equiv\text{S}_{\text{off}}$.

The switching boundary between two regions $\Lambda_{\text{A}}$ and $\Lambda_{\text{B}}$, composed of two lines $\theta=0$ and $\omega=a\theta$, is denoted by $\Gamma^{\text{s}}$. See Figs.~\ref{fig:IntermittentControl.eps}(b). Specifically, we denote the border of $\Lambda_{\text{A}}$ for switching as $\Gamma_{\text{A}}^{\text{s}}$, and that of $\Lambda_{\text{B}}$ as $\Gamma_{\text{B}}^{\text{s}}$. Note that $\Gamma^{\text{s}}=\Gamma_{\text{A}}^{\text{s}}=\Gamma_{\text{B}}^{\text{s}}$, and any of those representations are used at our convenience. $\Lambda=\Lambda_{\text{A}}\cup\Lambda_{\text{B}}$ is the entire domain of the finite element analysis, which would be defined later as a rectangular region with a finite area for numerical analysis. Outer edge of the entire domain $\Lambda$ is the boundary other than the switching boundary $\Gamma^{\text{s}}$, referred to as the domain boundary and denoted by $\Gamma^{\text{d}}$. Specifically, we denote the domain boundary for $\Lambda_{k}$ for $k\in\{\text{A},\text{B}\}$ as $\Gamma_{k}^{\text{d}}$. Then, $\Gamma^{\text{d}}=\Gamma_{\text{A}}^{\text{d}}+\Gamma_{\text{B}}^{\text{d}}$. Moreover, $\Gamma_{k}=\Gamma_{k}^{\text{d}}+\Gamma_{k}^{\text{s}}$, where $\Gamma_{k}$ is the entire boundary surrounding the region $\Lambda_{k}$.

In this section, we develop a comprehensive numerical recipe for approximating the ODE-approximated intermittent control model by a finite state Markov-chain using the FEM. Deriving such a Markov chain model is of critical importance to our motive for associating the Markov chain model with a Markov decision process and a reinforcement learning of the postural control strategy in our future study. Although the overall recipe developed here might not necessarily be novel, we decided to describe detailed instructions on how we derive a state transition probability matrix of the Markov chain model. Despite its lengthiness, we believe that a full description of the process could be beneficial, because our derivation process itself might be unique and has not been described in literatures. Note that the developed recipe can be applied generally, not only for the intermittent control model, but also for general switched-type hybrid stochastic dynamical systems that can be described by Eqs.~(\ref{eq:SDE_2D_hybrid_A}) and (\ref{eq:SDE_2D_hybrid_B}).

Let $\rho(\mathbf{x},t)$ be the probability density function (PDF) associated with the probability such that a state point that moves according to the SDE of Eqs.~(\ref{eq:SDE_2D_hybrid_A}) and (\ref{eq:SDE_2D_hybrid_B}) is located at a position $\mathbf{x}$ at time $t$. Time evolution of $\rho(\mathbf{x},t)$ is governed by the following hybrid FP equations:
\begin{align}
  \frac{\partial \rho(\mathbf{x}, t)}{\partial t} &= - \nabla \cdot \left(\mathbf{f}_{\text{A}}(\mathbf{x}) \rho(\mathbf{x}, t)\right) + \nabla \cdot \left(\mathbf{D}_{\text{A}} \nabla \rho(\mathbf{x}, t)\right) \equiv - \nabla \cdot \mathbf{j}_{\text{A}}(\mathbf{x}, t),  \quad  \text{if } \mathbf{x} \in \Lambda_{\text{A}},
\label{eq:FPE_2D_hybrid_A}\\
  \frac{\partial \rho(\mathbf{x}, t)}{\partial t} &= - \nabla \cdot \left(\mathbf{f}_{\text{B}}(\mathbf{x}) \rho(\mathbf{x}, t)\right) + \nabla \cdot \left(\mathbf{D}_{\text{B}} \nabla \rho(\mathbf{x}, t)\right) \equiv - \nabla \cdot \mathbf{j}_{\text{B}}(\mathbf{x}, t),  \quad  \text{if } \mathbf{x} \in \Lambda_{\text{B}},
\label{eq:FPE_2D_hybrid_B}
\end{align}
where $\mathbf{D}_{k}$ for $k\in\{\text{A},\text{B}\}$ is a diffusion coefficient matrix defined as $\mathbf{D}_{k}=(1/2)\bm{\Sigma}_{k}^{2}$, and $\mathbf{j}_{k}(\mathbf{x},t)$ is the probability current defined as
\begin{equation}
  \mathbf{j}_{k}(\mathbf{x},t)\equiv\mathbf{f}_{k}(\mathbf{x})\rho(\mathbf{x},t)-\mathbf{D}_{k}\nabla\rho(\mathbf{x},t).
\end{equation}

For the FEM-based analysis of Eqs.~(\ref{eq:FPE_2D_hybrid_A}) and (\ref{eq:FPE_2D_hybrid_B}), it is natural to impose the following boundary conditions on solutions of Eqs.~(\ref{eq:FPE_2D_hybrid_A}) and (\ref{eq:FPE_2D_hybrid_B}):
\begin{align}
  \mathbf{j}_{\text{A}}(\mathbf{x},t) \cdot \mathbf{n}_{\Gamma_{\text{A}}}(\mathbf{x}) = 0,&  \quad \mathbf{x}\in\Gamma_{\text{A}}^{\text{d}}
\label{eq:zero_flux_A}  \\
  \mathbf{j}_{\text{B}}(\mathbf{x},t) \cdot \mathbf{n}_{\Gamma_{\text{B}}}(\mathbf{x}) = 0,&  \quad \mathbf{x}\in\Gamma_{\text{B}}^{\text{d}}
\label{eq:zero_flux_B}  \\
  \mathbf{j}_{\text{A}}(\mathbf{x},t) \cdot \mathbf{n}_{\Gamma_{\text{A}}}(\mathbf{x})=\mathbf{j}_{\text{B}}(\mathbf{x},t) \cdot \left(- \mathbf{n}_{\Gamma_{\text{B}}}(\mathbf{x})\right),& \quad \mathbf{x}\in\Gamma^{\text{s}}
\label{eq:flux_conservation}
\end{align}
where $\mathbf{n}_{\Gamma_{k}}(\mathbf{x})$ for $k\in\{\text{A},\text{B}\}$ is a unit normal vector initiating from a point $\mathbf{x}$ on the boundary $\Gamma_{k}$ of $\Lambda_{k}$ and directing from inside to outside of $\Lambda_{k}$. Eqs.~(\ref{eq:zero_flux_A}) and (\ref{eq:zero_flux_B}) are conditions at the domain boundary $\Gamma^{\text{d}}$ for the conservation of total probability by setting the normal direction component of the probability current to zero at the domain boundary. This is plausible, because a solution of Eqs.~(\ref{eq:FPE_2D_hybrid_A}) and (\ref{eq:FPE_2D_hybrid_B}) should be bounded within the entire domain $\Lambda$, if $\Lambda$ defined for the FEM analysis is sufficiently (and practically) larger than diffusive motions of any stable solutions. Eq.~(\ref{eq:flux_conservation}) is a condition at the switching boundary $\Gamma^{\text{s}}$, imposing that the normal component of the probability current flowing out from the region A (flowing into the region A) is equal to the normal component of the probability current flowing into the region B (flowing out from the region B). Unlike more general switched systems with discontinuous jumps in deterministic solutions, such as a system describing an inelastic collision \cite{Burkhardt_2000,Burkhardt_2007,Bect_2006,Wang_2020}, the intermittent control model with Eq.~(\ref{eq:StateSpaceRep_int}) satisfies Eq.~(\ref{eq:flux_conservation}), because the model without noise exhibits $C^{0}$-continuity for solutions at the switching boundary. Moreover, by the $C^{0}$-continuity, we can also impose a continuity condition on the PDF at the switching boundary.

In this sequel, we develop a numerical recipe to represent the hybrid FP equations of Eqs.~(\ref{eq:FPE_2D_hybrid_A}) and (\ref{eq:FPE_2D_hybrid_B}) approximately as a finite state Markov-chain using the FEM. To this end, the entire domain $\Lambda$ is divided into a finite number of small elements, and approximate solutions $\tilde{\rho}(\mathbf{x},t)$ of Eqs.~(\ref{eq:FPE_2D_hybrid_A}) and (\ref{eq:FPE_2D_hybrid_B}) are represented by basis functions with their coefficient values at the nodes of finite elements.

\subsection{Preliminary remarks}
In this study, we simply use the method of weighted residuals to find an approximate solution $\tilde{\rho}(\mathbf{x},t)$. To this end, the true solution $\rho(\mathbf{x},t)$ of Eqs.~(\ref{eq:FPE_2D_hybrid_A}) and (\ref{eq:FPE_2D_hybrid_B}) is replaced by its approximate solution $\tilde{\rho}(\mathbf{x},t)$, and a sum of differences between the left- and the right-hand-sides of each of Eqs.~(\ref{eq:FPE_2D_hybrid_A}) and (\ref{eq:FPE_2D_hybrid_B}) is multiplied by the weighting function $w(\mathbf{x})$. Then, $\tilde{\rho}(\mathbf{x},t)$ is calculated so that the value of integral of the sum of weighted differences over the entire domain become identical to zero for any weighting functions.
\begin{align}
  &\int_{\Lambda_{\text{A}}} w(\mathbf{x}) \left\{\frac{\partial \tilde{\rho}(\mathbf{x}, t)}{\partial t} + \nabla^{\text{T}} \left(\mathbf{f}_{\text{A}}(\mathbf{x}) \tilde{\rho}(\mathbf{x}, t)\right) - \nabla^{\text{T}} \left(\mathbf{D}_{\text{A}} \nabla \tilde{\rho}(\mathbf{x}, t)\right)\right\} \mathrm{d}\Lambda  \nonumber\\
  &+ \int_{\Lambda_{\text{B}}} w(\mathbf{x}) \left\{\frac{\partial \tilde{\rho}(\mathbf{x}, t)}{\partial t} + \nabla^{\text{T}} \left(\mathbf{f}_{\text{B}}(\mathbf{x}) \tilde{\rho}(\mathbf{x}, t)\right) - \nabla^{\text{T}} \left(\mathbf{D}_{\text{B}} \nabla \tilde{\rho}(\mathbf{x}, t)\right)\right\}\mathrm{d}\Lambda  \nonumber\\
  &= 0.
\label{eq:FPE_2D_hybrid_2_pre}
\end{align}

In the typical FEM analysis, the second and third integrands of each integral in Eq.~(\ref{eq:FPE_2D_hybrid_2_pre}) are rewritten, respectively, as
\begin{align}
  w(\mathbf{x}) \nabla^{\text{T}} \left(\mathbf{f}_{k}(\mathbf{x}) \tilde{\rho}(\mathbf{x}, t)\right) &= \nabla^{\text{T}} \left\{w(\mathbf{x}) \left(\mathbf{f}_{k}(\mathbf{x}) \tilde{\rho}(\mathbf{x}, t)\right)\right\} - \left(\nabla^{\text{T}} w(\mathbf{x})\right) \left(\mathbf{f}_{k}(\mathbf{x}) \tilde{\rho}(\mathbf{x}, t)\right),
\label{eq:Nabla_formulate_f_rho}  \\
  w(\mathbf{x}) \nabla^{\text{T}} \left(\mathbf{D}_{k} \nabla \tilde{\rho}(\mathbf{x}, t)\right) &= \nabla^{\text{T}} \left\{w(\mathbf{x}) \left(\mathbf{D}_{k} \nabla \tilde{\rho}(\mathbf{x}, t)\right)\right\} - \left(\nabla^{\text{T}} w(\mathbf{x})\right) \left(\mathbf{D}_{k} \nabla \tilde{\rho}(\mathbf{x}, t)\right).
\label{eq:Nabla_formulate_D_nabla_rho}
\end{align}
Surface integrals of the first components of the right-hand-sides of Eq.~(\ref{eq:Nabla_formulate_f_rho}) and Eq.~(\ref{eq:Nabla_formulate_D_nabla_rho}) over $\Lambda_{k}$ for $k\in\{\text{A},\text{B}\}$ are represented by line integrals along the boundary $\Gamma_{k}$($=\Gamma_{k}^{\text{d}}+\Gamma_{k}^{\text{s}}$), respectively, as
\begin{align}
  \int_{\Lambda_{k}} \nabla^{\text{T}} \left\{w(\mathbf{x}) \left(\mathbf{f}_{k}(\mathbf{x}) \tilde{\rho}(\mathbf{x}, t)\right)\right\} \mathrm{d}\Lambda &= \oint_{\Gamma_{k}^{\text{d}}+\Gamma_{k}^{\text{s}}} w(\mathbf{x}) \left(\mathbf{n}_{\Gamma_{\text{A}}}(\mathbf{x})\right)^{\text{T}} \mathbf{f}_{k}(\mathbf{x}) \tilde{\rho}(\mathbf{x}, t) \mathrm{d}\Gamma,
\label{eq:DivTheorem_f_rho}  \\
  \int_{\Lambda_{k}} \nabla^{\text{T}} \left\{w(\mathbf{x}) \left(\mathbf{D}_{k} \nabla \tilde{\rho}(\mathbf{x}, t)\right)\right\} \mathrm{d}\Lambda &= \oint_{\Gamma_{k}^{\text{d}}+\Gamma_{k}^{\text{s}}} w(\mathbf{x}) \left(\mathbf{n}_{\Gamma_{k}}(\mathbf{x})\right)^{\text{T}} \mathbf{D}_{k} \bm{\upbeta}_{k}(\mathbf{x}, t) \mathrm{d}\Gamma,
\label{eq:DivTheorem_D_nabla_rho}
\end{align}
by the divergence theorem, where $\bm{\upbeta}_{k}(\mathbf{x},t)$ in the right-hand-side of Eq.~(\ref{eq:DivTheorem_D_nabla_rho}) is a formal spatial derivative of the approximate solution, i.e., $\nabla \tilde{\rho}(\mathbf{x},t)$ at a point $\mathbf{x}$ on the boundary $\Gamma_{k}$ of the region $\Lambda_{\text{k}}$. The term $\bm{\upbeta}_{k}(\mathbf{x},t)$ is defined formally, because the spatial derivative of the approximate solution $\tilde{\rho}(\mathbf{x},t)$ on the boundary is not necessarily well-defined. In general, the existence of this ill-defined term $\bm{\upbeta}_{k}(\mathbf{x},t)$ could be a major obstacle to develop the finite state Markov-chain of the stochastic hybrid system. Fortunately, the ill-defined term $\bm{\upbeta}_{k}(\mathbf{x},t)$ would disappear in the case if the domain and switching boundary conditions are described by Eqs.~(\ref{eq:zero_flux_A})-(\ref{eq:flux_conservation}). Namely, by substituting Eqs.~(\ref{eq:Nabla_formulate_f_rho}) and (\ref{eq:Nabla_formulate_D_nabla_rho}) into Eq.~(\ref{eq:FPE_2D_hybrid_2_pre}) and using Eqs.~(\ref{eq:DivTheorem_f_rho}) and (\ref{eq:DivTheorem_D_nabla_rho}), Eq.~(\ref{eq:FPE_2D_hybrid_2_pre}) is rewritten as
\begin{align}
  &\int_{\Lambda_{\text{A}}} \left\{w(\mathbf{x}) \frac{\partial \tilde{\rho}(\mathbf{x}, t)}{\partial t} - \left(\nabla w(\mathbf{x})\right)^{\text{T}} \left(\mathbf{f}_{\text{A}}(\mathbf{x}) \tilde{\rho}(\mathbf{x}, t) - \mathbf{D}_{\text{A}} \nabla \tilde{\rho}(\mathbf{x}, t)\right)\right\} \mathrm{d}\Lambda  \nonumber\\
  &+ \int_{\Lambda_{\text{B}}} \left\{w(\mathbf{x}) \frac{\partial \tilde{\rho}(\mathbf{x}, t)}{\partial t} - \left(\nabla w(\mathbf{x})\right)^{\text{T}} \left(\mathbf{f}_{\text{B}}(\mathbf{x}) \tilde{\rho}(\mathbf{x}, t) - \mathbf{D}_{\text{B}} \nabla \tilde{\rho}(\mathbf{x}, t)\right)\right\} \mathrm{d}\Lambda  \nonumber\\
  &+ \oint_{\Gamma_{\text{A}}^{\text{d}}+\Gamma_{\text{A}}^{\text{s}}} w(\mathbf{x}) \left(\mathbf{n}_{\Gamma_{\text{A}}}(\mathbf{x})\right)^{\text{T}} \left(\mathbf{f}_{\text{A}}(\mathbf{x}) \tilde{\rho}(\mathbf{x}, t) - \mathbf{D}_{\text{A}} \bm{\upbeta}_{\text{A}}(\mathbf{x}, t)\right) \mathrm{d}\Gamma  \nonumber\\
  &+ \oint_{\Gamma_{\text{B}}^{\text{d}}+\Gamma_{\text{B}}^{\text{s}}} w(\mathbf{x}) \left(\mathbf{n}_{\Gamma_{\text{B}}}(\mathbf{x})\right)^{\text{T}} \left(\mathbf{f}_{\text{B}}(\mathbf{x}) \tilde{\rho}(\mathbf{x}, t) - \mathbf{D}_{\text{B}} \bm{\upbeta}_{\text{B}}(\mathbf{x}, t)\right) \mathrm{d}\Gamma  \nonumber\\
  &= 0,
\label{eq:FPE_2D_hybrid_2_mod}
\end{align}
for which the boundary conditions of Eqs.~(\ref{eq:zero_flux_A})-(\ref{eq:flux_conservation}) are applied. Then, the third and fourth line integrals in the left-hand-side of Eq.~(\ref{eq:FPE_2D_hybrid_2_mod}) along $\Gamma_{\text{A}}^{\text{d}}$ and $\Gamma_{\text{B}}^{\text{d}}$ vanish due to the boundary conditions of Eqs.~(\ref{eq:zero_flux_A}) and (\ref{eq:zero_flux_B}). The remaining part of line integrals along $\Gamma_{\text{A}}^{\text{s}}$ and $\Gamma_{\text{B}}^{\text{s}}$ would be
\begin{align}
  &\int_{\Gamma_{\text{A}}^{\text{s}}} w(\mathbf{x}) \left(\mathbf{n}_{\Gamma_{\text{A}}}(\mathbf{x})\right)^{\text{T}} \left(\mathbf{f}_{\text{A}}(\mathbf{x}) \tilde{\rho}(\mathbf{x}, t) - \mathbf{D}_{\text{A}} \bm{\upbeta}_{\text{A}}(\mathbf{x}, t)\right) \mathrm{d}\Gamma  \nonumber\\
  &+ \int_{\Gamma_{\text{B}}^{\text{s}}} w(\mathbf{x}) \left(\mathbf{n}_{\Gamma_{\text{B}}}(\mathbf{x})\right)^{\text{T}} \left(\mathbf{f}_{\text{B}}(\mathbf{x}) \tilde{\rho}(\mathbf{x}, t) - \mathbf{D}_{\text{B}} \bm{\upbeta}_{\text{B}}(\mathbf{x}, t)\right) \mathrm{d}\Gamma,
\label{eq:FPE_2D_hybrid_2_mod_3rd_4th}
\end{align}
which is also zero by the boundary condition at the switching boundary of Eq.~(\ref{eq:flux_conservation}). In this way, the third and fourth line integrals in the left-hand-side of Eq.~(\ref{eq:FPE_2D_hybrid_2_mod}) vanish, by which the ill-defined term $\bm{\upbeta}_{k}(\mathbf{x},t)$ disappears, and we do not have to compute values of the term $\bm{\upbeta}_{k}(\mathbf{x},t)$.

However, there are more general cases, in which the term $\bm{\upbeta}_{k}(\mathbf{x},t)$ does not disappear. For example, if a stochastic hybrid system includes a mechanism that generates a discontinuity or a ``jump'' from a switching boundary to another switching boundary \cite{Bect_2006,Wang_2020}, the term $\bm{\upbeta}_{k}(\mathbf{x},t)$ does not necessarily disappear. In order to explain such a general situation, let us consider a stochastic hybrid system similar to the one we consider in this study, but with a switching boundary condition modified from Eq.~(\ref{eq:flux_conservation}) to consider discontinuities or jumps at the switching boundary. Specifically, we consider a stochastic hybrid system consisting of a subsystem A and a subsystem B, in which the boundary condition at the domain boundaries of the subsystems A and B are the same as Eqs.~(\ref{eq:zero_flux_A}) and (\ref{eq:zero_flux_B}), but the switching boundary has a mechanism such that a state point arrived at a point $\mathbf{x}$ on the switching boundary of the subsystem A jumps to a point $\mathbf{x}^{\prime}$ on the switching boundary of the subsystem B. Because of the jump, $\mathbf{x}$ is not equal to $\mathbf{x}^{\prime}$ in general. We denote such jump-type switching boundaries of the subsystems A and B as $\Gamma_{\text{A}}^{\text{JS}}$ and $\Gamma_{\text{B}}^{\text{JS}}$, respectively. Using a bijection $\Xi: \Gamma_{\text{A}}^{\text{JS}}\to\Gamma_{\text{B}}^{\text{JS}}$ between $\Gamma_{\text{A}}^{\text{JS}}$ and $\Gamma_{\text{B}}^{\text{JS}}$, we can consider the following boundary condition at the jump-type switching boundary:
\begin{equation}
  \mathbf{j}_{\text{A}}(\mathbf{x},t) \cdot \mathbf{n}_{\Gamma_{\text{A}}}(\mathbf{x}) = \left|\Xi^{\prime}\right| \left(\mathbf{j}_{\text{B}}(\Xi(\mathbf{x}),t) \cdot \left(- \mathbf{n}_{\Gamma_{\text{B}}}(\Xi(\mathbf{x}))\right)\right), \quad \mathbf{x}\in\Gamma_{\text{A}}^{\text{JS}}
\label{eq:flux_conservation_withXi}
\end{equation}
where $\left|\Xi^{\prime}\right|$ is the Jacobian determinant of $\Xi$. In this case, Eq.~(\ref{eq:FPE_2D_hybrid_2_mod_3rd_4th}) with the jump-type switching boundary condition would be
\begin{align}
  &\int_{\Gamma_{\text{A}}^{\text{JS}}} w(\mathbf{x}) \left(- \mathbf{n}_{\Gamma_{\text{B}}}(\Xi(\mathbf{x}))\right)^{\text{T}} \left(\mathbf{f}_{\text{B}}(\Xi(\mathbf{x})) \tilde{\rho}(\Xi(\mathbf{x}), t) - \mathbf{D}_{\text{B}} \bm{\upbeta}_{\text{B}}(\Xi(\mathbf{x}), t)\right) \left|\Xi^{\prime}\right| \mathrm{d}\Gamma  \nonumber\\
  &+ \int_{\Gamma_{\text{B}}^{\text{JS}}} w(\mathbf{x}) \left(\mathbf{n}_{\Gamma_{\text{B}}}(\mathbf{x})\right)^{\text{T}} \left(\mathbf{f}_{\text{B}}(\mathbf{x}) \tilde{\rho}(\mathbf{x}, t) - \mathbf{D}_{\text{B}} \bm{\upbeta}_{\text{B}}(\mathbf{x}, t)\right) \mathrm{d}\Gamma.
\label{eq:FPE_2D_hybrid_2_mod_3rd_4th_2}
\end{align}
Using the integration by substitution for the first line integral of Eq.~(\ref{eq:FPE_2D_hybrid_2_mod_3rd_4th_2}) for an inverse function of the bijection $\Xi$, Eq.~(\ref{eq:FPE_2D_hybrid_2_mod_3rd_4th_2}) can be rewritten as
\begin{equation}
  \int_{\Gamma_{\text{B}}^{\text{JS}}} \left(- w(\Xi^{-1}(\mathbf{x})) + w(\mathbf{x})\right) \left(\mathbf{n}_{\Gamma_{\text{B}}}(\mathbf{x})\right)^{\text{T}} \left(\mathbf{f}_{\text{B}}(\mathbf{x}) \tilde{\rho}(\mathbf{x}, t) - \mathbf{D}_{\text{B}} \bm{\upbeta}_{\text{B}}(\mathbf{x}, t)\right) \mathrm{d}\Gamma.
\label{eq:FPE_2D_hybrid_2_mod_3rd_4th_3}
\end{equation}
Unlike the case with the boundary condition of Eq.~(\ref{eq:flux_conservation}), Eq.~(\ref{eq:FPE_2D_hybrid_2_mod_3rd_4th_3}) cannot be zero, unless the equality $w(\Xi^{-1}(\mathbf{x}))=w(\mathbf{x})$ is satisfied at any point on $\Gamma_{\text{B}}^{\text{JS}}$. In other words, the third and fourth line integrals in the left-hand-side of Eq.~(\ref{eq:FPE_2D_hybrid_2_mod}) do not vanish and the ill-defined term $\bm{\upbeta}_{k}(\mathbf{x},t)$ in this general case does not disappear. Thus, although the intermittent control model that we consider in this study exhibits a simpler switching behavior with the $C^{0}$-continuity, we develop a numerical recipe extendable to more general stochastic hybrid systems exhibiting discontinuous jumps, which requires intricate evaluations of the values of $\bm{\upbeta}_{k}(\mathbf{x},t)$.

In summary, we develop a numerical recipe to derive a Markov-chain model of stochastic hybrid systems with $C^{0}$-continuity at the switching boundary, which can be extended to more general systems with discontinuous jumps at the switching boundary. To this end, we employ a weak form of the switched FP equations, in which we keep the line integrals until we derive the finite element equation later in this section. We begin our analysis by the use of the divergence theorem for Eq.~(\ref{eq:DivTheorem_D_nabla_rho}) to rewrite Eq.~(\ref{eq:FPE_2D_hybrid_2_pre}) as
\begin{align}
  &\int_{\Lambda_{\text{A}}} w(\mathbf{x}) \left\{\frac{\partial \tilde{\rho}(\mathbf{x}, t)}{\partial t} + \nabla^{\text{T}} \left(\mathbf{f}_{\text{A}}(\mathbf{x}) \tilde{\rho}(\mathbf{x}, t)\right)\right\} \mathrm{d}\Lambda + \int_{\Lambda_{\text{A}}} \left(\nabla w (\mathbf{x})\right)^{\text{T}} \mathbf{D}_{\text{A}} \nabla \tilde{\rho}(\mathbf{x}, t)\mathrm{d}\Lambda  \nonumber\\
  &+ \int_{\Lambda_{\text{B}}} w(\mathbf{x}) \left\{\frac{\partial \tilde{\rho}(\mathbf{x}, t)}{\partial t} + \nabla^{\text{T}} \left(\mathbf{f}_{\text{B}}(\mathbf{x}) \tilde{\rho}(\mathbf{x}, t)\right)\right\} \mathrm{d}\Lambda + \int_{\Lambda_{\text{B}}} \left(\nabla w(\mathbf{x})\right)^{\text{T}} \mathbf{D}_{\text{B}} \nabla \tilde{\rho}(\mathbf{x}, t)\mathrm{d}\Lambda  \nonumber\\
  &- \oint_{\Gamma_{\text{A}}^{\text{d}}+\Gamma_{\text{A}}^{\text{s}}} w(\mathbf{x}) \left(\mathbf{n}_{\Gamma_{\text{A}}}(\mathbf{x})\right)^{\text{T}} \mathbf{D}_{\text{A}} \bm{\upbeta}_{\text{A}}(\mathbf{x}, t) \mathrm{d}\Gamma - \oint_{\Gamma_{\text{B}}^{\text{d}}+\Gamma_{\text{B}}^{\text{s}}} w(\mathbf{x}) \left(\mathbf{n}_{\Gamma_{\text{B}}}(\mathbf{x})\right)^{\text{T}} \mathbf{D}_{\text{B}} \bm{\upbeta}_{\text{B}}(\mathbf{x}, t)\mathrm{d}\Gamma  \nonumber\\
  &=0.
\label{eq:FPE_2D_hybrid_2}
\end{align}
In this sequel, values of the ill-defined term $\bm{\upbeta}_{k}(\mathbf{x},t)$ for $k\in\{\text{A},\text{B}\}$ are determined by using the boundary conditions satisfied by spatial derivatives of the true solution $\rho(\mathbf{x},t)$ at the boundary, without using spatial derivatives of $\tilde{\rho}(\mathbf{x},t)$.

\subsection{Derivation of finite element equations}
For each region of $k\in\{\text{A},\text{B}\}$, $\Lambda_{k}$ is divided into finite elements, referred to as $\Lambda_{k,e_{k}}$ representing the $e_{k}$-th element of $\Lambda_{k}$, where $e_{k}$($=1,2,\ldots,E_{k}$) represents the serial number assigned to the finite elements of $\Lambda_{k}$ with no duplications with $E_{k}$ being the total number of the finite elements. Each $\Lambda_{k,e_{k}}$ is defined by using nodes located at $\mathbf{x}_{k,n_{k}}$ in $\Lambda_{k}$, where $n_{k}$($=1,2,\ldots,N_{k}$) is the serial number assigned to the nodes in $\Lambda_{k}$ with no duplications. We assume that all nodes located on the switching boundary separating $\Lambda_{\text{A}}$ and $\Lambda_{\text{B}}$ are shared commonly, where the number of such nodes on the switching boundary is $N^{\text{s}}$.

Any approximate solution $\tilde{\rho}(\mathbf{x},t)$ over the entire domain $\Lambda$ is represented by the sum of $(E_{\text{A}}+E_{\text{B}})$ local approximate solutions $\tilde{\rho}_{k,e_{k}}(\mathbf{x},t)$ obtained for each compact finite element $\Lambda_{k,e_{k}}$. In this study, we consider triangular finite elements, each of which is defined by three nodes. Then, each local approximate solution is represented by a linear sum of three trigonal pyramid shaped basis functions with three node-values (coefficients) for each basis function. Although we have defined the node position as $\mathbf{x}_{k,n_{k}}$, we alternatively represent the node position as $\mathbf{x}_{k,e_{k},n}$ in a way such that it can specify the finite element $\Lambda_{k,e_{k}}$, where $n$($=1,2,3$) is the number to represent one of the three nodes for the $e_{k}$-th finite element. Note that, if a node $\mathbf{x}_{k,n_{k}}$ is for the $n$-th vertex ($n=1,2,3$) of $\Lambda_{k,e_{k}}$ and also for the $n^{\prime}$-th vertex ($n^{\prime}=1,2,3$) of $\Lambda_{k,e_{k}^{\prime}}$ that shares a side with an adjacent triangular element $\Lambda_{k,e_{k}}$, then we have $\mathbf{x}_{k,n_{k}}=\mathbf{x}_{k,e_{k},n}=\mathbf{x}_{k,e_{k}^{\prime},n^{\prime}}$, i.e., distinct three representations for an identical node position.

We denote the three trigonal pyramid shaped basis functions for $\mathbf{x}\in\Lambda_{k,e_{k}}$ by $\phi_{k,e_{k},n}(\mathbf{x})$ with their node values $p_{k,e_{k},n}(t)$ at the nodes $\mathbf{x}_{k,e_{k},n}$ for $n=1$, $2$ and $3$. Each of the three basis functions $\phi_{k,e_{k},n}(\mathbf{x})$ for $n=1$, $2$ and $3$ is a map $\phi_{k,e_{k},n}:\Lambda_{k}\rightarrow\mathbb{R}$, which satisfies $\phi_{k,e_{k},n}(\mathbf{x})=0$ for any $\mathbf{x}\in\Lambda_{k}$, except for $\mathbf{x}\in\Lambda_{k,e_{k}}$. The basis function $\phi_{k,e_{k},n}(\mathbf{x})$ for a specific $n$-th vertex of the triangular element $\Lambda_{k,e_{k}}$ takes the unity value at $\mathbf{x}_{k,e_{k},n}$, and zero values at the remaining two vertices, which forms the trigonal pyramid with its undersurface of $\Lambda_{k,e_{k}}$. Then, an approximate solution $\tilde{\rho}(\mathbf{x},t)$ is expressed by the sum of $(E_{\text{A}}+E_{\text{B}})$ local approximate solutions $\tilde{\rho}_{k,e_{k}}(\mathbf{x},t)$ as
\begin{equation}
  \tilde{\rho}(\mathbf{x},t) = \sum_{k\in\{\text{A},\text{B}\}} \left(\sum_{e_{k}=1}^{E_{k}} \tilde{\rho}_{k,e_{k}}(\mathbf{x},t)\right) = \sum_{k\in\{\text{A},\text{B}\}} \left(\sum_{e_{k}=1}^{E_{k}}\sum_{n=1}^{3} p_{k,e_{k},n}(t) \phi_{k,e_{k},n}(\mathbf{x})\right).
\label{eq:apprx_pdf_2D_hybrid}
\end{equation}
In this study, we employ the Galerkin method, in which a linear combination of the basis functions $\phi_{k,e_{k},n}(\mathbf{x})$ is used also for the weight functions $w(\mathbf{x})$, i.e.,
\begin{equation}
  w(\mathbf{x}) = \sum_{k\in\{\text{A},\text{B}\}} \left(\sum_{e_{k}=1}^{E_{k}} \sum_{n=1}^{3} w_{k,e_{k},n} \phi_{k,e_{k},n}(\mathbf{x})\right),
\label{eq:weightF_2D_hybrid}
\end{equation}
where $w_{k,e_{k},n}$ is the constant value of the weight function (weight coefficient) at the node $\mathbf{x}_{k,e_{k},n}$.

Determining the unknown node values $p_{k,e_{k},n}(t)$ for all nodes $\mathbf{x}_{k,e_{k},n}$ for $k\in\{\text{A},\text{B}\}$ as functions of time $t$ is equivalent to obtaining the approximate solution of the hybrid FP equations in Eqs.~(\ref{eq:FPE_2D_hybrid_A}) and (\ref{eq:FPE_2D_hybrid_B}). Cumbersomeness in determining $p_{k,e_{k},n}(t)$ for Eq.~(\ref{eq:apprx_pdf_2D_hybrid}) arises from the fact that the representation of node values $p_{k,e_{k},n}(t)$ for all nodes $\mathbf{x}_{k,e_{k},n}$ for $k\in\{\text{A},\text{B}\}$ are duplicated as discussed above, while such a duplicated indexing based on the finite element number $e_{k}$ is easy to enumerate the finite elements and their nodes in a program code and thus preferable to represent expansion of the PDF using the basis functions. We therefore introduce alternative notation of a node value $p_{k,n_{k}}(t)$ with the index of the non-duplicated serial node number $n_{k}$ for the node $\mathbf{x}_{k,n_{k}}$, i.e., the node value $p_{k,e_{k},n}(t)$ is alternatively denoted by $p_{k,n_{k}}(t)$ for the identical but differently represented node $\mathbf{x}_{k,e_{k},n}=\mathbf{x}_{k,n_{k}}$. Similarly, the weight coefficient $w_{k,e_{k},n}$ at $\mathbf{x}_{k,e_{k},n}$ is denoted alternatively by $w_{k,n_{k}}$ for the identical but differently represented node $\mathbf{x}_{k,n_{k}}=\mathbf{x}_{k,e_{k},n}$. Moreover, we define vectors of the node values and the weight coefficients indexed by the serial number $n_{k}$ in the region $\Lambda_{k}$ for $k\in\{\text{A},\text{B}\}$ as
\begin{align}
  \mathbf{p}_{k}(t) &= (p_{k,1}(t),\ldots,p_{k,n_{k}}(t),\cdots,p_{k,N_{k}}(t))^{\text{T}},
\label{eq:p_k_def}\\
  \mathbf{w}_{k} &= (w_{k,1},\ldots,w_{k,n_{k}},\cdots,w_{k,N_{k}})^{\text{T}},
\label{eq:w_k_def}
\end{align}
which do not include duplications in counting the nodes within each region. Once again, determining the unknown vectors $\mathbf{p}_{\text{A}}(t)$ and $\mathbf{p}_{\text{B}}(t)$ is equivalent to obtaining the approximate solution of the hybrid FP equations in Eqs.~(\ref{eq:FPE_2D_hybrid_A}) and (\ref{eq:FPE_2D_hybrid_B}). Note that the nodes located on the switching boundary are shared by the regions $\Lambda_{\text{A}}$ and $\Lambda_{\text{B}}$, which causes another type of duplications in $\mathbf{p}_{\text{A}}(t)$ and $\mathbf{p}_{\text{B}}(t)$. This type of duplication will be taken care of appropriately later.

In this sequel, we rewrite Eq.~(\ref{eq:apprx_pdf_2D_hybrid}) using $\mathbf{p}_{\text{A}}(t)$ and $\mathbf{p}_{\text{B}}(t)$. We begin by defining the operator
\begin{equation}
  \begin{array}{cl}
    \multirow{2}{*}{$\bm{\mathcal{L}}_{\mathcal{S}}^{\mathcal{U}}:$}  &  \text{node (or weight) value vectors for a set of nodes in }\mathcal{U}  \\
    & \hspace{1cm}\rightarrow \text{node (or weight) value vectors for a set of nodes in }\mathcal{S}
  \end{array}
\label{eq:L_def}
\end{equation}
that extracts node (or weight) values of all nodes included in a set of nodes, referred to as $\mathcal{S}$, from a set of nodes included in a larger set of nodes, referred to as $\mathcal{U}\supset\mathcal{S}$. For the definition of $\bm{\mathcal{L}}_{\mathcal{S}}^{\mathcal{U}}$, the nodes in $\mathcal{U}$ should be indexed uniquely. For example, in the case of $\mathcal{U}=\Lambda_{k}$, the non-redundant index $\{k,n_{k}\}$ can specify the node $\mathbf{x}_{k,n_{k}}$ uniquely. The node sets of $\mathcal{U}$ and $\mathcal{S}$ will be defined in a variety of ways later, depending on the purpose of the node extraction. Specification of a single node to be extracted its node value will be made by performing an inner product operation between $\boldsymbol{\ell}_{\mathbf{x}_{\mathcal{S}_{\sharp}}}^{\mathcal{U}}$ and the node or the weight value vector, where the dimension of $\boldsymbol{\ell}_{\mathbf{x}_{\mathcal{S}_{\sharp}}}^{\mathcal{U}}$ is equal to the number of the nodes included in $\mathcal{U}$, and $\mathcal{S}_{\sharp}$ is the index of a node in the set of nodes in $\mathcal{S}$, i.e., $\mathbf{x}_{\mathcal{S}_{\sharp}}\in \mathcal{S}$. $\boldsymbol{\ell}_{\mathbf{x}_{\mathcal{S}_{\sharp}}}^{\mathcal{U}}$ is the column vector with its dimension equal to the number of nodes in $\mathcal{U}$ and all components are zero, except a specific component with a unique index number to be extracted is unity. In this way, the node at $\mathbf{x}_{\mathcal{S}_{\sharp}}$, which might be indexed non-uniquely because of a duplicated indexing for $\mathcal{S}_{\sharp}$, is uniquely specified and extracted by using $\boldsymbol{\ell}_{\mathbf{x}_{\mathcal{S}_{\sharp}}}^{\mathcal{U}}$.

For the first use of the operator $\bm{\mathcal{L}}_{\mathcal{S}}^{\mathcal{U}}$, we consider the case with $\mathcal{U}=\Lambda_{k}$ that includes $N_{k}$ nodes and $\mathcal{S}=\Lambda_{k,e_{k}}$ that includes three nodes. In this case, $\mathcal{S}_{\sharp}=\{k,e_{k},n\}$ for $n=1,2,3$. More specifically, the operator $\bm{\mathcal{L}}_{\Lambda_{k,e_{k}}}^{\Lambda_{k}}$ extracts three node values $p_{k,e_{k},1}(t)$, $p_{k,e_{k},2}(t)$ and $p_{k,e_{k},3}(t)$ for a specific finite element $\Lambda_{k,e_{k}}$ from the node value vector $\mathbf{p}_{k}$ of the region $\Lambda_{k}$. This is done by using $\boldsymbol{\ell}_{\mathbf{x}_{k,e_{k},n}}^{\Lambda_{k}}$ for $n=1,2,3$, according to the equality $\mathbf{x}_{k,n_{k}}=\mathbf{x}_{k,e_{k},n}$. That is,
\begin{align}
  \left(\begin{array}{c}
    p_{k,e_{k},1}(t)  \\
    p_{k,e_{k},2}(t)  \\
    p_{k,e_{k},3}(t)
  \end{array}\right) &= \left(\begin{array}{c}
    \left(\boldsymbol{\ell}_{\mathbf{x}_{k,e_{k},1}}^{\Lambda_{k}}\right)^{\text{T}}  \\
    \left(\boldsymbol{\ell}_{\mathbf{x}_{k,e_{k},2}}^{\Lambda_{k}}\right)^{\text{T}}  \\
    \left(\boldsymbol{\ell}_{\mathbf{x}_{k,e_{k},3}}^{\Lambda_{k}}\right)^{\text{T}}
  \end{array}\right) \mathbf{p}_{k}(t) \nonumber\\
  &\equiv \bm{\mathcal{L}}_{\Lambda_{k,e_{k}}}^{\Lambda_{k}} \mathbf{p}_{k}(t).
\label{eq:def_Lp}
\end{align}
In this case, the $N_{k}$-dimensional column vector $\boldsymbol{\ell}_{\mathbf{x}_{k,e_{k},1}}^{\Lambda_{k}}$ has a non-zero unity value at the $n_{k}$-th component, since we extract the node value of the node $\mathbf{x}_{k,e_{k},1}$ that is uniquely specified by the identity of $\mathbf{x}_{k,n_{k}}=\mathbf{x}_{k,e_{k},1}$. Similarly, the $N_{k}$-dimensional column vectors $\boldsymbol{\ell}_{\mathbf{x}_{k,e_{k},2}}^{\Lambda_{k}}$ and $\boldsymbol{\ell}_{\mathbf{x}_{k,e_{k},3}}^{\Lambda_{k}}$ have non-zero unity values at the $n_{k}^{\prime}$-th and the $n_{k}^{\prime\prime}$-th components, respectively, for extracting the node values of the nodes $\mathbf{x}_{k,e_{k},2}$ and $\mathbf{x}_{k,e_{k},3}$ that are uniquely specified by the identities of $\mathbf{x}_{k,n_{k}^{\prime}}=\mathbf{x}_{k,e_{k},2}$ and $\mathbf{x}_{k,n_{k}^{\prime\prime}}=\mathbf{x}_{k,e_{k},3}$. The resultant operator $\bm{\mathcal{L}}_{\Lambda_{k,e_{k}}}^{\Lambda_{k}}$ is represented by the $3 \times N_{k}$ matrix. 

Using $\mathbf{p}_{k}(t)$ and $\bm{\mathcal{L}}_{\Lambda_{k,e_{k}}}^{\Lambda_{k}}$, the local estimate solution $\tilde{\rho}_{k,e_{k}}(\mathbf{x},t)$ in Eq.~(\ref{eq:apprx_pdf_2D_hybrid}) can be rewritten as
\begin{equation}
  \tilde{\rho}_{k,e_{k}}(\mathbf{x},t) = \sum_{n=1}^{3} p_{k,e_{k},n}(t) \phi_{k,e_{k},n}(\mathbf{x}) = \left(\bm{\upphi}_{k,e_{k}}(\mathbf{x})\right)^{\text{T}} \bm{\mathcal{L}}_{\Lambda_{k,e_{k}}}^{\Lambda_{k}} \mathbf{p}_{k}(t),
\end{equation}
where $\bm{\upphi}_{k,e_{k}}(\mathbf{x})$ is the column vector, whose components are the three basis functions, i.e.,
\begin{equation}
  \bm{\upphi}_{k,e_{k}}(\mathbf{x})=(\phi_{k,e_{k},1}(\mathbf{x}), \phi_{k,e_{k},2}(\mathbf{x}), \phi_{k,e_{k},3}(\mathbf{x}))^{\text{T}}
\end{equation}
for the element $\Lambda_{k,e_{k}}$. Then, Eq.~(\ref{eq:apprx_pdf_2D_hybrid}) is rewritten as
\begin{equation}
  \tilde{\rho}(\mathbf{x},t) = \sum_{k\in\{\text{A},\text{B}\}} \left(\sum_{e_{k}=1}^{E_{k}} \left(\bm{\upphi}_{k,e_{k}}(\mathbf{x})\right)^{\text{T}} \bm{\mathcal{L}}_{\Lambda_{k,e_{k}}}^{\Lambda_{k}} \mathbf{p}_{k}(t) \right).
\label{eq:def_Phi_L_p}
\end{equation}
Similarly, using the operator $\bm{\mathcal{L}}_{\Lambda_{k,e_{k}}}^{\Lambda_{k}}$, the weight value $w_{k,e_{k},n}$ is extracted from $\mathbf{w}_{k}$ as
\begin{align}
  \left(\begin{array}{c}
    w_{k,e_{k},1}  \\
    w_{k,e_{k},2}  \\
    w_{k,e_{k},3}
  \end{array}\right) &= \bm{\mathcal{L}}_{\Lambda_{k,e_{k}}}^{\Lambda_{k}} \mathbf{w}_{k},
\label{eq:def_Lw}
\end{align}
and Eq.~(\ref{eq:weightF_2D_hybrid}) is rewritten as
\begin{equation}
  w(\mathbf{x}) = \sum_{k\in\{\text{A},\text{B}\}} \left(\sum_{e_{k}=1}^{E_{k}} \left(\bm{\upphi}_{k,e_{k}}(\mathbf{x})\right)^{\text{T}} \bm{\mathcal{L}}_{\Lambda_{k,e_{k}}}^{\Lambda_{k}} \mathbf{w}_{k} \right).
\label{eq:def_Phi_L_w}
\end{equation}

By substituting Eqs.~(\ref{eq:apprx_pdf_2D_hybrid}) and (\ref{eq:weightF_2D_hybrid}) into Eq.~(\ref{eq:FPE_2D_hybrid_2}), and rearranging Eq.~(\ref{eq:FPE_2D_hybrid_2}) using Eqs.~(\ref{eq:def_Phi_L_p}) and (\ref{eq:def_Phi_L_w}), we have
\begin{align}
  &\sum_{k\in\{\text{A},\text{B}\}} \left\{\sum_{e_{k}=1}^{E_{k}} \int_{\Lambda_{k,e_{k}}} \left(\left(\bm{\upphi}_{k,e_{k}}(\mathbf{x})\right)^{\text{T}} \bm{\mathcal{L}}_{\Lambda_{k,e_{k}}}^{\Lambda_{k}} \mathbf{w}_{k}\right) \frac{\partial}{\partial t} \left(\left(\bm{\upphi}_{k,e_{k}}(\mathbf{x})\right)^{\text{T}} \bm{\mathcal{L}}_{\Lambda_{k,e_{k}}}^{\Lambda_{k}} \mathbf{p}_{k}(t)\right)\mathrm{d}\Lambda\right.  \nonumber\\
  &\hspace{1.5cm}+ \sum_{e_{k}=1}^{E_{k}} \int_{\Lambda_{k,e_{k}}} \left(\left(\bm{\upphi}_{k,e_{k}}(\mathbf{x})\right)^{\text{T}} \bm{\mathcal{L}}_{\Lambda_{k,e_{k}}}^{\Lambda_{k}} \mathbf{w}_{k}\right) \nabla^{\text{T}} \left(\mathbf{f}_{k}(\mathbf{x}) \left(\left(\bm{\upphi}_{k,e_{k}}(\mathbf{x})\right)^{\text{T}} \bm{\mathcal{L}}_{\Lambda_{k,e_{k}}}^{\Lambda_{k}} \mathbf{p}_{k}(t)\right)\right)\mathrm{d}\Lambda  \nonumber\\
  &\hspace{1.5cm}\left.+ \sum_{e_{k}=1}^{E_{k}} \int_{\Lambda_{k,e_{k}}} \left(\nabla \left(\bm{\upphi}_{k,e_{k}}(\mathbf{x})\right)^{\text{T}} \bm{\mathcal{L}}_{\Lambda_{k,e_{k}}}^{\Lambda_{k}} \mathbf{w}_{k}\right)^{\text{T}} \mathbf{D}_{k} \left(\nabla \left(\bm{\upphi}_{k,e_{k}}(\mathbf{x})\right)^{\text{T}} \bm{\mathcal{L}}_{\Lambda_{k,e_{k}}}^{\Lambda_{k}} \mathbf{p}_{k}(t)\right)\mathrm{d}\Lambda\right\}  \nonumber\\
  &= \sum_{k\in\{\text{A},\text{B}\}}\left\{\sum_{e_{k}=1}^{E_{k}} \int_{\Gamma_{k}^{\text{d}}} \left(\left(\bm{\upphi}_{k,e_{k}}(\mathbf{x})\right)^{\text{T}} \bm{\mathcal{L}}_{\Lambda_{k,e_{k}}}^{\Lambda_{k}} \mathbf{w}_{k}\right) \left(\mathbf{n}_{\Gamma_{k}}(\mathbf{x})\right)^{\text{T}} \mathbf{D}_{k} \bm{\upbeta}_{k}(\mathbf{x},t)\mathrm{d}\Gamma\right.  \nonumber\\
  &\hspace{2cm}\left.+ \sum_{e_{k}=1}^{E_{k}} \int_{\Gamma_{k}^{\text{s}}} \left(\left(\bm{\upphi}_{k,e_{k}}(\mathbf{x})\right)^{\text{T}} \bm{\mathcal{L}}_{\Lambda_{k,e_{k}}}^{\Lambda_{k}} \mathbf{w}_{k}\right) \left(\mathbf{n}_{\Gamma_{k}}(\mathbf{x})\right)^{\text{T}} \mathbf{D}_{k} \bm{\upbeta}_{k}(\mathbf{x},t)\mathrm{d}\Gamma\right\}.
\label{eq:FPE_2D_hybrid_4}
\end{align}
The right-hand-side of Eq.~(\ref{eq:FPE_2D_hybrid_4}) is the fifth and sixth terms of Eq.~(\ref{eq:FPE_2D_hybrid_2}) as the line integrals along the boundary $\Gamma_{k}^{\text{d}}+\Gamma_{k}^{\text{s}}$ for $k\in\{\text{A},\text{B}\}$.

We further rewrite the right-hand-side of Eq.~(\ref{eq:FPE_2D_hybrid_4}) as follows:
\begin{align}
  &\sum_{k\in\{\text{A},\text{B}\}} \left\{\sum_{e_{k}^{\text{d}}} \sum_{h_{k,e_{k}^{\text{d}}}^{\text{d}}=1}^{H_{k,e_{k}^{\text{d}}}^{\text{d}}} \int_{\Gamma_{k,e_{k}^{\text{d}},h_{k,e_{k}^{\text{d}}}^{\text{d}}}^{\text{d}}} \left(\left(\bm{\upphi}_{k,e_{k}^{\text{d}}}(\mathbf{x})\right)^{\text{T}} \bm{\mathcal{L}}_{\Lambda_{k,e_{k}^{\text{d}}}}^{\Lambda_{k}} \mathbf{w}_{k}\right) \left(\mathbf{n}_{\Gamma_{k}}(\mathbf{x})\right)^{\text{T}} \mathbf{D}_{k} \bm{\upbeta}_{k}(\mathbf{x},t)\mathrm{d}\Gamma\right.  \nonumber\\
  &\hspace{1.5cm}\left.+ \sum_{e_{k}^{\text{s}}} \sum_{h_{k,e_{k}^{\text{s}}}^{\text{s}}=1}^{H_{k,e_{k}^{\text{s}}}^{\text{s}}} \int_{\Gamma_{k,e_{k}^{\text{s}},h_{k,e_{k}^{\text{s}}}^{\text{s}}}^{\text{s}}} \left(\left(\bm{\upphi}_{k,e_{k}^{\text{s}}}(\mathbf{x})\right)^{\text{T}} \bm{\mathcal{L}}_{\Lambda_{k,e_{k}^{\text{s}}}}^{\Lambda_{k}} \mathbf{w}_{k}\right) \left(\mathbf{n}_{\Gamma_{k}}(\mathbf{x})\right)^{\text{T}} \mathbf{D}_{k} \bm{\upbeta}_{k}(\mathbf{x},t)\mathrm{d}\Gamma\right\},
\label{eq:FPE_2D_hybrid_4_right}
\end{align}
where new indices $e_{k}^{\text{d}}$ and $e_{k}^{\text{s}}$ represent the index numbers for finite elements that provide at least one of the three edges of the triangular elements, respectively, to the domain boundary $\Gamma_{k}^{\text{d}}$ and to the switching boundary $\Gamma_{k}^{\text{s}}$ of the region $\Lambda_{k}$. The use of index $e_{k}^{\text{d}}$ is beneficial, because the first line integral in the right-hand-side of Eq.~(\ref{eq:FPE_2D_hybrid_4}) for each finite element that does not provide any edge to the domain boundary is zero, and summation of the line integrals for all elements in region $\Lambda_{k}$ is equal to summation of the line integrals only for the elements, denoted by $e_{k}^{\text{d}}$, that provide edges to the domain boundary. Similarly, the use of $e_{k}^{\text{s}}$ is beneficial, because the second line integral in the right-hand-side of Eq.~(\ref{eq:FPE_2D_hybrid_4}) for each finite element that does not provide any edge to the switching boundary is zero, and summation of the line integrals for all elements in region $\Lambda_{k}$ is equal to summation of the line integrals only for the elements, denoted by $e_{k}^{\text{s}}$, that provide edges to the switching boundary. For the elements at the domain boundary, the path $\Gamma_{k,e_{k}^{\text{d}},h_{k,e_{k}^{\text{d}}}^{\text{d}}}^{\text{d}}$ of the first line integral is the $h_{k,e_{k}^{\text{d}}}^{\text{d}}$-th edge of the element $\Lambda_{k,e_{k}^{\text{d}}}$ that provides at least one edge to the domain boundary $\Gamma_{k}^{\text{d}}$ of the region $\Lambda_{k}$, where the index $h_{k,e_{k}^{\text{d}}}^{\text{d}}$ for such an edge runs through $h_{k,e_{k}^{\text{d}}}^{\text{d}}=1,\ldots,H_{k,e_{k}^{\text{d}}}^{\text{d}}\le 3$ with $H_{k,e_{k}^{\text{d}}}^{\text{d}}$ being the total number of edges contributed from three edges of the element $\Lambda_{k,e_{k}^{\text{d}}}$ along the domain boundary $\Gamma_{k}^{\text{d}}$ to the domain boundary. Similarly, the path $\Gamma_{k,e_{k}^{\text{s}},h_{k,e_{k}^{\text{s}}}^{\text{s}}}^{\text{s}}$ of the second line integral of Eq.~(\ref{eq:FPE_2D_hybrid_4_right}) is the $h_{k,e_{k}^{\text{s}}}^{\text{s}}$-th edge of the element $\Lambda_{k,e_{k}^{\text{s}}}$ that provides at least one edge to the switching boundary $\Gamma_{k}^{\text{s}}$ of the region $\Lambda_{k}$, where the index $h_{k,e_{k}^{\text{s}}}^{\text{s}}$ for such an edge runs through $h_{k,e_{k}^{\text{s}}}^{\text{s}}=1,\ldots,H_{k,e_{k}^{\text{s}}}^{\text{s}}\le 3$ with $H_{k,e_{k}^{\text{s}}}^{\text{s}}$ being the total number of edges contributed from three edges of the element $\Lambda_{k,e_{k}^{\text{s}}}$ along the switching boundary $\Gamma_{k}^{\text{s}}$ to the switching boundary.

Using the equality
\begin{equation}
  \left(\bm{\upphi}_{k,e_{k}}(\mathbf{x})\right)^{\text{T}} \bm{\mathcal{L}}_{\Lambda_{k,e_{k}}}^{\Lambda_{k}} \mathbf{w}_{k} = \left(\left(\bm{\upphi}_{k,e_{k}}(\mathbf{x})\right)^{\text{T}} \bm{\mathcal{L}}_{\Lambda_{k,e_{k}}}^{\Lambda_{k}} \mathbf{w}_{k}\right)^{\text{T}},
\end{equation}
which holds by the fact that a transpose of a scalar value leaves its value unchanged, Eq.~(\ref{eq:FPE_2D_hybrid_4}) with the right-hand-side replaced by Eq.~(\ref{eq:FPE_2D_hybrid_4_right}) can be rearranged as follows:
\begin{align}
  &\left(\begin{array}{c}
    \mathbf{w}_{\text{A}}  \\
    \mathbf{w}_{\text{B}}
  \end{array}\right)^{\text{T}}  \left\{\left(\begin{array}{cc}
    \mathbf{M}_{\text{A}}  &       \mathbf{0}  \\
         \mathbf{0}        &  \mathbf{M}_{\text{B}}
  \end{array}\right) \frac{\partial}{\partial t} \left(\begin{array}{c}
    \mathbf{p}_{\text{A}}(t)  \\
    \mathbf{p}_{\text{B}}(t)
  \end{array}\right) + \left(\!\begin{array}{cc}
    \mathbf{K}_{\text{A}}  &       \mathbf{0}  \\
         \mathbf{0}        &  \mathbf{K}_{\text{B}}
  \end{array}\right) \left(\begin{array}{c}
    \mathbf{p}_{\text{A}}(t)  \\
    \mathbf{p}_{\text{B}}(t)
  \end{array}\right)\right\}  \nonumber\\
  &= \left(\begin{array}{c}
    \mathbf{w}_{\text{A}}  \\
    \mathbf{w}_{\text{B}}
  \end{array}\right)^{\text{T}} \left(\begin{array}{c}
    \mathbf{b}_{\text{A}}^{\text{d}}(t) + \mathbf{b}_{\text{A}}^{\text{s}}(t)  \\
    \mathbf{b}_{\text{B}}^{\text{d}}(t) + \mathbf{b}_{\text{B}}^{\text{s}}(t)
  \end{array}\right),
\label{eq:appFPE_wf_MatrixVector_2D_hybrid}
\end{align}
where
\begin{align}
  \mathbf{M}_{k} &= \sum_{e_{k}=1}^{E_{k}} \left(\bm{\mathcal{L}}_{\Lambda_{k,e_{k}}}^{\Lambda_{k}}\right)^{\text{T}} \int_{\Lambda_{k,e_{k}}} \bm{\upphi}_{k,e_{k}}(\mathbf{x}) \left(\bm{\upphi}_{k,e_{k}}(\mathbf{x})\right)^{\text{T}}\mathrm{d}\Lambda \bm{\mathcal{L}}_{\Lambda_{k,e_{k}}}^{\Lambda_{k}},
\label{eq:def_M_2D_hybrid}\\
  \mathbf{K}_{k} &= \sum_{e_{k}=1}^{E_{k}} \left(\bm{\mathcal{L}}_{\Lambda_{k,e_{k}}}^{\Lambda_{k}}\right)^{\text{T}} \int_{\Lambda_{k,e_{k}}} \left\{\bm{\upphi}_{k,e_{k}}(\mathbf{x}) \nabla^{\text{T}} \left(\mathbf{f}_{k}(\mathbf{x}) \left(\bm{\upphi}_{k,e_{k}}(\mathbf{x})\right)^{\text{T}}\right)  \right.\nonumber\\
  &\left.\hspace{4.5cm} + \left(\nabla \left(\bm{\upphi}_{k,e_{k}}(\mathbf{x})\right)^{\text{T}}\right)^{\text{T}} \mathbf{D}_{k} \left(\nabla \left(\bm{\upphi}_{k,e_{k}}(\mathbf{x})\right)^{\text{T}}\right)\right\}\mathrm{d}\Lambda \bm{\mathcal{L}}_{\Lambda_{k,e_{k}}}^{\Lambda_{k}},
\label{eq:def_K_2D_hybrid}\\
  \mathbf{b}_{k}^{\text{d}}(t) &= \sum_{e_{k}^{\text{d}}} \sum_{h_{k,e_{k}^{\text{d}}}^{\text{d}}=1}^{H_{k,e_{k}^{\text{d}}}^{\text{d}}} \left(\bm{\mathcal{L}}_{\Lambda_{k,e_{k}^{\text{d}}}}^{\Lambda_{k}}\right)^{\text{T}} \int_{\Gamma_{k,e_{k}^{\text{d}},h_{k,e_{k}^{\text{d}}}^{\text{d}}}^{\text{d}}} \hspace{-5mm} \bm{\upphi}_{k,e_{k}^{\text{d}}}(\mathbf{x}) \left(\mathbf{n}_{\Gamma_{k}}(\mathbf{x})\right)^{\text{T}} \mathbf{D}_{k} \bm{\upbeta}_{k}(\mathbf{x},t)\mathrm{d}\Gamma,
\label{eq:def_bd_2D_hybrid}\\
  \mathbf{b}_{k}^{\text{s}}(t) &= \sum_{e_{k}^{\text{s}}} \sum_{h_{k,e_{k}^{\text{s}}}^{\text{s}}=1}^{H_{k,e_{k}^{\text{s}}}^{\text{s}}} \left(\bm{\mathcal{L}}_{\Lambda_{k,e_{k}^{\text{s}}}}^{\Lambda_{k}}\right)^{\text{T}} \int_{\Gamma_{k,e_{k}^{\text{s}},h_{k,e_{k}^{\text{s}}}^{\text{s}}}^{\text{s}}} \hspace{-5mm} \bm{\upphi}_{k,e_{k}^{\text{s}}}(\mathbf{x}) \left(\mathbf{n}_{\Gamma_{k}}(\mathbf{x})\right)^{\text{T}} \mathbf{D}_{k} \bm{\upbeta}_{k}(\mathbf{x},t)\mathrm{d}\Gamma.
\label{eq:def_bs_2D_hybrid}
\end{align}
$\mathbf{M}_{k}$ and $\mathbf{K}_{k}$ are the $N_{k} \times N_{k}$ constant matrices that can be computed using $\bm{\mathcal{L}}_{\Lambda_{k,e_{k}}}^{\Lambda_{k}}$, $\bm{\upphi}_{k,e_{k}}(\mathbf{x})$, $\mathbf{f}_{k}(\mathbf{x})$, and $\mathbf{D}_{k}$. $\mathbf{b}_{k}^{\text{d}}(t)$ and $\mathbf{b}_{k}^{\text{s}}(t)$ are the $N_{k}$ dimensional column vectors that are represented by using the formal spatial derivative $\bm{\upbeta}_{k}(\mathbf{x},t)$ of the approximate solution $\tilde{\rho}(\mathbf{x},t)$ at $\mathbf{x}$ on the boundary. Most components of $\mathbf{b}_{k}^{\text{d}}(t)$ and $\mathbf{b}_{k}^{\text{s}}(t)$ are zero, and only the components for the nodes located on the domain boundary and the switching boundary, respectively, take non-zero values. For example, if the node $\mathbf{x}_{k,n_{k}^{\text{s}}}$ is located at the switching boundary, $n_{k}^{\text{s}}$-th component of $\mathbf{b}_{k}^{\text{s}}(t)$ takes non-zero value. In the later section, we obtain numerically time evolution and steady-state of the PDF that satisfies the boundary conditions at the domain and switching boundaries (Eqs.~(\ref{eq:zero_flux_A})-(\ref{eq:flux_conservation})), without evaluating values of the ill-defined spatial derivatives $\bm{\upbeta}_{k}(\mathbf{x},t)$, but instead, by using the values of $\bm{\upbeta}_{k}(\mathbf{x},t)$ that are represented by the non-zero components of $\mathbf{b}_{k}^{\text{d}}(t)$ and $\mathbf{b}_{k}^{\text{s}}(t)$.

Requiring that Eq.~(\ref{eq:FPE_2D_hybrid_2}) holds for any weighting function and requiring that Eq.~(\ref{eq:FPE_2D_hybrid_4}) holds for any weight coefficient vector $\mathbf{w}_{k}$ is equivalent to requiring that Eq.~(\ref{eq:appFPE_wf_MatrixVector_2D_hybrid}) holds for any weight value vectors $\mathbf{w}_{\text{A}}$ and $\mathbf{w}_{\text{B}}$, we have the following finite element equations to be solved for unknown vectors $\mathbf{p}_{\text{A}}(t)$ and $\mathbf{p}_{\text{B}}(t)$:
\begin{equation}
  \left(\begin{array}{cc}
    \mathbf{M}_{\text{A}}  &        \mathbf{0}       \\
          \mathbf{0}       &  \mathbf{M}_{\text{B}}  \\
  \end{array}\right) \frac{\partial}{\partial t} \left(\begin{array}{c}
    \mathbf{p}_{\text{A}}(t)  \\
    \mathbf{p}_{\text{B}}(t)
  \end{array}\right) + \left(\begin{array}{cc}
    \mathbf{K}_{\text{A}}  &        \mathbf{0}       \\
          \mathbf{0}       &  \mathbf{K}_{\text{B}}
  \end{array}\right) \left(\begin{array}{c}
    \mathbf{p}_{\text{A}}(t)  \\
    \mathbf{p}_{\text{B}}(t)
  \end{array}\right) = \left(\begin{array}{c}
    \mathbf{b}_{\text{A}}^{\text{d}}(t) + \mathbf{b}_{\text{A}}^{\text{s}}(t)  \\
    \mathbf{b}_{\text{B}}^{\text{d}}(t) + \mathbf{b}_{\text{B}}^{\text{s}}(t)
  \end{array}\right).
\label{eq:FEEq_2D_hybrid}
\end{equation}

\subsection{Markov chain approximation of the finite element equations}
To obtain time evolution and steady-state solutions of the PDF of Eqs.~(\ref{eq:FPE_2D_hybrid_A}) and (\ref{eq:FPE_2D_hybrid_B}), we discretize Eq.~(\ref{eq:FEEq_2D_hybrid}) in time using the forward Euler method with a time step $\delta t$ as follows:
\begin{align}
  &\left(\begin{array}{c}
    \mathbf{p}_{\text{A}}(t+\delta t)  \\
    \mathbf{p}_{\text{B}}(t+\delta t)
  \end{array}\right)  \nonumber\\
  &= \left(\begin{array}{c}
    \mathbf{p}_{\text{A}}(t)  \\
    \mathbf{p}_{\text{B}}(t)
  \end{array}\right) + \delta t \left(\begin{array}{cc}
    \mathbf{M}_{\text{A}}  &       \mathbf{0}  \\
         \mathbf{0}        &  \mathbf{M}_{\text{B}}
  \end{array}\right)^{-1} \left\{- \left(\begin{array}{cc}
    \mathbf{K}_{\text{A}}  &       \mathbf{0}  \\
         \mathbf{0}        &  \mathbf{K}_{\text{B}}
  \end{array}\right) \left(\begin{array}{c}
    \mathbf{p}_{\text{A}}(t)  \\
    \mathbf{p}_{\text{B}}(t)
  \end{array}\right) + \left(\begin{array}{c}
    \mathbf{b}_{\text{A}}^{\text{d}}(t) + \mathbf{b}_{\text{A}}^{\text{s}}(t)  \\
    \mathbf{b}_{\text{B}}^{\text{d}}(t) + \mathbf{b}_{\text{B}}^{\text{s}}(t)
  \end{array}\right)\right\}.
\label{eq:FEEq_Euler_2D_hybrid}
\end{align}
As mentioned above, $\mathbf{b}_{k}^{\text{d}}(t)$ and $\mathbf{b}_{k}^{\text{s}}(t)$ for $k\in\{\text{A},\text{B}\}$ in the right-hand-side of Eq.~(\ref{eq:FEEq_Euler_2D_hybrid}) include $\bm{\upbeta}_{k}(\mathbf{x},t)$, the formal spatial derivatives of approximated PDF, on the boundary, which are not necessarily well-defined. Thus, we evaluate values of $\mathbf{b}_{k}^{\text{d}}(t)$ and $\mathbf{b}_{k}^{\text{s}}(t)$ by expressing them by using $\mathbf{p}_{k}(t)$, without evaluating $\bm{\upbeta}_{k}(\mathbf{x},t)$ on the boundary by taking into account the boundary conditions described in Eqs.~(\ref{eq:zero_flux_A})-(\ref{eq:flux_conservation}).

First, we express $\mathbf{b}_{k}^{\text{d}}(t)$ by using $\mathbf{p}_{k}(t)$, according to the boundary conditions Eqs.~(\ref{eq:zero_flux_A}) and (\ref{eq:zero_flux_B}) satisfied by the true solution $\rho(\mathbf{x},t)$ of Eqs.~(\ref{eq:FPE_2D_hybrid_A}) and (\ref{eq:FPE_2D_hybrid_B}) on the domain boundary $\Gamma_{k}^{\text{d}}$, which can be rewritten as
\begin{equation}
  \left(\mathbf{f}_{k}(\mathbf{x}) \rho(\mathbf{x},t) - \mathbf{D}_{k} \nabla \rho(\mathbf{x},t)\right) \cdot \mathbf{n}_{\Gamma_{k}}(\mathbf{x}) = 0, \quad \mathbf{x}\in\Gamma_{k}^{\text{d}}.
\label{eq:zero_flux_mod}
\end{equation}
By substituting approximate solution $\tilde{\rho}(\mathbf{x},t)$ into $\rho(\mathbf{x},t)$ of Eq.~(\ref{eq:zero_flux_mod}), we have
\begin{equation}
  \left(\mathbf{D}_{k}\bm{\upbeta}_{k}(\mathbf{x},t)\right) \cdot \mathbf{n}_{\Gamma_{k}}(\mathbf{x}) = \left(\mathbf{f}_{k}(\mathbf{x})\tilde{\rho}(\mathbf{x},t)\right) \cdot \mathbf{n}_{\Gamma_{k}}(\mathbf{x}).
\end{equation}
Then, this relationship is substituted into Eq.~(\ref{eq:def_bd_2D_hybrid}), and together with Eq.~(\ref{eq:def_Phi_L_p}), we have
\begin{align}
  \mathbf{b}_{k}^{\text{d}}(t) &= \sum_{e_{k}^{\text{d}}} \sum_{h_{k,e_{k}^{\text{d}}}^{\text{d}}=1}^{H_{k,e_{k}}^{\text{d}}} \left(\bm{\mathcal{L}}_{\Lambda_{k,e_{k}^{\text{d}}}}^{\Lambda_{k}}\right)^{\text{T}} \int_{\Gamma_{k,e_{k}^{\text{d}},h_{k,e_{k}}^{\text{d}}}^{\text{d}}} \hspace{-5mm} \bm{\upphi}_{k,e_{k}^{\text{d}}}(\mathbf{x}) \left(\mathbf{n}_{\Gamma_{k}}(\mathbf{x})\right)^{\text{T}} \mathbf{f}_{k}(\mathbf{x}) \tilde{\rho}(\mathbf{x},t) \mathrm{d}\Gamma  \nonumber\\
  &= \sum_{e_{k}^{\text{d}}} \sum_{h_{k,e_{k}^{\text{d}}}^{\text{d}}=1}^{H_{k,e_{k}}^{\text{d}}} \left(\bm{\mathcal{L}}_{\Lambda_{k,e_{k}^{\text{d}}}}^{\Lambda_{k}}\right)^{\text{T}} \left\{\int_{\Gamma_{k,e_{k}^{\text{d}},h_{k,e_{k}}^{\text{d}}}^{\text{d}}} \hspace{-5mm} \bm{\upphi}_{k,e_{k}^{\text{d}}}(\mathbf{x}) \left(\mathbf{n}_{\Gamma_{k}}(\mathbf{x})\right)^{\text{T}} \mathbf{f}_{k}(\mathbf{x}) \left(\bm{\upphi}_{k,e_{k}^{\text{d}}}(\mathbf{x})\right)^{\text{T}}\mathrm{d}\Gamma\right\} \bm{\mathcal{L}}_{\Lambda_{k,e_{k}^{\text{d}}}}^{\Lambda_{k}} \mathbf{p}_{k}(t)  \nonumber\\
  &\equiv \mathbf{K}_{k}^{\text{d}} \mathbf{p}_{k}(t),
\label{eq:zero_flux_mod_2}
\end{align}
where $\mathbf{K}_{k}^{\text{d}}$ is the $N_{k} \times N_{k}$ constant matrix defined by $\bm{\mathcal{L}}_{\Lambda_{k,e_{k}^{\text{d}}}}^{\Lambda_{k}}$, $\bm{\upphi}_{k,e_{k}^{\text{d}}}(\mathbf{x})$, $\mathbf{n}_{\Gamma_{k}}(\mathbf{x})$, and $\mathbf{f}_{k}(\mathbf{x})$. In this way, $\mathbf{b}_{k}^{\text{d}}(t)$ is expressed by using $\mathbf{p}_{k}(t)$.

Next, we express $\mathbf{b}_{k}^{\text{s}}(t)$ by using $\mathbf{p}_{k}(t)$, according to the boundary condition Eq.~(\ref{eq:flux_conservation}) and the continuity condition on the PDF satisfied by the true solution $\rho(\mathbf{x},t)$ of Eqs.~(\ref{eq:FPE_2D_hybrid_A}) and (\ref{eq:FPE_2D_hybrid_B}) on the switching boundary $\Gamma^{\text{s}}$. Eq.~(\ref{eq:flux_conservation}) can be rewritten as
\begin{equation}
  \left(\mathbf{f}_{\text{A}}(\mathbf{x}) \rho(\mathbf{x},t) - \mathbf{D}_{\text{A}} \nabla \rho(\mathbf{x},t)\right) \cdot \mathbf{n}_{\Gamma_{\text{A}}}(\mathbf{x}) = \left(\mathbf{f}_{\text{B}}(\mathbf{x}) \rho(\mathbf{x},t) - \mathbf{D}_{\text{B}} \nabla \rho(\mathbf{x},t)\right) \cdot \left(- \mathbf{n}_{\Gamma_{\text{B}}}(\mathbf{x})\right),  \quad \mathbf{x} \in \Gamma^{\text{s}}.
\label{eq:flux_conservation_mod}
\end{equation}
Substituting the approximate solution $\tilde{\rho}(\mathbf{x},t)$ into $\rho(\mathbf{x},t)$ of Eq.~(\ref{eq:flux_conservation_mod}), and requiring weighted residuals, i.e., the line integration of product between the weighting function $w(\mathbf{x})$ and the difference between left- and right-hand-sides of Eq.~(\ref{eq:flux_conservation_mod}) along the switching boundary $\Gamma_{k}^{\text{s}}$, to be zero, we have
\begin{align}
  &\int_{\Gamma^{\text{s}}} w(\mathbf{x}) \left\{\left(\mathbf{n}_{\Gamma_{\text{A}}}(\mathbf{x})\right)^{\text{T}} \mathbf{D}_{\text{A}} \bm{\upbeta}_{\text{A}}(\mathbf{x},t) + \left(\mathbf{n}_{\Gamma_{\text{B}}}(\mathbf{x})\right)^{\text{T}} \mathbf{D}_{\text{B}} \bm{\upbeta}_{\text{B}}(\mathbf{x},t)\right\} \mathrm{d}\Gamma  \nonumber\\
  &= \int_{\Gamma^{\text{s}}} w(\mathbf{x}) \left\{\left(\mathbf{n}_{\Gamma_{\text{A}}}(\mathbf{x})\right)^{\text{T}} \left(\mathbf{f}_{\text{A}}(\mathbf{x}) \tilde{\rho}(\mathbf{x},t)\right) + \left(\mathbf{n}_{\Gamma_{\text{B}}}(\mathbf{x})\right)^{\text{T}} \left(\mathbf{f}_{\text{B}}(\mathbf{x}) \tilde{\rho}(\mathbf{x},t)\right)\right\} \mathrm{d}\Gamma.
\label{eq:flux_conservation_withW}
\end{align}
As we performed for the right-hand-side of Eq.~(\ref{eq:FPE_2D_hybrid_4}) to obtain Eq.~(\ref{eq:FPE_2D_hybrid_4_right}), the line integrations along the switching boundary $\Gamma^{\text{s}}=\Gamma_{\text{A}}^{\text{s}}=\Gamma_{\text{B}}^{\text{s}}$ in Eq.~(\ref{eq:flux_conservation_withW}) can be rewritten as the sums of line integrations along edges $\Gamma_{k,e_{k}^{\text{s}},h_{k,e_{k}^{\text{s}}}^{\text{s}}}^{\text{s}}$ of the elements $\Lambda_{k,e_{k}^{\text{s}}}$ that provide at least one of three edges of each element to the boundary:
\begin{align}
  &\sum_{k\in\{\text{A},\text{B}\}} \sum_{e_{k}^{\text{s}}} \sum_{h_{k,e_{k}^{\text{s}}}^{\text{s}}=1}^{H_{k,e_{k}^{\text{s}}}^{\text{s}}} \int_{\Gamma_{k,e_{k}^{\text{s}},h_{k,e_{k}^{\text{s}}}^{\text{s}}}^{\text{s}}} \hspace{-5mm} w(\mathbf{x}) \left(\mathbf{n}_{\Gamma_{k}}(\mathbf{x})\right)^{\text{T}} \mathbf{D}_{k} \bm{\upbeta}_{k}(\mathbf{x},t) \mathrm{d}\Gamma  \nonumber\\
  &= \sum_{k\in\{\text{A},\text{B}\}} \sum_{e_{k}^{\text{s}}} \sum_{h_{k,e_{k}^{\text{s}}}^{\text{s}}=1}^{H_{k,e_{k}^{\text{s}}}^{\text{s}}} \int_{\Gamma_{k,e_{k}^{\text{s}},h_{k,e_{k}^{\text{s}}}^{\text{s}}}^{\text{s}}} \hspace{-5mm} w(\mathbf{x}) \left(\mathbf{n}_{\Gamma_{k}}(\mathbf{x})\right)^{\text{T}} \left(\mathbf{f}_{k}(\mathbf{x}) \tilde{\rho}(\mathbf{x},t)\right) \mathrm{d}\Gamma.
\label{eq:flux_conservation_withW_2}
\end{align}
Then, substituting Eqs.~(\ref{eq:def_Phi_L_p}) and (\ref{eq:def_Phi_L_w}) into Eq.~(\ref{eq:flux_conservation_withW_2}) and rearranging it using $\mathbf{p}_{k}(t)$, $\mathbf{w}_{k}$, $\bm{\upphi}_{k,e_{k}^{\text{s}}}(\mathbf{x})$, and $\bm{\mathcal{L}}_{\Lambda_{k,e_{k}^{\text{s}}}}^{\Lambda_{k}}$, we have
\begin{align}
  &\sum_{k\in\{\text{A},\text{B}\}} \sum_{e_{k}^{\text{s}}} \sum_{h_{k,e_{k}^{\text{s}}}^{\text{s}}=1}^{H_{k,e_{k}^{\text{s}}}^{\text{s}}} \int_{\Gamma_{k,e_{k}^{\text{s}},h_{k,e_{k}^{\text{s}}}^{\text{s}}}^{\text{s}}} \hspace{-5mm} \left(\mathbf{w}_{k}^{\text{T}} \left(\bm{\mathcal{L}}_{\Lambda_{k,e_{k}^{\text{s}}}}^{\Lambda_{k}}\right)^{\text{T}} \bm{\upphi}_{k,e_{k}^{\text{s}}}(\mathbf{x})\right) \left(\mathbf{n}_{\Gamma_{k}}(\mathbf{x})\right)^{\text{T}} \mathbf{D}_{k} \bm{\upbeta}_{k}(\mathbf{x},t) \mathrm{d}\Gamma  \nonumber\\
  &= \sum_{k\in\{\text{A},\text{B}\}} \sum_{e_{k}^{\text{s}}} \sum_{h_{k,e_{k}^{\text{s}}}^{\text{s}}=1}^{H_{k,e_{k}^{\text{s}}}^{\text{s}}} \int_{\Gamma_{k,e_{k}^{\text{s}},h_{k,e_{k}^{\text{s}}}^{\text{s}}}^{\text{s}}} \hspace{-5mm} \left(\mathbf{w}_{k}^{\text{T}} \left(\bm{\mathcal{L}}_{\Lambda_{k,e_{k}^{\text{s}}}}^{\Lambda_{k}}\right)^{\text{T}} \bm{\upphi}_{k,e_{k}^{\text{s}}}(\mathbf{x})\right) \left(\mathbf{n}_{\Gamma_{k}}(\mathbf{x})\right)^{\text{T}}  \nonumber\\
  &\hspace{7cm}\left(\mathbf{f}_{k}(\mathbf{x}) \left(\bm{\upphi}_{k,e_{k}^{\text{s}}}(\mathbf{x})\right)^{\text{T}} \bm{\mathcal{L}}_{\Lambda_{k,e_{k}^{\text{s}}}}^{\Lambda_{k}} \mathbf{p}_{k}(t)\right) \mathrm{d}\Gamma.
\label{eq:flux_conservation_withW_3}
\end{align}

Here, we introduce another index number to specify each of the $N^{\text{s}}$ nodes located on the switching boundary in $\Lambda_{k}$, denoted by $n_{k,i}^{\text{s}}$ for $k\in\{\text{A},\text{B}\}$, where $i=1,2,\ldots,N^{\text{s}}$ is the serial number assigned to each node commonly for $n_{\text{A},i}^{\text{s}}$ and $n_{\text{B},i}^{\text{s}}$ on the switching boundary. That is, $n_{\text{A},i}^{\text{s}}$ and $n_{\text{B},i}^{\text{s}}$ for a given index number $i$ indicate that two nodes with the same index number $i$ are identical, i.e., $\mathbf{x}_{\text{A},n_{\text{A},i}^{\text{s}}}=\mathbf{x}_{\text{B},n_{\text{B},i}^{\text{s}}}$, where $\mathbf{x}_{k,n_{k,i}^\text{s}}$ represents the node position indexed by $n_{k,i}^\text{s}$, i.e., the $i$-th node on the switching boundary in $\Lambda_{k}$. This serial numbering by $n_{k,i}^{\text{s}}$ is possible, because the nodes on the switching boundary are shared by the regions $\Lambda_{\text{A}}$ and $\Lambda_{\text{B}}$. On the other hand, as mentioned above, all nodes of the region $\Lambda_{k}$, including the nodes on the switching boundary, have been indexed by the serial number $n_{k}$, in which the numbering of the nodes on the switching has been achieved independently for $k=\text{A}$ and $k=\text{B}$. Therefore, an index number $n_{\text{A}}$ assigned for a node on the switching boundary as a node of $\Lambda_{\text{A}}$ and an index number $n_{\text{B}}$ assigned for an identical node on the switching boundary as a node of $\Lambda_{\text{B}}$ are different in general. In other words, $\mathbf{x}_{\text{A},n_{\text{A}}}=\mathbf{x}_{\text{B},n_{\text{B}}}$ does not mean $n_{\text{A}}=n_{\text{B}}$ in general. The index $n_{k,i}^{\text{s}}$ for the nodes on the switching boundary is introduced, because we need a table that represents the identicality between any two identical nodes with different indices $n_{\text{A}}$ and $n_{\text{B}}$, by which we can simplify Eq.~(\ref{eq:flux_conservation_withW_3}). To establish such a table, we define another operator based on the $\bm{\mathcal{L}}_{\mathcal{S}}^{\mathcal{U}}$ defined in Eq.~(\ref{eq:L_def}), in this case, for the sets $\mathcal{U}=\Lambda_{k}$, $\mathcal{S}=\Gamma_{k}^{\text{s}}$ and $\mathcal{S}_{\sharp}=\{k,n_{k,i}^\text{s}\}$. More specifically, we extracts $N^\text{s}$ node values ($p_{k,n_{k,1}^{\text{s}}},p_{k,n_{k,2}^{\text{s}}},\ldots,p_{k,n_{k,N^{\text{s}}}^{\text{s}}}$) for all nodes on the switching boundary $\Gamma_{k}^{\text{s}}$ from the node value vector $\mathbf{p}_{k}$ of the region $\Lambda_{k}$, by using the $N_{k}$-dimensional column vector $\boldsymbol{\ell}_{\mathbf{x}_{k,n_{k,i}^\text{s}}}^{\Lambda_{k}}$ for $i=1,2,\ldots,N^\text{s}$, according to the equality $\mathbf{x}_{k,n_{k}}=\mathbf{x}_{k, n_{k,i}^\text{s}}$. That is,
\begin{align}
  \left(\begin{array}{c}
    p_{k,n_{k,1}^{\text{s}}}  \\
    p_{k,n_{k,2}^{\text{s}}}  \\
    \vdots  \\
    p_{k,n_{k,N^{\text{s}}}^{\text{s}}}
  \end{array}\right) &= \left(\begin{array}{c}
    \left(\boldsymbol{\ell}_{\mathbf{x}_{k,n_{k,1}^{\text{s}}}}^{\Lambda_{k}}\right)^{\text{T}}  \\
    \left(\boldsymbol{\ell}_{\mathbf{x}_{k,n_{k,2}^{\text{s}}}}^{\Lambda_{k}}\right)^{\text{T}}  \\
    \vdots  \\
    \left(\boldsymbol{\ell}_{\mathbf{x}_{k,n_{k,N^{\text{s}}}^{\text{s}}}}^{\Lambda_{k}}\right)^{\text{T}}
  \end{array}\right) \mathbf{p}_{k}(t)  \nonumber\\
  &\equiv \bm{\mathcal{L}}_{\Gamma_{k}^{\text{s}}}^{\Lambda_{k}} \mathbf{p}_{k}(t),
\label{eq:def_L_Gamma_pre}
\end{align}
where $\bm{\mathcal{L}}_{\Gamma_{k}^{\text{s}}}^{\Lambda_{k}}$ is the $N^{\text{s}}\times N_{k}$ matrix.

It is easy to show, from the identicality of $\mathbf{x}_{\text{A},n_{\text{A},i}^{\text{s}}}=\mathbf{x}_{\text{B},n_{\text{B},i}^{\text{s}}}$ for any $i=1,2,\ldots,N^{\text{s}}$ on the switching boundary, that the following equalities for the node and weight values hold:
\begin{align}
  \bm{\mathcal{L}}_{\Gamma_{\text{A}}^{\text{s}}}^{\Lambda_{\text{A}}} \mathbf{p}_{\text{A}} &= \bm{\mathcal{L}}_{\Gamma_{\text{B}}^{\text{s}}}^{\Lambda_{\text{B}}} \mathbf{p}_{\text{B}},
\label{eq:L_gamma_p}  \\
  \bm{\mathcal{L}}_{\Gamma_{\text{A}}^{\text{s}}}^{\Lambda_{\text{A}}} \mathbf{w}_{\text{A}} &= \bm{\mathcal{L}}_{\Gamma_{\text{B}}^{\text{s}}}^{\Lambda_{\text{B}}} \mathbf{w}_{\text{B}}.
\label{eq:L_gamma_w}
\end{align}
Eq.~(\ref{eq:L_gamma_w}) is used to simplify Eq.~(\ref{eq:flux_conservation_withW_3}) as follows. First, we consider left-multiplication by the transpose of $\bm{\mathcal{L}}_{\Gamma_{\text{B}}^{\text{s}}}^{\Lambda_{\text{B}}}$ on both sides of Eq.~(\ref{eq:L_gamma_w}) as
\begin{equation}
  \left(\bm{\mathcal{L}}_{\Gamma_{\text{B}}^{\text{s}}}^{\Lambda_{\text{B}}}\right)^{\text{T}} \bm{\mathcal{L}}_{\Gamma_{\text{B}}^{\text{s}}}^{\Lambda_{\text{B}}} \mathbf{w}_{\text{B}} = \left(\bm{\mathcal{L}}_{\Gamma_{\text{B}}^{\text{s}}}^{\Lambda_{\text{B}}}\right)^{\text{T}} \bm{\mathcal{L}}_{\Gamma_{\text{A}}^{\text{s}}}^{\Lambda_{\text{A}}} \mathbf{w}_{\text{A}},
\label{eq:L_gamma_p_w_2}
\end{equation}
and, by using Eq.~(\ref{eq:L_gamma_p_w_2}), we can derive the following equalities:
\begin{align}
  \mathbf{w}_{\text{B}}^{\text{T}} \left(\bm{\mathcal{L}}_{\Lambda_{\text{B},e_{\text{B}}^{\text{s}}}}^{\Lambda_{\text{B}}}\right)^{\text{T}} \bm{\upphi}_{e_{\text{B}}^{\text{s}}}(\mathbf{x}) &= \mathbf{w}_{\text{B}}^{\text{T}} \left(\bm{\mathcal{L}}_{\Gamma_{\text{B}}^{\text{s}}}^{\Lambda_{\text{B}}}\right)^{\text{T}} \bm{\mathcal{L}}_{\Gamma_{\text{B}}^{\text{s}}}^{\Lambda_{\text{B}}} \left(\bm{\mathcal{L}}_{\Lambda_{\text{B},e_{\text{B}}^{\text{s}}}}^{\Lambda_{\text{B}}}\right)^{\text{T}} \bm{\upphi}_{e_{\text{B}}^{\text{s}}}(\mathbf{x})  \nonumber\\
  &= \mathbf{w}_{\text{A}}^{\text{T}} \mathbf{R} \left(\bm{\mathcal{L}}_{\Lambda_{\text{B},e_{\text{B}}^{\text{s}}}}^{\Lambda_{\text{B}}}\right)^{\text{T}} \bm{\upphi}_{e_{\text{B}}^{\text{s}}}(\mathbf{x}),
\label{eq:NodeRelationship_onSwitchingBoundary}
\end{align}
for $\mathbf{x}\in\Gamma_{\text{A}}^{\text{s}}=\Gamma_{\text{B}}^{\text{s}}$, where we define the matrix $\mathbf{R}$ as
\begin{equation}
  \mathbf{R} \equiv \left(\bm{\mathcal{L}}_{\Gamma_{\text{A}}^{\text{s}}}^{\Lambda_{\text{A}}}\right)^{\text{T}} \bm{\mathcal{L}}_{\Gamma_{\text{B}}^{\text{s}}}^{\Lambda_{\text{B}}}.
\label{eq:def_R}
\end{equation}
The first equality of Eq.~(\ref{eq:NodeRelationship_onSwitchingBoundary}) is derived from the facts that a matrix obtained by multiplication of $\bm{\mathcal{L}}_{\Lambda_{\text{B},e_{\text{B}}^{\text{s}}}}^{\Lambda_{\text{B}}}$ and the transpose of $\bm{\mathcal{L}}_{\Lambda_{\text{B},e_{\text{B}}^{\text{s}}}}^{\Lambda_{\text{B}}}$ is a $N_{\text{B}}\times N_{\text{B}}$ diagonal matrix, whose diagonal components are mostly zero, except that the $n_{\text{B},i}^{\text{s}}$-th ($i=1,2,\ldots,N^{\text{s}}$) diagonal component, corresponding to the node located on the switching boundary, is unity, and component of $\bm{\upphi}_{e_{\text{B}}^{\text{s}}}(\mathbf{x})$ that does not correspond to the nodes located on the switching boundary is zero for $\mathbf{x}\in\Gamma_{\text{A}}^{\text{s}}=\Gamma_{\text{B}}^{\text{s}}$. The matrix $\mathbf{R}$ is the $N_{\text{A}}\times N_{\text{B}}$ matrix, where $N_{\text{A}}$ and $N_{\text{B}}$ are the total numbers of nodes in the regions $\Lambda_{\text{A}}$ and $\Lambda_{\text{B}}$, respectively. Only $N^{\text{s}}$ components of $\mathbf{R}$ are non-zero and unity, and they are the $(n_{\text{A},i}^{\text{s}}, n_{\text{B},i}^{\text{s}})$ component for $i=1,2,\ldots,N^{\text{s}}$. In this way, the matrix $\mathbf{R}$ represents the identicality between the duplicated nodes in $\mathbf{p}_{\text{A}}$ and $\mathbf{p}_{\text{B}}$.

Then, the left-hand-side of Eq.~(\ref{eq:flux_conservation_withW_3}) is simplified by substituting Eq.~(\ref{eq:NodeRelationship_onSwitchingBoundary}) as
\begin{align}
  &\mathbf{w}_{\text{A}}^{\text{T}} \left(\sum_{e_{\text{A}}^{\text{s}}} \sum_{h_{\text{A},e_{\text{A}}^{\text{s}}}^{\text{s}}=1}^{H_{\text{A},e_{\text{A}}^{\text{s}}}^{\text{s}}} \left(\bm{\mathcal{L}}_{\Lambda_{\text{A},e_{\text{A}}^{\text{s}}}}^{\Lambda_{\text{A}}}\right)^{\text{T}}\int_{\Gamma_{\text{A},e_{\text{A}}^{\text{s}},h_{\text{A},e_{\text{A}}^{\text{s}}}^{\text{s}}}^{\text{s}}} \hspace{-5mm} \bm{\upphi}_{\text{A},e_{\text{A}}^{\text{s}}}(\mathbf{x}) \left(\mathbf{n}_{\Gamma_{\text{A}}}(\mathbf{x})\right)^{\text{T}} \mathbf{D}_{\text{A}} \bm{\upbeta}_{\text{A}}(\mathbf{x},t) \mathrm{d}\Gamma  \right.\nonumber\\
  &\hspace{8mm}\left. + \mathbf{R} \sum_{e_{\text{B}}^{\text{s}}} \sum_{h_{\text{B},e_{\text{B}}^{\text{s}}}^{\text{s}}=1}^{H_{\text{B},e_{\text{B}}^{\text{s}}}^{\text{s}}} \left(\bm{\mathcal{L}}_{\Lambda_{\text{B},e_{\text{B}}^{\text{s}}}}^{\Lambda_{\text{B}}}\right)^{\text{T}} \int_{\Gamma_{\text{B},e_{\text{B}}^{\text{s}},h_{\text{B},e_{\text{B}}^{\text{s}}}^{\text{s}}}^{\text{s}}} \hspace{-5mm} \bm{\upphi}_{\text{B},e_{\text{B}}^{\text{s}}}(\mathbf{x}) \left(\mathbf{n}_{\Gamma_{\text{B}}}(\mathbf{x})\right)^{\text{T}} \mathbf{D}_{\text{B}} \bm{\upbeta}_{\text{B}}(\mathbf{x},t) \mathrm{d}\Gamma\right)  \nonumber\\
  &= \mathbf{w}_{\text{A}}^{\text{T}} \left(\mathbf{b}_{\text{A}}^{\text{s}}(t) + \mathbf{R} \mathbf{b}_{\text{B}}^{\text{s}}(t)\right).
\label{eq:flux_conservation_withW_left}
\end{align}
Similarly, the right-hand-side of Eq.~(\ref{eq:flux_conservation_withW_3}) can be simplified as
\begin{align}
  &\mathbf{w}_{\text{A}}^{\text{T}} \left\{\left(\sum_{e_{\text{A}}^{\text{s}}} \sum_{h_{\text{A},e_{\text{A}}^{\text{s}}}^{\text{s}}=1}^{H_{\text{A},e_{\text{A}}^{\text{s}}}^{\text{s}}} \left(\bm{\mathcal{L}}_{\Lambda_{\text{A},e_{\text{A}}^{\text{s}}}}^{\Lambda_{\text{A}}}\right)^{\text{T}} \int_{\Gamma_{\text{A},e_{\text{A}}^{\text{s}},h_{\text{A},e_{\text{A}}^{\text{s}}}^{\text{s}}}^{\text{s}}} \hspace{-5mm} \bm{\upphi}_{\text{A},e_{\text{A}}^{\text{s}}}(\mathbf{x}) \left(\mathbf{n}_{\Gamma_{\text{A}}}(\mathbf{x})\right)^{\text{T}} \mathbf{f}_{\text{A}}(\mathbf{x}) \left(\bm{\upphi}_{\text{A},e_{\text{A}}^{\text{s}}}(\mathbf{x})\right)^{\text{T}} \mathrm{d}\Gamma \bm{\mathcal{L}}_{\Lambda_{\text{A},e_{\text{A}}^{\text{s}}}}^{\Lambda_{\text{A}}}\right) \mathbf{p}_{\text{A}}(t)  \right.\nonumber\\
  &\hspace{8mm}\left.+ \left(\mathbf{R} \sum_{e_{\text{B}}^{\text{s}}} \sum_{h_{\text{B},e_{\text{B}}^{\text{s}}}^{\text{s}}=1}^{H_{\text{B},e_{\text{B}}^{\text{s}}}^{\text{s}}} \left(\bm{\mathcal{L}}_{\Lambda_{\text{B},e_{\text{B}}^{\text{s}}}}^{\Lambda_{\text{B}}}\right)^{\text{T}} \int_{\Gamma_{\text{B},e_{\text{B}}^{\text{s}},h_{\text{B},e_{\text{B}}^{\text{s}}}^{\text{s}}}} \hspace{-5mm} \bm{\upphi}_{\text{B},e_{\text{B}}^{\text{s}}}(\mathbf{x}) \left(\mathbf{n}_{\Gamma_{\text{B}}}(\mathbf{x})\right)^{\text{T}} \mathbf{f}_{\text{B}}(\mathbf{x}) \left(\bm{\upphi}_{\text{B},e_{\text{B}}^{\text{s}}}(\mathbf{x})\right)^{\text{T}} \mathrm{d}\Gamma \bm{\mathcal{L}}_{\Lambda_{\text{B},e_{\text{B}}^{\text{s}}}}^{\Lambda_{\text{B}}}\right) \mathbf{p}_{\text{B}}(t)\right\}  \nonumber\\
  &\equiv \mathbf{w}_{\text{A}}^{\text{T}} \left(\begin{array}{cc}
    \mathbf{K}_{\text{A}}^{\text{s}}  &  \mathbf{R} \mathbf{K}_{\text{B}}^{\text{s}}
  \end{array}\right) \left(\begin{array}{c}
    \mathbf{p}_{\text{A}}(t)  \\
    \mathbf{p}_{\text{B}}(t)
  \end{array}\right),
\label{eq:flux_conservation_withW_right}
\end{align}
where the constant matrices $\mathbf{K}_{\text{A}}^{\text{s}}$ and $\mathbf{K}_{\text{B}}^{\text{s}}$ are defined by using $\bm{\mathcal{L}}_{\Lambda_{k,e_{k}^{\text{s}}}}^{\Lambda_{k}}$, $\bm{\upphi}_{k,e_{k}^{\text{s}}}(\mathbf{x})$, $\mathbf{n}_{\Gamma_{k}}(\mathbf{x})$, and $\mathbf{f}_{k}(\mathbf{x})$, where $\mathbf{K}_{\text{A}}^{\text{s}}$ and $\mathbf{K}_{\text{B}}^{\text{s}}$ are $N_{\text{A}}\times N_{\text{A}}$ and $N_{\text{B}}\times N_{\text{B}}$ matrices, respectively. In this way, by equalizing Eqs.~(\ref{eq:flux_conservation_withW_left}) and (\ref{eq:flux_conservation_withW_right}), the boundary condition Eq.~(\ref{eq:flux_conservation}) or Eq.~(\ref{eq:flux_conservation_mod}) at the switching boundary is expressed as
\begin{equation}
  \mathbf{b}_{\text{A}}^{\text{s}}(t) + \mathbf{R} \mathbf{b}_{\text{B}}^{\text{s}}(t) = \left(\begin{array}{cc}
    \mathbf{K}_{\text{A}}^{\text{s}}  &  \mathbf{R} \mathbf{K}_{\text{B}}^{\text{s}}
  \end{array}\right) \left(\begin{array}{c}
    \mathbf{p}_{\text{A}}(t)  \\
    \mathbf{p}_{\text{B}}(t)
  \end{array}\right),
\label{eq:flux_conservation_mod_2}
\end{equation}
which relates $\mathbf{b}_{k}^{\text{s}}(t)$ and $\mathbf{p}_{k}(t)$. Substituting Eqs.~(\ref{eq:zero_flux_mod_2}) and (\ref{eq:flux_conservation_mod_2}) into Eq.~(\ref{eq:FEEq_Euler_2D_hybrid}), we have
\begin{align}
  &\left(\begin{array}{c}
    \mathbf{p}_{\text{A}}(t+\delta t)  \\
    \mathbf{p}_{\text{B}}(t+\delta t)
  \end{array}\right)  \nonumber\\
  &= \left(\begin{array}{c}
    \mathbf{p}_{\text{A}}(t)  \\
    \mathbf{p}_{\text{B}}(t)
  \end{array}\right) + \delta t \left(\begin{array}{cc}
    \mathbf{M}_{\text{A}}  &       \mathbf{0}  \\
         \mathbf{0}        &  \mathbf{M}_{\text{B}}
  \end{array}\right)^{-1} \left\{- \left(\begin{array}{cc}
    \mathbf{K}_{\text{A}} - \mathbf{K}_{\text{A}}^{\text{d}} - \mathbf{K}_{\text{A}}^{\text{s}}  &  - \mathbf{R} \mathbf{K}_{\text{B}}^{\text{s}}  \\
         \mathbf{0}  &  \mathbf{K}_{\text{B}} - \mathbf{K}_{\text{B}}^{\text{d}}
  \end{array}\right) \left(\begin{array}{c}
    \mathbf{p}_{\text{A}}(t)  \\
    \mathbf{p}_{\text{B}}(t)
  \end{array}\right)  \right.\nonumber\\
  &\hspace{7cm}\left. + \left(\begin{array}{c}
    -\mathbf{R}  \\
    \mathbf{I}_{N_{\text{B}}\times N_{\text{B}}}
  \end{array}\right) \mathbf{b}_{\text{B}}^{\text{s}}(t)\right\}.
\label{eq:FEEq_Euler_2D_hybrid_mod}
\end{align}

We further rewrite Eq.~(\ref{eq:FEEq_Euler_2D_hybrid_mod}) to express $\mathbf{b}_{\text{B}}^{\text{s}}(t)$ by using $\mathbf{p}_{k}(t)$. First, the last term of Eq.~(\ref{eq:FEEq_Euler_2D_hybrid_mod}) is rewritten as
\begin{equation}
  \left(\begin{array}{c}
    -\mathbf{R}  \\
    \mathbf{I}_{N_{\text{B}}\times N_{\text{B}}}
  \end{array}\right) \mathbf{b}_{\text{B}}^{\text{s}}(t) = \left(\begin{array}{c}
    -\mathbf{R}  \\
    \mathbf{I}_{N_{\text{B}}\times N_{\text{B}}}
  \end{array}\right) \left(\bm{\mathcal{L}}_{\Gamma_{\text{B}}^{\text{s}}}^{\Lambda_{\text{B}}}\right)^{\text{T}} \bm{\mathcal{L}}_{\Gamma_{\text{B}}^{\text{s}}}^{\Lambda_{\text{B}}} \mathbf{b}_{\text{B}}^{\text{s}}(t),
\label{eq:preparation_contract}
\end{equation}
which is derived using the fact that only the $(n_{\text{B},i}^{\text{s}},n_{\text{B},i}^{\text{s}})$ components of the $N_{\text{B}}\times N_{\text{B}}$ matrix $(\bm{\mathcal{L}}_{\Gamma_{\text{B}}^{\text{s}}}^{\Lambda_{\text{B}}})^{\text{T}}\bm{\mathcal{L}}_{\Gamma_{\text{B}}^{\text{s}}}^{\Lambda_{\text{B}}}$ are unity values with all the others being zeros. Then, using Eq.~(\ref{eq:preparation_contract}), we rewrite Eq.~(\ref{eq:FEEq_Euler_2D_hybrid_mod}) as
\begin{align}
  &\left(\begin{array}{c}
    \mathbf{p}_{\text{A}}(t+\delta t)  \\
    \mathbf{p}_{\text{B}}(t+\delta t)
  \end{array}\right)  \nonumber\\
  &= \left(\begin{array}{c}
    \mathbf{p}_{\text{A}}(t)  \\
    \mathbf{p}_{\text{B}}(t)
  \end{array}\right) + \delta t \left(\begin{array}{cc}
    \mathbf{M}_{\text{A}}  &       \mathbf{0}  \\
         \mathbf{0}        &  \mathbf{M}_{\text{B}}
  \end{array}\right)^{-1} \left\{- \left(\begin{array}{cc}
    \mathbf{K}_{\text{A}} - \mathbf{K}_{\text{A}}^{\text{d}} - \mathbf{K}_{\text{A}}^{\text{s}}  &  - \mathbf{R} \mathbf{K}_{\text{B}}^{\text{s}}  \\
         \mathbf{0}        &  \mathbf{K}_{\text{B}} - \mathbf{K}_{\text{B}}^{\text{d}}
  \end{array}\right) \left(\begin{array}{c}
    \mathbf{p}_{\text{A}}(t)  \\
    \mathbf{p}_{\text{B}}(t)
  \end{array}\right)  \right.\nonumber\\
  &\hspace{7.0cm}\left. + \left(\begin{array}{c}
    -\mathbf{R}  \\
    \mathbf{I}_{N_{\text{B}}\times N_{\text{B}}}
  \end{array}\right) \left(\bm{\mathcal{L}}_{\Gamma_{\text{B}}^{\text{s}}}^{\Lambda_{\text{B}}}\right)^{\text{T}} \bm{\mathcal{L}}_{\Gamma_{\text{B}}^{\text{s}}}^{\Lambda_{\text{B}}} \mathbf{b}_{\text{B}}^{\text{s}}(t)\right\}  \nonumber\\
  &\equiv \left(\begin{array}{c}
    \mathbf{p}_{\text{A}}(t)  \\
    \mathbf{p}_{\text{B}}(t)
  \end{array}\right) + \delta t \left\{\left(\begin{array}{c}
    \mathbf{q}_{\text{A}}(t)  \\
    \mathbf{q}_{\text{B}}(t)
  \end{array}\right) + \left(\begin{array}{c}
    \mathbf{U}_{\text{A}}  \\
    \mathbf{U}_{\text{B}}
  \end{array}\right) \mathbf{r}_{\text{B}}^{\text{s}}(t)\right\},
\label{eq:FEEq_Euler_2D_hybrid_pre_4}
\end{align}
where $\mathbf{q}_{k}(t)$, $\mathbf{U}_{k}$, and $\mathbf{r}_{\text{B}}^{\text{s}}(t)$ are defined as follows:
\begin{align}
  \left(\begin{array}{c}
    \mathbf{q}_{\text{A}}(t)  \\
    \mathbf{q}_{\text{B}}(t)
  \end{array}\right) &= \left(\begin{array}{cc}
    \mathbf{M}_{\text{A}}  &       \mathbf{0}  \\
         \mathbf{0}        &  \mathbf{M}_{\text{B}}
  \end{array}\right)^{-1} \left\{- \left(\begin{array}{cc}
    \mathbf{K}_{\text{A}} - \mathbf{K}_{\text{A}}^{\text{d}} - \mathbf{K}_{\text{A}}^{\text{s}}  &  -\mathbf{R} \mathbf{K}_{\text{B}}^{\text{s}}  \\
         \mathbf{0}        &  \mathbf{K}_{\text{B}} - \mathbf{K}_{\text{B}}^{\text{d}}
  \end{array}\right)\right\} \left(\begin{array}{c}
    \mathbf{p}_{\text{A}}(t)  \\
    \mathbf{p}_{\text{B}}(t)
  \end{array}\right),
\label{eq:def_q_k}  \\
  \left(\begin{array}{c}
    \mathbf{U}_{\text{A}}  \\
    \mathbf{U}_{\text{B}}
  \end{array}\right) &= \left(\begin{array}{cc}
    \mathbf{M}_{\text{A}}  &       \mathbf{0}  \\
         \mathbf{0}        &  \mathbf{M}_{\text{B}}
  \end{array}\right)^{-1} \left(\begin{array}{c}
    -\mathbf{R}  \\
    \mathbf{I}_{N_{\text{B}}\times N_{\text{B}}}
  \end{array}\right) \left(\bm{\mathcal{L}}_{\Gamma_{\text{B}}^{\text{s}}}^{\Lambda_{\text{B}}}\right)^{\text{T}},
\label{eq:def_U_k}  \\
  \mathbf{r}_{\text{B}}^{\text{s}}(t) &= \bm{\mathcal{L}}_{\Gamma_{\text{B}}^{\text{s}}}^{\Lambda_{\text{B}}} \mathbf{b}_{\text{B}}^{\text{s}}(t).
\label{eq:def_r_B}
\end{align}
Moving the first term on the right-hand-side of Eq.~(\ref{eq:FEEq_Euler_2D_hybrid_pre_4}) to the left-hand-side, and multiplying the upper and lower terms of both sides by $\bm{\mathcal{L}}_{\Gamma_{\text{A}}^{\text{s}}}^{\Lambda_{\text{A}}}$ and $\bm{\mathcal{L}}_{\Gamma_{\text{B}}^{\text{s}}}^{\Lambda_{\text{B}}}$, respectively, and then dividing both sides by $\delta t$, we have
\begin{align}
  \bm{\mathcal{L}}_{\Gamma_{\text{A}}^{\text{s}}}^{\Lambda_{\text{A}}} \left(\frac{\mathbf{p}_{\text{A}}(t+\delta t) -\mathbf{p}_{\text{A}}(t)}{\delta t}\right) &= \bm{\mathcal{L}}_{\Gamma_{\text{A}}^{\text{s}}}^{\Lambda_{\text{A}}} \left\{\mathbf{q}_{\text{A}}(t) +  \mathbf{U}_{\text{A}}\mathbf{r}_{\text{B}}^{\text{s}}(t) \right\},
\label{eq:velocity_equalityA} \\
  \bm{\mathcal{L}}_{\Gamma_{\text{B}}^{\text{s}}}^{\Lambda_{\text{B}}} \left(\frac{\mathbf{p}_{\text{B}}(t+\delta t) -\mathbf{p}_{\text{B}}(t)}{\delta t}\right) &= \bm{\mathcal{L}}_{\Gamma_{\text{B}}^{\text{s}}}^{\Lambda_{\text{B}}} \left\{\mathbf{q}_{\text{B}}(t)  + \mathbf{U}_{\text{B}}\mathbf{r}_{\text{B}}^{\text{s}}(t) \right\}.
\label{eq:velocity_equalityB}
\end{align}
Here, we consider the continuity condition on the PDF on the switching boundary. Because all of $N^{\text{s}}$ nodes located at the switching boundary are shared by the regions $\Lambda_{\text{A}}$ and $\Lambda_{\text{B}}$, values of the shared nodes must be identical always at any time $t$, including at $t=0$, which requires Eq.~(\ref{eq:L_gamma_p}) and the following equality:
\begin{equation}
  \bm{\mathcal{L}}_{\Gamma_{\text{A}}^{\text{s}}}^{\Lambda_{\text{A}}} \frac{\mathrm{d}\mathbf{p}_{\text{A}}(t)}{\mathrm{d}t} = \bm{\mathcal{L}}_{\Gamma_{\text{B}}^{\text{s}}}^{\Lambda_{\text{B}}} \frac{\mathrm{d}\mathbf{p}_{\text{B}}(t)}{\mathrm{d}t}.
\label{eq:velocity_equality2}
\end{equation}
Because Eq.~(\ref{eq:velocity_equality2}) indicates that the right-hand-side of Eq.~(\ref{eq:velocity_equalityA}) is equal to the right-hand-side of Eq.~(\ref{eq:velocity_equalityB}) for $\delta t\rightarrow 0$, we have
\begin{equation}
  \left(\bm{\mathcal{L}}_{\Gamma_{\text{A}}^{\text{s}}}^{\Lambda_{\text{A}}} \mathbf{U}_{\text{A}} - \bm{\mathcal{L}}_{\Gamma_{\text{B}}^{\text{s}}}^{\Lambda_{\text{B}}} \mathbf{U}_{\text{B}}\right) \mathbf{r}_{\text{B}}^{\text{s}}(t) = - \bm{\mathcal{L}}_{\Gamma_{\text{A}}^{\text{s}}}^{\Lambda_{\text{A}}} \mathbf{q}_{\text{A}} + \bm{\mathcal{L}}_{\Gamma_{\text{B}}^{\text{s}}}^{\Lambda_{\text{B}}} \mathbf{q}_{\text{B}},
\end{equation}
which can be rearranged as
\begin{align}
  \mathbf{r}_{\text{B}}^{\text{s}}(t) = \left\{\left(\begin{array}{cc}
    \bm{\mathcal{L}}_{\Gamma_{\text{A}}^{\text{s}}}^{\Lambda_{\text{A}}}  & - \bm{\mathcal{L}}_{\Gamma_{\text{B}}^{\text{s}}}^{\Lambda_{\text{B}}}
  \end{array}\right) \left(\begin{array}{c}
    \mathbf{U}_{\text{A}}  \\
    \mathbf{U}_{\text{B}}
  \end{array}\right)\right\}^{-1} \left(\begin{array}{cc}
    - \bm{\mathcal{L}}_{\Gamma_{\text{A}}^{\text{s}}}^{\Lambda_{\text{A}}}  &  \bm{\mathcal{L}}_{\Gamma_{\text{B}}^{\text{s}}}^{\Lambda_{\text{B}}}
  \end{array}\right) \left(\begin{array}{c}
    \mathbf{q}_{\text{A}}(t)  \\
    \mathbf{q}_{\text{B}}(t)
  \end{array}\right).
\label{eq:r_B^s}
\end{align}
Substituting Eq.~(\ref{eq:def_q_k}) into Eq.~(\ref{eq:r_B^s}), and then substituting the resultant Eq.~(\ref{eq:r_B^s}) into Eq.~(\ref{eq:FEEq_Euler_2D_hybrid_pre_4}), we have the following update equation of $\mathbf{p}_{k}(t)$:
\begin{equation}
  \left(\begin{array}{c}
    \mathbf{p}_{\text{A}}(t+\delta t)  \\
    \mathbf{p}_{\text{B}}(t+\delta t)
  \end{array}\right) = \left(\begin{array}{c}
    \mathbf{p}_{\text{A}}(t)  \\
    \mathbf{p}_{\text{B}}(t)
  \end{array}\right) + \delta t \mathbf{V} \left(\begin{array}{c}
    \mathbf{p}_{\text{A}}(t)  \\
    \mathbf{p}_{\text{B}}(t)
  \end{array}\right),
\label{eq:FEEq_Euler_2D_hybrid_2} 
\end{equation}
where $\mathbf{V}$ is the constant matrix defined as
\begin{align}
  \mathbf{V} &= - \left[\mathbf{I}_{N\times N} + \left(\begin{array}{c}
    \mathbf{U}_{\text{A}}  \\
    \mathbf{U}_{\text{B}}
  \end{array}\right) \left\{\left(\begin{array}{cc}
    \bm{\mathcal{L}}_{\Gamma_{\text{A}}^{\text{s}}}^{\Lambda_{\text{A}}}  & - \bm{\mathcal{L}}_{\Gamma_{\text{B}}^{\text{s}}}^{\Lambda_{\text{B}}}
  \end{array}\right) \left(\begin{array}{c}
    \mathbf{U}_{\text{A}}  \\
    \mathbf{U}_{\text{B}}
  \end{array}\right)\right\}^{-1} \left(\begin{array}{cc}
    - \bm{\mathcal{L}}_{\Gamma_{\text{A}}^{\text{s}}}^{\Lambda_{\text{A}}}  &  \bm{\mathcal{L}}_{\Gamma_{\text{B}}^{\text{s}}}^{\Lambda_{\text{B}}}
  \end{array}\right)\right]  \nonumber\\
  &\hspace{1cm}\cdot \left(\begin{array}{cc}
    \mathbf{M}_{\text{A}}  &       \mathbf{0}  \\
         \mathbf{0}        &  \mathbf{M}_{\text{B}}
  \end{array}\right)^{-1} \left(\begin{array}{cc}
    \mathbf{K}_{\text{A}} - \mathbf{K}_{\text{A}}^{\text{d}} - \mathbf{K}_{\text{A}}^{\text{s}}  &  - \mathbf{R} \mathbf{K}_{\text{B}}^{\text{s}}  \\
         \mathbf{0}        &  \mathbf{K}_{\text{B}} - \mathbf{K}_{\text{B}}^{\text{d}}
  \end{array}\right),
\label{eq:def_V}
\end{align}
where $N=N_{\text{A}}+N_{\text{B}}$. Eq.~(\ref{eq:FEEq_Euler_2D_hybrid_2}) can be rewritten as
\begin{equation}
  \left(\begin{array}{c}
    \mathbf{p}_{\text{A}}(t+\delta t)  \\
    \mathbf{p}_{\text{B}}(t+\delta t)
  \end{array}\right)=\bm{\Psi}^{\text{rdn}} \left(\begin{array}{c}
    \mathbf{p}_{\text{A}}(t)  \\
    \mathbf{p}_{\text{B}}(t)
  \end{array}\right),
\label{eq:Markov_pre}
\end{equation}
where
\begin{equation}
  \mathbf{\Psi}^{\text{rdn}} = \mathbf{I}_{N\times N} + \delta t \mathbf{V}.
\label{eq:def_Psi_rdn}
\end{equation}
Note that the superscript `$\text{rdn}$' of the matrix $\mathbf{\Psi}^{\text{rdn}}$ is given, because $\mathbf{\Psi}^{\text{rdn}}$ includes a redundancy due to duplications in $\mathbf{p}_{\text{A}}$ and $\mathbf{p}_{\text{B}}$, as noted previously.

In the remaining of this subsection, we take care of this node duplications caused by the definition of $\mathbf{p}_{\text{A}}$ and $\mathbf{p}_{\text{B}}$. Namely, the nodes located on the switching boundary have been doubly indexed for the region $\Lambda_{\text{A}}$ and also for the region $\Lambda_{\text{B}}$. Let us consider the $i$-th node on the switching boundary, whose position is represented by $\mathbf{x}_{\text{A},n_{\text{A},i}^{\text{s}}}$ and also by $\mathbf{x}_{\text{B},n_{\text{B},i}^{\text{s}}}$ with the serial node indexing number $i$ commonly for both the region $\Lambda_{\text{A}}$ and $\Lambda_{\text{B}}$. The node value of this node appears in both of $\mathbf{p}_{\text{A}}$ and $\mathbf{p}_{\text{B}}$ as $p_{\text{A},n_{\text{A},i}^{\text{s}}}$ and $p_{\text{B},n_{\text{B},i}^{\text{s}}}$, respectively, as if they were the values of different nodes. If we do not take care of this duplicated probabilities, the matrix $\mathbf{\Psi}^{\text{rdn}}$ in Eq.~(\ref{eq:Markov_pre}) does not satisfy the properties required for a Markov state transition probability matrix.

\subsubsection{Obtaining a proper Markov state transition probability matrix: A toy model}
In order to explain a main idea of how we treat the matrix $\mathbf{\Psi}^{\text{rdn}}$ to make it a proper state transition probability matrix, let us consider a simple toy Markov chain model that consists of four states $\text{s}_{1}$, $\text{s}_{2}$, $\text{s}_{3}$, and $\text{s}_{4}$. First, we define the probability $q_{j}(t)$ such that the state of the model is located at $\text{s}_{j}$ ($j=1,\ldots,4$) at time $t$, and consider the following state transition probability from time $t$ to $t+\delta t$:
\begin{equation}
  \mathbf{q}(t+\delta t) = \left(\begin{array}{c}
    q_{1}(t+\delta t)  \\
    q_{2}(t+\delta t)  \\
    q_{3}(t+\delta t)  \\
    q_{4}(t+\delta t)
  \end{array}\right) = \left(\begin{array}{cccc}
    \psi_{1\to1}  &  \psi_{2\to1}  &  \psi_{3\to1}  &  \psi_{4\to1}  \\
    \psi_{1\to2}  &  \psi_{2\to2}  &  \psi_{3\to2}  &  \psi_{4\to2}  \\
    \psi_{1\to3}  &  \psi_{2\to3}  &  \psi_{3\to3}  &  \psi_{4\to3}  \\
    \psi_{1\to4}  &  \psi_{2\to4}  &  \psi_{3\to4}  &  \psi_{4\to4}
  \end{array}\right) \left(\begin{array}{c}
    q_{1}(t)  \\
    q_{2}(t)  \\
    q_{3}(t)  \\
    q_{4}(t)
  \end{array}\right) \equiv \mathbf{\Psi}_{\text{toy}} \mathbf{q}(t),
\label{eq:simpleMarkov}
\end{equation}
where $\mathbf{q}(t)$ is the state vector representing the probabilities that the state at time $t$ is located at each of the four states. $\mathbf{\Psi}_{\text{toy}}$ is the state transition probability matrix of this simple toy Markov chain model, with the elements $\psi_{j\to j^{\prime}}$ representing the transition probability from the state $j$ to the state $j^{\prime}$ for $j,j^\prime=1,\ldots,4$.

Now, let us consider a situation such that the state $\text{s}_{2}$ of the toy model is represented with a duplication by $\text{s}_{2\text{A}}$ and $\text{s}_{2\text{B}}$, and the probability for the state of the model to be located at $\text{s}_{2}$ is represented in two ways as $q_{2\text{A}}$ and $q_{2\text{B}}$. Because $\text{s}_{2\text{A}}$ and $\text{s}_{2\text{B}}$ are the identical state in reality, the equality $q_{2\text{A}}=q_{2\text{B}}=q_{2}$ must hold. Moreover, to make the toy model analogous to the switched hybrid model of this study, let us define $\mathbf{q}_{\text{A}}$ and $\mathbf{q}_{\text{B}}$ as
\begin{equation}
  \mathbf{q}_{\text{A}}(t) = \left(\begin{array}{c}
        q_{1}(t)  \\
    q_{2\text{A}}(t)
  \end{array}\right),\hspace{1cm} \mathbf{q}_{\text{B}}(t) = \left(\begin{array}{c}
    q_{2\text{B}}(t)  \\
        q_{3}(t)      \\
        q_{4}(t)
  \end{array}\right),
\end{equation}
which represent a situation such that the states $\text{s}_{1}$ is included in the region A, $\text{s}_{3}$ and $\text{s}_{4}$ are in the region B, and $\text{s}_{2}$ is in the border of A and B. In this case, the probability for $\text{s}_{2}$ is duplicated as $q_{2\text{A}}(t)$ and $q_{2\text{B}}(t)$, i.e., the second component of $\mathbf{q}_{\text{A}}(t)$ and the first component of $\mathbf{q}_{\text{B}}(t)$ are duplicated. For this toy model, if we distinguish between $\text{s}_{2\text{A}}$ and $\text{s}_{2\text{B}}$ with their distinguished probabilities $q_{2\text{A}}$ and $q_{2\text{B}}$, the Markov chain model is represented as
\begin{align}
  \left(\begin{array}{c}
    \mathbf{q}_{\text{A}}(t+\delta t)  \\
    \mathbf{q}_{\text{B}}(t+\delta t)
  \end{array}\right) &= \left(\begin{array}{ccccc}
        \psi_{1\to1}      &      \psi_{2\text{A}\to1}      &      \psi_{2\text{B}\to1}      &      \psi_{3\to1}      &      \psi_{4\to1}      \\
    \psi_{1\to2\text{A}}  &  \psi_{2\text{A}\to2\text{A}}  &  \psi_{2\text{B}\to2\text{A}}  &  \psi_{3\to2\text{A}}  &  \psi_{4\to2\text{A}}  \\
    \psi_{1\to2\text{B}}  &  \psi_{2\text{A}\to2\text{B}}  &  \psi_{2\text{B}\to2\text{B}}  &  \psi_{3\to2\text{B}}  &  \psi_{4\to2\text{B}}  \\
        \psi_{1\to3}      &      \psi_{2\text{A}\to3}      &      \psi_{2\text{B}\to3}      &      \psi_{3\to3}      &      \psi_{4\to3}      \\
        \psi_{1\to4}      &      \psi_{2\text{A}\to4}      &      \psi_{2\text{B}\to4}      &      \psi_{3\to4}      &      \psi_{4\to4}
  \end{array}\right) \left(\begin{array}{c}
    \mathbf{q}_{\text{A}}(t)  \\
    \mathbf{q}_{\text{B}}(t)
  \end{array}\right)  \nonumber\\
  &\equiv \mathbf{\Psi}_{\text{toy}}^{\text{rdn}} \left(\begin{array}{c}
    \mathbf{q}_{\text{A}}(t)  \\
    \mathbf{q}_{\text{B}}(t)
  \end{array}\right),
\label{eq:simpleMarkov_2}
\end{align}
instead of Eq.~(\ref{eq:simpleMarkov}). The superscript `$\text{rdn}$' is used for the matrix $\mathbf{\Psi}_{\text{toy}}^{\text{rdn}}$, because $\mathbf{\Psi}_{\text{toy}}^{\text{rdn}}$ is redundant as in the case of $\mathbf{\Psi}^{\text{rdn}}$ in Eq.~(\ref{eq:Markov_pre}), due to the duplicated representations of probabilities for the state $\text{s}_{2}$, i.e., $q_{2\text{A}}$ and $q_{2\text{B}}$. In this sequel, we describe properties of the redundant matrix $\mathbf{\Psi}_{\text{toy}}^{\text{rdn}}$, and then illustrate how we can obtain the non-redundant matrix $\mathbf{\Psi}_{\text{toy}}$ from the redundant matrix $\mathbf{\Psi}_{\text{toy}}^{\text{rdn}}$.

First, we describe the redundancy of the matrix $\mathbf{\Psi}_{\text{toy}}^{\text{rdn}}$. As mentioned above, because $\text{s}_{2\text{A}}$ and $\text{s}_{2\text{B}}$ must be identical as $\text{s}_{2}$, $q_{2\text{A}}=q_{2\text{B}}=q_{2}$ must hold. This means that the following equalities for the state transition probabilities must hold:
\begin{align}
  \psi_{1\to2\text{A}} &= \psi_{1\to2\text{B}} = \psi_{1\to2},
\label{eq:psi_1_to_2}  \\
  \psi_{3\to2\text{A}} &= \psi_{3\to2\text{B}} = \psi_{3\to2},
\label{eq:psi_3_to_2}  \\
  \psi_{4\to2\text{A}} &= \psi_{4\to2\text{B}} = \psi_{4\to2}.
\label{eq:psi_4_to_2}
\end{align}
By substituting the equality of $q_{2\text{A}}=q_{2\text{B}}=q_{2}$ and Eqs.~(\ref{eq:psi_1_to_2})-(\ref{eq:psi_4_to_2}) into Eq.~(\ref{eq:simpleMarkov_2}), we obtain the following equation.
\begin{equation}
  \left(\begin{array}{c}
    q_{1}(t+\delta t)  \\
    q_{2}(t+\delta t)  \\
    q_{2}(t+\delta t)  \\
    q_{3}(t+\delta t)  \\
    q_{4}(t+\delta t)
  \end{array}\right) = \left(\begin{array}{cccc}
    \psi_{1\to1}  &          \psi_{2\text{A}\to1} + \psi_{2\text{B}\to1}          &  \psi_{3\to1}  &  \psi_{4\to1}  \\
    \psi_{1\to2}  &  \psi_{2\text{A}\to2\text{A}} + \psi_{2\text{B}\to2\text{A}}  &  \psi_{3\to2}  &  \psi_{4\to2}  \\
    \psi_{1\to2}  &  \psi_{2\text{A}\to2\text{B}} + \psi_{2\text{B}\to2\text{B}}  &  \psi_{3\to2}  &  \psi_{4\to2}  \\
    \psi_{1\to3}  &          \psi_{2\text{A}\to3} + \psi_{2\text{B}\to3}          &  \psi_{3\to3}  &  \psi_{4\to3}  \\
    \psi_{1\to4}  &          \psi_{2\text{A}\to4} + \psi_{2\text{B}\to4}          &  \psi_{3\to4}  &  \psi_{4\to4}
  \end{array}\right) \left(\begin{array}{c}
    q_{1}(t)  \\
    q_{2}(t)  \\
    q_{3}(t)  \\
    q_{4}(t)
  \end{array}\right).
\label{eq:simpleMarkov_3}
\end{equation}
One can observe that the state transition probabilities from each node to state $\text{s}_{2}$ is duplicately written in the second and third rows of Eq.~(\ref{eq:simpleMarkov_3}). In other words, the third (or the second) row of Eq.~(\ref{eq:simpleMarkov_3}) is redundant. In this way, the third (or the second) row of the matrix $\mathbf{\Psi}_{\text{toy}}^{\text{rdn}}$ in Eq.~(\ref{eq:simpleMarkov_2}) is also redundant, because the second and the third rows of the matrix in Eq.~(\ref{eq:simpleMarkov_3}) are equivalent to the second and the third rows of the matrix $\mathbf{\Psi}_{\text{toy}}^{\text{rdn}}$. Note that the second and the third rows of the matrix $\mathbf{\Psi}_{\text{toy}}^{\text{rdn}}$ describe the state transition probabilities into the states $\text{s}_{2\text{A}}$ and $\text{s}_{2\text{B}}$ from the other states, both of which are the state transition probabilities into the state $\text{s}_{2}$ in a duplicated manner in each column of $\mathbf{\Psi}_{\text{toy}}^{\text{rdn}}$. This duplication is the major issue, by which $\mathbf{\Psi}_{\text{toy}}^{\text{rdn}}$ does not satisfy the properties required for the state transition probability matrix. Therefore, we should eliminate the redundant row of $\mathbf{\Psi}_{\text{toy}}^{\text{rdn}}$ so that $\mathbf{\Psi}_{\text{toy}}^{\text{rdn}}$ can satisfy the property of the state transition matrix as the matrix $\mathbf{\Psi}_{\text{toy}}$.

We can obtain the state transition matrix $\mathbf{\Psi}_{\text{toy}}$ from the redundant matrix $\mathbf{\Psi}_{\text{toy}}^{\text{rdn}}$ as follows. As explained above, the third (or the second) row of Eq.~(\ref{eq:simpleMarkov_3}) is redundant. Thus, we simply eliminate the third row of Eq.~(\ref{eq:simpleMarkov_3}) to obtain
\begin{equation}
  \left(\begin{array}{c}
    q_{1}(t+\delta t)  \\
    q_{2}(t+\delta t)  \\
    q_{3}(t+\delta t)  \\
    q_{4}(t+\delta t)
  \end{array}\right) = \left(\begin{array}{cccc}
    \psi_{1\to1}  &          \psi_{2\text{A}\to1} + \psi_{2\text{B}\to1}          &  \psi_{3\to1}  &  \psi_{4\to1}  \\
    \psi_{1\to2}  &  \psi_{2\text{A}\to2\text{A}} + \psi_{2\text{B}\to2\text{A}}  &  \psi_{3\to2}  &  \psi_{4\to2}  \\
    \psi_{1\to3}  &          \psi_{2\text{A}\to3} + \psi_{2\text{B}\to3}          &  \psi_{3\to3}  &  \psi_{4\to3}  \\
    \psi_{1\to4}  &          \psi_{2\text{A}\to4} + \psi_{2\text{B}\to4}          &  \psi_{3\to4}  &  \psi_{4\to4}
  \end{array}\right) \left(\begin{array}{c}
    q_{1}(t)  \\
    q_{2}(t)  \\
    q_{3}(t)  \\
    q_{4}(t)
  \end{array}\right).
\label{eq:simpleMarkov_4}
\end{equation}
By equating the right-hand-sides of Eq.~(\ref{eq:simpleMarkov}) and Eq.~(\ref{eq:simpleMarkov_4}), we find the following relationships:
\begin{align}
  \psi_{2\to1} &= \psi_{2\text{A}\to1} + \psi_{2\text{B}\to1},
\label{eq:psi_2_to_1}  \\
  \psi_{2\to3} &= \psi_{2\text{A}\to3} + \psi_{2\text{B}\to3},
\label{eq:psi_2_to_3}  \\
  \psi_{2\to4} &= \psi_{2\text{A}\to4} + \psi_{2\text{B}\to4},
\label{eq:psi_2_to_4}  \\
  \psi_{2\to2} &= \psi_{2\text{A}\to2\text{A}} + \psi_{2\text{B}\to2\text{A}}.
\label{eq:psi_2_to_2}
\end{align}
Eqs.~(\ref{eq:psi_2_to_1})-(\ref{eq:psi_2_to_2}) show that the state transition probabilities from $\text{s}_{2}$ to the other states in the matrix $\mathbf{\Psi}_{\text{toy}}$ are the simple sum of the state transition probability from $\text{s}_{2\text{A}}$ and that from $\text{s}_{2\text{B}}$ to the other states, which is split into the second and third columns of $\mathbf{\Psi}_{\text{toy}}^{\text{rdn}}$ in Eq.~(\ref{eq:simpleMarkov_2}). This means that the elements of the second column of the non-redundant matrix $\mathbf{\Psi}_{\text{toy}}$ can be obtained by combining the second and third columns of the redundant matrix $\mathbf{\Psi}_{\text{toy}}^{\text{rdn}}$. In summary, the non-redundant Markov state transition matrix $\mathbf{\Psi}_{\text{toy}}$ can be obtained from the redundant matrix $\mathbf{\Psi}_{\text{toy}}^{\text{rdn}}$ by eliminating the repeated rows (i.e., by eliminating the third row), and then by combining the split columns (i.e., by summing the second and third columns). This process can be generally formulated as
\begin{equation}
  \mathbf{\Psi}_{\text{toy}} = \bm{\mathcal{Z}}_{\text{toy}}^{\text{elm}} \mathbf{\Psi}_{\text{toy}}^{\text{rdn}} \bm{\mathcal{Z}}_{\text{toy}}^{\text{cmb}},
\label{eq:derivation_Psi_toy}
\end{equation}
where $\bm{\mathcal{Z}}_{\text{toy}}^{\text{elm}}$ and $\bm{\mathcal{Z}}_{\text{toy}}^{\text{cmb}}$ are the eliminating and the combining operations as described above. These two operations can be detailed by using matrix operations as follows:
\begin{equation}
  \bm{\mathcal{Z}}_{\text{toy}}^{\text{elm}} = \left(\begin{array}{cc|ccc}
    1  &  0  &  0  &  0  &  0  \\
    0  &  1  &  0  &  0  &  0  \\
    \hline
    0  &  0  &  0  &  1  &  0  \\
    0  &  0  &  0  &  0  &  1
  \end{array}\right),\hspace{1cm} \bm{\mathcal{Z}}_{\text{toy}}^{\text{cmb}} = \left(\begin{array}{cc|cc}
    1  &  0  &  0  &  0  \\
    0  &  1  &  0  &  0  \\
    \hline
    0  &  1  &  0  &  0  \\
    0  &  0  &  1  &  0  \\
    0  &  0  &  0  &  1
  \end{array}\right).
\label{eq:L_elm_toy_L_cmb_toy}
\end{equation}

Eq.~(\ref{eq:derivation_Psi_toy}) can be validated easily as follows:
\begin{equation}
  \mathbf{q}(t+\delta t) = \bm{\mathcal{Z}}_{\text{toy}}^{\text{elm}} \left(\begin{array}{c}
    \mathbf{q}_{\text{A}}(t+\delta t)  \\
    \mathbf{q}_{\text{B}}(t+\delta t)
  \end{array}\right) = \bm{\mathcal{Z}}_{\text{toy}}^{\text{elm}} \mathbf{\Psi}_{\text{toy}}^{\text{rdn}} \left(\begin{array}{c}
    \mathbf{q}_{\text{A}}(t)  \\
    \mathbf{q}_{\text{B}}(t)
  \end{array}\right) = \bm{\mathcal{Z}}_{\text{toy}}^{\text{elm}} \mathbf{\Psi}_{\text{toy}}^{\text{rdn}} \bm{\mathcal{Z}}_{\text{toy}}^{\text{cmb}} \mathbf{q}(t) = \mathbf{\Psi}_{\text{toy}} \mathbf{q}(t),
\label{eq:derivation_Psi_toy_equation}
\end{equation}
where $\mathbf{q}$ is the non-redundant state vector that can be represented by eliminating the redundant component from the redundant state vector $(\mathbf{q}_{\text{A}}^{\text{T}},\mathbf{q}_{\text{B}}^{\text{T}})^{\text{T}}$.

The construction of the eliminating matrix $\bm{\mathcal{Z}}_{\text{toy}}^{\text{elm}}$ and the combining matrix $\bm{\mathcal{Z}}_{\text{toy}}^{\text{cmb}}$ and the use of these matrices to obtain the non-redundant state transition matrix $\mathbf{\Psi}_{\text{toy}}$ from the redundant matrix $\mathbf{\Psi}_{\text{toy}}^{\text{rdn}}$ can be generalized based on the following formulation. We notice first that both $\bm{\mathcal{Z}}_{\text{toy}}^{\text{elm}}$ and $\bm{\mathcal{Z}}_{\text{toy}}^{\text{cmb}}$ consist of four block matrices. The upper left block of $\bm{\mathcal{Z}}_{\text{toy}}^{\text{elm}}$ is the identity matrix, the size of which is the same as the number of components of the vector $\mathbf{q}_{\text{A}}$, which is $2$ in the toy model. Both upper right and lower left blocks are the zero matrices. Components of the lower right block are mostly zero, except two unity values at $(1, 2)$ and $(2, 3)$, respectively. The lower right block extracts two components of $\mathbf{q}_{\text{B}}$, $q_{3}(t)$ and $q_{4}(t)$, that are not duplicated with any components of $\mathbf{q}_{\text{A}}$, from $(\mathbf{q}_{\text{A}}^{\text{T}},\mathbf{q}_{\text{B}}^{\text{T}})^{\text{T}}$. The lower right block can be represented formally using a type of the operator $\bm{\mathcal{L}}_{\mathcal{S}}^{\mathcal{U}}$ defined by Eq.~(\ref{eq:L_def}) with $\mathcal{U}=\mathbf{q}_{\text{B}}$ and $\mathcal{S}=\mathbf{q}_{\text{B}}\backslash q_{2\text{B}}$. The new operator $\bm{\mathcal{L}}_{\mathbf{q}_{\text{B}}\backslash q_{2\text{B}}}^{\mathbf{q}_{\text{B}}}$ acts on the three dimensional node vector $\mathbf{q}_{\text{B}}=(q_{2\text{B}},q_{3},q_{4})^{\text{T}}$ and extracts only the non-redundant components, i.e., $(q_{3},q_{4})^{\text{T}}$. In this way, the eliminating matrix $\bm{\mathcal{Z}}_{\text{toy}}^{\text{elm}}$ is rewritten as follows.
\begin{equation}
  \bm{\mathcal{Z}}_{\text{toy}}^{\text{elm}} = \left(\begin{array}{cc}
    \mathbf{I}_{2\times 2}  &  \mathbf{0}  \\
            \mathbf{0}          &  \bm{\mathcal{L}}_{\mathbf{q}_{\text{B}}\backslash q_{2\text{B}}}^{\mathbf{q}_{\text{B}}}
  \end{array}\right).
\label{eq:def_Z_toy_elm}
\end{equation}
Since each row of the eliminating matrix $\bm{\mathcal{Z}}_{\text{toy}}^{\text{elm}}$ includes only one unity component for the toy model, each row of the matrix $\bm{\mathcal{Z}}_{\text{toy}}^{\text{elm}}$ extracts only one row of the matrix $\mathbf{\Psi}_{\text{toy}}^{\text{rdn}}$ by multiplying from the left as in Eq.~(\ref{eq:derivation_Psi_toy}). In summary, the upper two rows of the matrix $\bm{\mathcal{Z}}_{\text{toy}}^{\text{elm}}$, which consists of the identity matrix and the zero matrix, extracts the first and second rows of the matrix $\mathbf{\Psi}_{\text{toy}}^{\text{rdn}}$, and the lower two rows, which consists of the zero matrix and the operator $\bm{\mathcal{L}}_{\mathbf{q}_{\text{B}}\backslash q_{2\text{B}}}^{\mathbf{q}_{\text{B}}}$, extracts the fourth and fifth rows. In this way, the eliminating matrix $\bm{\mathcal{Z}}_{\text{toy}}^{\text{elm}}$ eliminates the redundant row(s) of the matrix $\mathbf{\Psi}_{\text{toy}}^{\text{rdn}}$.

As in the eliminating matrix $\bm{\mathcal{Z}}_{\text{toy}}^{\text{elm}}$, the upper left block of the combining matrix $\bm{\mathcal{Z}}_{\text{toy}}^{\text{cmb}}$ is the identity matrix, the size of which is the same as the number of components of the vector $\mathbf{q}_{\text{A}}$, $2$ for the toy model, and the upper right block is the zero matrix. The lower left block, mostly composed of zeros as described above, represents a table of duplicated components between $\mathbf{q}_{\text{A}}$ and $\mathbf{q}_{\text{B}}$, which corresponds indeed to a transpose of the matrix $\mathbf{R}$ defined by Eq.~(\ref{eq:def_R}) for the original hybrid S-ODE model. Namely, we can define the $\mathbf{R}$ matrix for the toy model, referred to as $\mathbf{R}_{\text{toy}}$, by considering the operators corresponding to Eq.~(\ref{eq:def_L_Gamma_pre}) for the toy model, denoted by $\bm{\mathcal{L}}_{q_{2\text{A}}}^{\mathbf{q}_{\text{A}}}$ and $\bm{\mathcal{L}}_{q_{2\text{B}}}^{\mathbf{q}_{\text{B}}}$ that extract the duplicated values $q_{2\text{A}}$ and $q_{2\text{B}}$, respectively, from the vector $\mathbf{q}_{\text{A}}$ and $\mathbf{q}_{\text{B}}$. Specifically, $\mathbf{R}_{\text{toy}}$ matrix is defined as
\begin{equation}
  \mathbf{R}_{\text{toy}} = \left(\bm{\mathcal{L}}_{q_{2\text{A}}}^{\mathbf{q}_{\text{A}}}\right)^{\text{T}} \bm{\mathcal{L}}_{q_{2\text{B}}}^{\mathbf{q}_{\text{B}}} = \left(\begin{array}{c}
    0  \\
    1
  \end{array}\right) \left(\begin{array}{ccc}
    1  &  0  &  0
  \end{array}\right) = \left(\begin{array}{ccc}
    0  &  0  &  0  \\
    1  &  0  &  0
  \end{array}\right).
\end{equation}
Moreover, the lower right block of the matrix $\bm{\mathcal{Z}}_{\text{toy}}^{\text{cmb}}$ can be represented by the transpose of the operator $\bm{\mathcal{L}}_{\mathbf{q}_{\text{B}}\backslash q_{2\text{B}}}^{\mathbf{q}_{\text{B}}}$. Thus, the combining matrix $\bm{\mathcal{Z}}_{\text{toy}}^{\text{cmb}}$ can be rewritten as
\begin{equation}
  \bm{\mathcal{Z}}_{\text{toy}}^{\text{cmb}} = \left(\begin{array}{cc}
              \mathbf{I}_{2\times 2}             &  \mathbf{0}  \\
    \left(\mathbf{R}_{\text{toy}}\right)^{\text{T}}  &  \left(\bm{\mathcal{L}}_{\mathbf{q}_{\text{B}}\backslash q_{2\text{B}}}^{\mathbf{q}_{\text{B}}}\right)^{\text{T}}
  \end{array}\right).
\label{eq:def_Z_toy_cmb}
\end{equation}
Right multiplication by the combining matrix $\bm{\mathcal{Z}}_{\text{toy}}^{\text{cmb}}$ causes a linear combination of columns of the matrix $\mathbf{\Psi}_{\text{toy}}^{\text{rdn}}$ in Eq.~(\ref{eq:derivation_Psi_toy}).

\subsubsection{Markov chain approximation with a proper state transition probability matrix}
As in the simple toy model, the matrix $\mathbf{\Psi}^{\text{rdn}}$ in Eq.~(\ref{eq:Markov_pre}) for the hybrid S-ODE model is redundant, and it does not satisfy the properties required for a state transition probability matrix. As discussed above, it is due to the nodes located on the switching boundary with duplicated indexing and their duplicated probabilities $\mathbf{p}_{\text{A}}$ and $\mathbf{p}_{\text{B}}$. More specifically, state transition probabilities from a node on the switching boundary, whose position is represented with duplication by $\mathbf{x}_{\text{A},n_{\text{A},i}^{\text{s}}}$ and $\mathbf{x}_{\text{B},n_{\text{B},i}^{\text{s}}}$ ($i=1,2,\ldots,N^{\text{s}}$), are split into the $n_{\text{A},i}^{\text{s}}$-th column and the $(N_{\text{A}}+n_{\text{B},i}^{\text{s}})$-th column of $\mathbf{\Psi}^{\text{rdn}}$. Moreover, state transition probabilities into the node $\mathbf{x}_{\text{A},n_{\text{A},i}^{\text{s}}}$ and those into the node $\mathbf{x}_{\text{B},n_{\text{B},i}^{\text{s}}}$ from the other nodes are repeatedly represented as the $n_{\text{A},i}^{\text{s}}$-th row and the $(N_{\text{A}}+n_{\text{B},i}^{\text{s}})$-th row of $\mathbf{\Psi}^{\text{rdn}}$.

In order to eliminate those duplications and obtain a Markov chain model represented by a proper state transition probability matrix, we introduce once again another notation $\mathbf{x}_{n}$ for the node position located over the whole domain $\Lambda$, where $n$($=1,2,\ldots,(N-N^{\text{s}})$) is the serial number assigned to each node with no duplications, where $N=N_{\text{A}}+N_{\text{B}}$. The number of node for the new serial number index $n$ is not $N$, but it is $(N-N^{\text{s}})$, because there is no duplication for indexing the nodes on the switching boundary in this case. Moreover, we denote the value of the node $\mathbf{x}_{n}$ at time $t$ by $p_{n}(t)$, and define the node value vector indexed by $n$ as
\begin{equation}
  \mathbf{p} = \left(p_{1},\ldots,p_{n},\ldots,p_{N-N^{\text{s}}}\right)^{\text{T}}.
\label{eq:def_p}
\end{equation}
We then define the eliminating matrix $\bm{\mathcal{Z}}^{\text{elm}}$ and the combining matrix $\bm{\mathcal{Z}}^{\text{cmb}}$ for the original hybrid S-ODE model as follows:
\begin{equation}
  \bm{\mathcal{Z}}^{\text{elm}} = \left(\begin{array}{cc}
    \mathbf{I}_{N_{\text{A}}\times N_{\text{A}}}  &  \mathbf{0}  \\
                  \mathbf{0}                      &  \bm{\mathcal{L}}_{\Lambda_{\text{B}}\backslash\Gamma_{\text{B}}^{\text{s}}}^{\Lambda_{\text{B}}}
  \end{array}\right),\hspace{1cm}\bm{\mathcal{Z}}^{\text{cmb}} = \left(\begin{array}{cc}
    \mathbf{I}_{N_{\text{A}}\times N_{\text{A}}}  &  \mathbf{0}  \\
                \mathbf{R}^{\text{T}}             &  \left(\bm{\mathcal{L}}_{\Lambda_{\text{B}}\backslash\Gamma_{\text{B}}^{\text{s}}}^{\Lambda_{\text{B}}}\right)^{\text{T}}
  \end{array}\right).
\label{eq:def_Z_elm_cmb}
\end{equation}
In this case, $\bm{\mathcal{L}}_{\Lambda_{k}\backslash\Gamma_{k}^{\text{s}}}^{\Lambda_{k}}$ for $k\in\{\text{A},\text{B}\}$ is the operator $\bm{\mathcal{L}}_{\mathcal{S}}^{\mathcal{U}}$ defined in Eq.~(\ref{eq:L_def}) for the case with $\mathcal{U}=\Lambda_{k}$ and $\mathcal{S}=\Lambda_{k}\backslash\Gamma_{k}^{\text{s}}$ to extract values of nodes that are not located on the switching boundary $\Gamma_{k}^{\text{s}}$ from $\mathbf{p}_{k}(t)$, and the size of the operator $\bm{\mathcal{L}}_{\Lambda_{k}\backslash\Gamma_{k}^{\text{s}}}^{\Lambda_{k}}$ is $(N_{k}-N^{\text{s}})\times N_{k}$. Therefore, $\bm{\mathcal{Z}}^{\text{elm}}$ is the $(N-N^{\text{s}})\times N$ matrix and $\bm{\mathcal{Z}}^{\text{cmb}}$ is the $N\times(N-N^{\text{s}})$ matrix. Note that the operation of $\bm{\mathcal{L}}_{\Lambda_{k}\backslash\Gamma_{k}^{\text{s}}}^{\Lambda_{k}}$ is complementary to that of $\bm{\mathcal{L}}_{\Gamma_{k}^{\text{s}}}^{\Lambda_{k}}$ defined by Eq.~(\ref{eq:def_L_Gamma_pre}).

Using the operators of $\bm{\mathcal{Z}}^{\text{elm}}$ and $\bm{\mathcal{Z}}^{\text{cmb}}$, as in the simple toy model case, we can obtain the proper state transition probability matrix of the original hybrid S-ODE model, referred to as $\mathbf{\Psi}$, from the redundant matrix $\mathbf{\Psi}^{\text{rdn}}$ as
\begin{equation}
  \mathbf{\Psi} = \bm{\mathcal{Z}}^{\text{elm}} \mathbf{\Psi}^{\text{rdn}} \bm{\mathcal{Z}}^{\text{cmb}},
\label{eq:def_Psi}
\end{equation}
which can be validated as follows:
\begin{equation}
  \mathbf{p}(t+\delta t) = \bm{\mathcal{Z}}^{\text{elm}} \left(\begin{array}{c}
    \mathbf{p}_{\text{A}}(t+\delta t)  \\
    \mathbf{p}_{\text{B}}(t+\delta t)
  \end{array}\right) = \bm{\mathcal{Z}}^{\text{elm}} \mathbf{\Psi}^{\text{rdn}} \left(\begin{array}{c}
    \mathbf{p}_{\text{A}}(t)  \\
    \mathbf{p}_{\text{B}}(t)
  \end{array}\right) = \bm{\mathcal{Z}}^{\text{elm}} \mathbf{\Psi}^{\text{rdn}} \bm{\mathcal{Z}}^{\text{cmb}} \mathbf{p}(t) \equiv \mathbf{\Psi} \mathbf{p}(t).
\label{eq:Markov}
\end{equation}

The initial value problem of Eq.~(\ref{eq:Markov}) can be solved numerically, based on the iterative multiplications of the matrix $\mathbf{\Psi}$, by which time evolution of PDF from a given initial distribution can be simulated.

\subsection{Solving stationary PDF of the Markov chain}
The steady-state solution of Eq.~(\ref{eq:FEEq_2D_hybrid}), referred to as $(\bar{\mathbf{p}}_{\text{A}}^{\text{T}},\bar{\mathbf{p}}_{\text{B}}^{\text{T}})^{\text{T}}$, is a stationary PDF of the FP equations, which can be obtained by putting $\partial \mathbf{p}_{k}(t)/\partial t=\mathbf{0}$ in Eq.~(\ref{eq:FEEq_2D_hybrid}) and solving the following algebraic equation:
\begin{equation}
  \left(\begin{array}{cc}
    \mathbf{K}_{\text{A}}  &        \mathbf{0}  \\
          \mathbf{0}       &  \mathbf{K}_{\text{B}}  \\
  \end{array}\right) \left(\begin{array}{c}
    \bar{\mathbf{p}}_{\text{A}}  \\
    \bar{\mathbf{p}}_{\text{B}}
  \end{array}\right) = \left(\begin{array}{c}
    \bar{\mathbf{b}}_{\text{A}}^{\text{d}} + \bar{\mathbf{b}}_{\text{A}}^{\text{s}}  \\
    \bar{\mathbf{b}}_{\text{B}}^{\text{d}} + \bar{\mathbf{b}}_{\text{B}}^{\text{s}}
  \end{array}\right),
\label{eq:FEEq_2D_hybrid_SS}
\end{equation}
where $\bar{\mathbf{b}}_{k}^{\text{d}}$ and $\bar{\mathbf{b}}_{k}^{\text{s}}$ are $\mathbf{b}_{k}^{\text{d}}(t)$ and $\mathbf{b}_{k}^{\text{s}}(t)$ in the steady state. Substituting Eq.~(\ref{eq:zero_flux_mod_2}) and Eq.~(\ref{eq:flux_conservation_mod_2}) into Eq.~(\ref{eq:FEEq_2D_hybrid_SS}), we have
\begin{equation}
  \left(\begin{array}{cc}
    \mathbf{K}_{\text{A}} - \mathbf{K}_{\text{A}}^{\text{d}} - \mathbf{K}_{\text{A}}^{\text{s}}  & \mathbf{R} \left(\mathbf{K}_{\text{B}} - \mathbf{K}_{\text{B}}^{\text{d}} - \mathbf{K}_{\text{B}}^{\text{s}}\right)  \\
          \mathbf{0}       &  \mathbf{K}_{\text{B}} - \mathbf{K}_{\text{B}}^{\text{d}}  \\
  \end{array}\right) \left(\begin{array}{c}
    \bar{\mathbf{p}}_{\text{A}}  \\
    \bar{\mathbf{p}}_{\text{B}}
  \end{array}\right) = \left(\begin{array}{c}
    \mathbf{0}  \\
    \bar{\mathbf{b}}_{\text{B}}^{\text{s}}
  \end{array}\right).
\label{eq:FEEq_2D_hybrid_SS_2}
\end{equation}
Once again, as in the previous section, $\bar{\mathbf{b}}_{\text{B}}^{\text{s}}$ should be expressed by using $\bar{\mathbf{p}}_{\text{A}}$ and $\bar{\mathbf{p}}_{\text{B}}$. To this end, we rewrite Eq.~(\ref{eq:L_gamma_p}) for $\bar{\mathbf{b}}_{\text{B}}^{\text{s}}$ as
\begin{equation}
  \left(\begin{array}{cc}
    \bm{\mathcal{L}}_{\Gamma_{\text{A}}^{\text{s}}}^{\Lambda_{\text{A}}}  &  -\bm{\mathcal{L}}_{\Gamma_{\text{B}}^{\text{s}}}^{\Lambda_{\text{B}}}
  \end{array}\right) \left(\begin{array}{c}
    \bar{\mathbf{p}}_{\text{A}}  \\
    \bar{\mathbf{p}}_{\text{B}}
  \end{array}\right) = \mathbf{0}.
\label{eq:value_atSwitchingBoundary_set}
\end{equation}
Operating $\bm{\mathcal{L}}_{\Lambda_{\text{B}}\backslash\Gamma_{\text{B}}^{\text{s}}}^{\Lambda_{\text{B}}}$ on both sides of the second row of Eq.~(\ref{eq:FEEq_2D_hybrid_SS_2}), we have 
\begin{equation}
  \bm{\mathcal{L}}_{\Lambda_{\text{B}}\backslash\Gamma_{\text{B}}^{\text{s}}}^{\Lambda_{\text{B}}} \left(\mathbf{K}_{\text{B}} - \mathbf{K}_{\text{B}}^{\text{d}}\right) \bar{\mathbf{p}}_{\text{B}} = \bm{\mathcal{L}}_{\Lambda_{\text{B}}\backslash\Gamma_{\text{B}}^{\text{s}}}^{\Lambda_{\text{B}}} \bar{\mathbf{b}}_{\text{B}}^{\text{s}} = \mathbf{0},
\label{eq:bar_pB_bar_bBs}
\end{equation}
where the second equality holds, because the components of $\bar{\mathbf{b}}_{\text{B}}^{\text{s}}$ (and $\mathbf{b}_{\text{B}}^{\text{s}}$) corresponding to the nodes located not on the switching boundary are zero. In this way, Eq.~(\ref{eq:FEEq_2D_hybrid_SS_2}) together with Eq.~(\ref{eq:value_atSwitchingBoundary_set}) is rewritten as
\begin{equation}
  \left(\begin{array}{cc}
    \mathbf{K}_{\text{A}} - \mathbf{K}_{\text{A}}^{\text{d}} - \mathbf{K}_{\text{A}}^{\text{s}}  & \mathbf{R} \left(\mathbf{K}_{\text{B}} - \mathbf{K}_{\text{B}}^{\text{d}} - \mathbf{K}_{\text{B}}^{\text{s}}\right)  \\
          \mathbf{0}       &  \bm{\mathcal{L}}_{\Lambda_{\text{B}}\backslash\Gamma_{\text{B}}^{\text{s}}}^{\Lambda_{\text{B}}} \left(\mathbf{K}_{\text{B}} - \mathbf{K}_{\text{B}}^{\text{d}}\right)  \\
    \bm{\mathcal{L}}_{\Gamma_{\text{A}}^{\text{s}}}^{\Lambda_{\text{A}}}  &  - \bm{\mathcal{L}}_{\Gamma_{\text{B}}^{\text{s}}}^{\Lambda_{\text{B}}}
  \end{array}\right) \left(\begin{array}{c}
    \bar{\mathbf{p}}_{\text{A}}  \\
    \bar{\mathbf{p}}_{\text{B}}
  \end{array}\right) \equiv \mathbf{K}^{\dagger} \left(\begin{array}{c}
    \bar{\mathbf{p}}_{\text{A}}  \\
    \bar{\mathbf{p}}_{\text{B}}
  \end{array}\right) = \mathbf{0}.
\label{eq:FEEq_2D_hybrid_SS_3}
\end{equation}
One can show that $\mathbf{K}^{\dagger}$ is singular, and thus, a solution $(\bar{\mathbf{p}}_{\text{A}}^{\text{T}},\bar{\mathbf{p}}_{\text{B}}^{\text{T}})^{\text{T}}$ of the Eq.~(\ref{eq:FEEq_2D_hybrid_SS_3}), i.e., the stationary PDF of Eq.~(\ref{eq:FEEq_2D_hybrid}), can be obtained as the unit vector spanned by the basis vectors of the kernel of $\mathbf{K}^{\dagger}$. Then, the vector $\bar{\mathbf{p}}$ without duplication is obtained as
\begin{equation}
  \bar{\mathbf{p}} = \bm{\mathcal{Z}}^{\text{elm}} \left(\begin{array}{c}
    \bar{\mathbf{p}}_{\text{A}}  \\
    \bar{\mathbf{p}}_{\text{B}}
  \end{array}\right).
\label{eq:SteadyState_withoutDuplication}
\end{equation}

\subsection{Power spectral density of postural sway}
We calculate the power spectral density (PSD) of postural sway during stationary quiet stance generated by the ODE-approximated intermittent control model by using $\bar{\mathbf{p}}$ as the stationary PDF obtained by solving Eq.~(\ref{eq:SteadyState_withoutDuplication}). To this end, we consider the stochastic process $\mathbf{X}[i]=(\Theta[i],\Omega[i])^{\text{T}}$ that takes the two-dimensional state $\mathbf{x}_{n[i]}=(\theta_{n[i]},\omega_{n[i]})^{\text{T}}$ from a set of the finite states $\{\mathbf{x}_{n}\}=\{(\theta_{n},\omega_{n})^{\text{T}}\}$ for $n=1,\ldots,(N-N^{\text{s}})$ as the function of the discrete time $t=i\delta t$ for $i\in\mathbb{Z}$. Consider a sample stationary time series of the length $j$, denoted by
\begin{equation}
  \mathbf{x}_{n[i]},\ \mathbf{x}_{n[i+1]},\ \cdots,\ \mathbf{x}_{n[i+j]},
\end{equation}
which represents a sample trajectory of
\begin{equation}
  \left(\begin{array}{c}
    \theta_{n[i]} \\
    \omega_{n[i]}
  \end{array}\right) \to \left(\begin{array}{c}
    \theta_{n[i+1]} \\
    \omega_{n[i+1]}
  \end{array}\right) \to\cdots\to \left(\begin{array}{c}
    \theta_{n[i+j]} \\
    \omega_{n[i+j]}
  \end{array}\right)
\end{equation}
in the discretized $\theta$-$\omega$ plane, and the postural sway time series of
\begin{equation}
  \theta_{n[i]},\ \theta_{n[i+1]},\ \cdots,\ \theta_{n[i+j]}.
\end{equation}
By the stationarity assumption,
\begin{equation}
  \text{Prob}(\mathbf{X}[i]=\mathbf{x}_{n[i]}) = \left(\boldsymbol{\ell}_{\mathbf{x}_{n[i]}}^{\Lambda}\right)^{\text{T}} \bar{\mathbf{p}},
\label{eq:Prob_i}
\end{equation}
where $\boldsymbol{\ell}_{\mathbf{x}_{n[i]}}^{\Lambda}$ is the operator $\boldsymbol{\ell}_{\mathbf{x}_{\mathcal{S}_{\sharp}}}^{\mathcal{U}}$ for the case with $\mathcal{U}=\Lambda$ and $\mathcal{S}_{\sharp}=n[i]$ to extract the value of node $\mathbf{x}_{n[i]}$ from the vector $\mathbf{p}$. More specifically, $\boldsymbol{\ell}_{\mathbf{x}_{n[i]}}^{\Lambda}$ is a $(N-N^{\text{s}})$-dimensional vector, in which all components are zero except that the $n[i]$-th component is unity. Moreover, by the Markov property of the process,
\begin{equation}
  \text{Prob}(\mathbf{X}[i+j]=\mathbf{x}_{n[i+j]} \mid \mathbf{X}[i]=\mathbf{x}_{n[i]}) = \left(\boldsymbol{\ell}_{\mathbf{x}_{n[i+j]}}^{\Lambda}\right)^{\text{T}} \bm{\Psi}^{j} \boldsymbol{\ell}_{\mathbf{x}_{n[i]}}^{\Lambda},
\label{eq:Prob_i+j}
\end{equation}
using the state transition probability matrix $\mathbf{\Psi}$ in Eq.~(\ref{eq:def_Psi}) and the operator $\boldsymbol{\ell}_{\mathbf{x}_{n[i]}}^{\Lambda}$.

Then, the autocorrelation function $R_{\theta}(j)$ of the stochastic process $\Theta[i]$ for the time lag $j$ is defined as
\begin{equation}
  R_{\Theta}(j) = E\left[\Theta[i] \Theta[i+j]\right],
\label{eq:R_theta}
\end{equation}
where $E$ represents the expectation operator. Using Eqs.~(\ref{eq:Prob_i}) and (\ref{eq:Prob_i+j}), Eq.~(\ref{eq:R_theta}) can be detailed as
\begin{align}
  R_{\theta}(j) &= \sum_{n=1}^{N-N^{\text{s}}} \theta_{n} P(\mathbf{X}[i]=\mathbf{x}_{n}) \sum_{n^{\prime}=1}^{N-N^{\text{s}}} \theta_{n^{\prime}} P(\mathbf{X}[i+j]=\mathbf{x}_{n^{\prime}} \mid \mathbf{X}[i]=\mathbf{x}_{n}) \nonumber \\
  &= \sum_{n=1}^{N-N^{\text{s}}} \theta_{n} \left(\boldsymbol{\ell}_{\mathbf{x}_{n}}^{\Lambda}\right)^{\text{T}} \bar{\mathbf{p}} \sum_{n^{\prime}=1}^{N-N^{\text{s}}} \theta_{n^{\prime}} \left(\boldsymbol{\ell}_{\mathbf{x}_{n^{\prime}}}^{\Lambda}\right)^{\text{T}} \bm{\Psi}^{j} \boldsymbol{\ell}_{\mathbf{x}_{n}}^{\Lambda}  \nonumber\\
  &= \bm{\uptheta}^{\text{T}} \bm{\Psi}^{j} \sum_{n=1}^{N-N^{\text{s}}} \boldsymbol{\ell}_{\mathbf{x}_{n}}^{\Lambda} \left(\theta_{n} \left(\boldsymbol{\ell}_{\mathbf{x}_{n}}^{\Lambda}\right)^{\text{T}} \bar{\mathbf{p}}\right)  \nonumber\\
  &= \bm{\uptheta}^{\text{T}} \bm{\Psi}^{j} \left(\bm{\uptheta} \circ \bar{\mathbf{p}}\right)
\label{eq:AutoCorr}
\end{align}
where $\bm{\uptheta}$ is the $(N-N^{\text{s}})$-dimensional column vector defined as
\begin{equation}
  \bm{\uptheta} = \left(\begin{array}{cccc}
    \theta_{1},  &  \theta_{2},  &  \cdots,  &  \theta_{N-N^{\text{s}}}
  \end{array}\right)^{\text{T}}
\end{equation}
with $\theta_{n}$ represents the value of $\theta$ for the node $\mathbf{x}_{n}$ in the $\theta$-$\omega$ plane. Moreover, The operate $\circ$ represents the Hadamard product.

PSD of the stochastic process of $\Theta$ in its stationary regime is then calculated by performing the fast Fourier transform of $R_{\theta}(j)$ in Eq.~(\ref{eq:AutoCorr}).

\section{Dynamics of the Markov chain model: comparison with Monte Carlo simulations of S-DDE and S-ODE models}
\label{sec:ComparisonWithMonteCarloSimulations}
\subsection{Numerical setup for FEM}
We numerically calculated time evolution and steady-state solutions of the Markov chain approximate of the hybrid FP equations for the ODE-approximate of the intermittent control model in Eq.~(\ref{eq:StateSpaceRep_int}), using Eq.~(\ref{eq:Markov}) and Eq.~(\ref{eq:SteadyState_withoutDuplication}), respectively. Basically, parameters of the model were set as values listed in Table \ref{tab:ModelParameters}, but values of some parameters are altered when we investigate how the PDF changes as a function of those parameters. The entire domain $\Lambda$ was defined as the rectangular area with $-0.1\le\theta\le0.1$ and $-0.02\le\omega\le0.02$, which was divided into finite elements, according to the nodes aligned as small rectangular grids at every 0.0005 rad in $\theta$-direction ($\mathrm{d}\theta=0.0005$) and 0.0002 rad/s in $\omega$-direction ($\mathrm{d}\omega=0.0002$). Then, each finite element was defined as a right triangle composed of nearest three nodes. For a node $(\tilde{\theta},\tilde{\omega})$, the other nodes of the corresponding triangle element are expressed as $(\tilde{\theta}+\mathrm{d}\theta,\tilde{\omega})$ and $(\tilde{\theta},\tilde{\omega}+\mathrm{d}\omega)$ or $(\tilde{\theta}-\mathrm{d}\theta,\tilde{\omega})$ and $(\tilde{\theta},\tilde{\omega}-\mathrm{d}\omega)$. The intermittent control model analyzed in this study has two switching boundaries of straight lines that separate $\Lambda_{\text{A}}=\text{S}_{\text{on}}$ and $\Lambda_{\text{B}}=\text{S}_{\text{off}}$, i.e., $\theta=0$ and $\omega=a\theta$. The boundary with $\theta=0$ was simply expressed by the nodes located on $\theta=0$. The boundary with $\omega=a\theta$ was expressed by the polygonal line, composed of connected side lines of finite elements along the line of $\omega=a\theta$.

For analyzing time evolution of PDF, an initial PDF was set as a two-dimensional Gaussian distribution with the mean located at $(\theta,\omega)=(0.01,0)$ and the standard deviations $0.001$ rad in $\theta$-direction and 0.001 rad/s in $\omega$-direction. Time step $\delta t$ to define the Markov chain model in Eq.~(\ref{eq:Markov}) was set as $1.0\times10^{-4}$ s.

\subsection{Numerical setup for Monte Carlo simulations of S-DDE and S-ODE models}
Alternative to solving the FP equations, a time evolution of PDF of the model can also be obtained based on the ensemble average of Monte Carlo simulations of the S-DDE in Eq.~(\ref{eq:MotionEquation}) with Eqs.~(\ref{eq:AnkleTorque}) and (\ref{eq:ActiveTorque_Int}) as well as the S-ODE in Eq.~(\ref{eq:StateSpaceRep_int}) as the original and the ODE-approximated intermittent control models, respectively. For both cases, the S-DDE and S-ODE were numerically integrated using Euler-Maruyama method \cite{Kloeden_2011} with time step $\delta t=1.0\times10^{-4}$ s. We simulated 50,000 sample paths of Monte Carlo simulation, each of which spans for 50 s. For every time step at $t$, a state of $(\theta(t),\omega(t))$, referred to here as a ``particle'', for each of 50,000 sample paths on the $\theta$-$\omega$ plane was allocated into one of the finite elements used for the FEM, and the number of particle in each element was counted to obtain a histogram of the particles for their spatial distribution. The histogram normalized by the total number of the particles (50,000) for each time step corresponds to the PDF of the FP equations. In this way, we validate the PDF of the Markov chain model by comparing it with the PDFs obtained by Monte Carlo simulations of the S-DDE and the S-ODE models.

\subsection{Time evolutions of PDF}
\begin{figure}[htbp]
  \includegraphics[keepaspectratio=true,width=130mm]{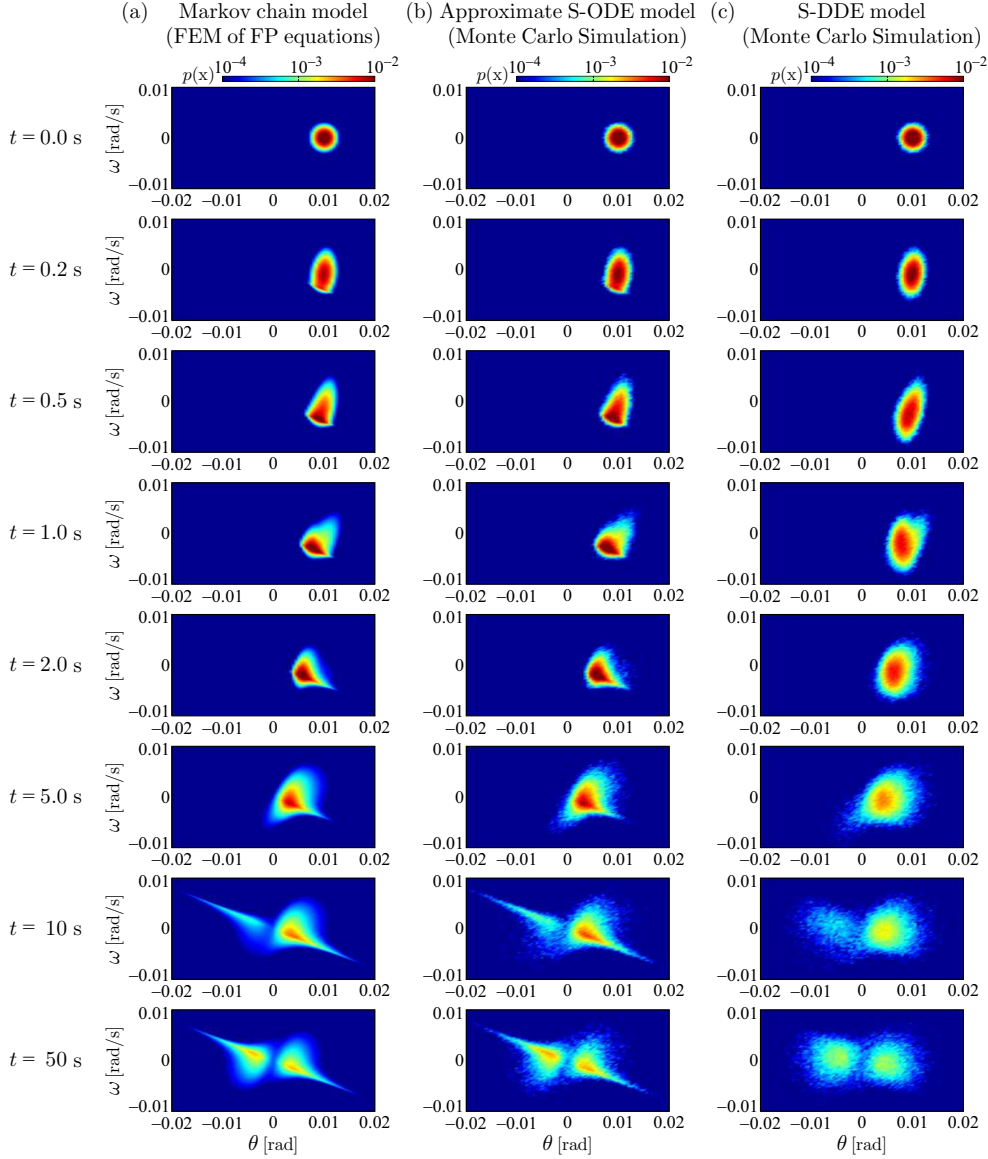}
  \caption{Time evolutions of PDFs of the Markov chain model (FEM simulation of the hybrid FP equations) in comparison with those obtained by Monte Carlo simulations of the S-ODE and S-DDE of the intermittent control model. In each panel, values of probability $\mathbf{p}$ are color-coded. Column (a) A time evolution of PDF of the Markov chain model, FEM-based model of the switched FP equations for the approximate S-ODE model using Eq.~(\ref{eq:Markov}). Column (b) A time evolution of PDF of approximate S-ODE model, obtained by Monte Carlo simulation. Column (c) A time evolution of PDF of the original S-DDE model, obtained by Monte Carlo simulation. Numerical simulations for each model was performed with time step $\delta t=1.0\times10^{-4}$. In (b) and (c), the histograms (PDFs) were obtained based on $50,000$ sample paths of Monte Carlo simulations.}
  \label{fig:Fig_03.eps}
\end{figure}

Figure \ref{fig:Fig_03.eps} exemplifies a time evolution of PDF for the Markov chain model (FEM simulation of the hybrid FP equations) in Eq.~(\ref{eq:Markov}), in comparison with the corresponding time evolutions of PDFs obtained by Monte Carlo simulations of the S-ODE and S-DDE of the intermittent control model. One can observe similarities between three columns of panels that show dynamic changes in the PDFs of three models. Particularly, time evolutions of the PDFs obtained by the Markov chain model and Monte Carlo simulations of the S-ODE matched very well.

For all of three models, the initial PDFs at $t=0$ s are prepared as the two-dimensional Gaussian distribution centered at $(\theta,\omega)=(0.01,0)$. They moves down (rotates clockwise) toward the switching boundary with elongating vertically ($t=0.5$ s), according to the focal vector filed around the unstable equilibrium point at the origin for the on-subsystem with the parameter values in Table \ref{tab:ModelParameters}. Then, the PDF for each model begins to stay around the switching boundary between the on- and off-subsystems ($t=1.0$-$5.0$ s). This is especially true in the approximate ODE model and the corresponding Markov chain model. After a while, probability begins to grow into the left half of the $\theta$-$\omega$ plane along the switching boundary ($t=10.0$ s). As time passes, the PDF for each model converges to a stationary distribution with two peaks symmetrically located at the left and right half planes. Although shape of the PDF for the S-DDE model is more rounded than the approximate ODE model and the corresponding Markov chain model, the stationary PDFs for three models show a good agreement.

\subsection{Stationary PDF}
\begin{figure}[htbp]
  \includegraphics[keepaspectratio=true,width=\textwidth]{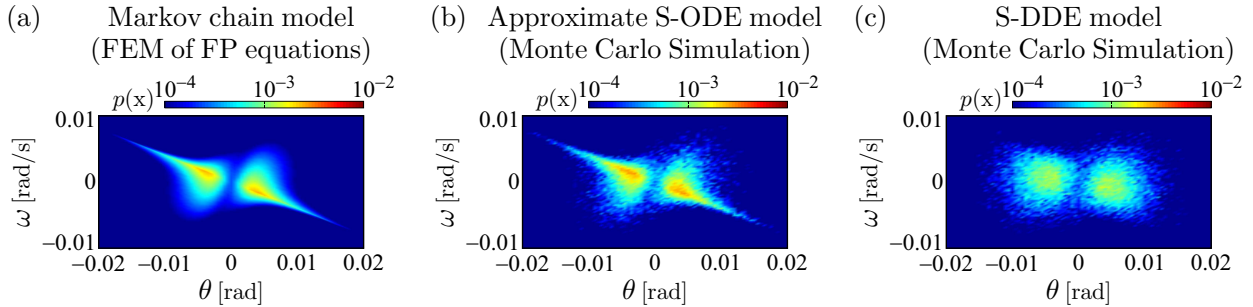}
  \caption{Stationary PDFs for the Markov chain model (FEM simulation of the hybrid FP equations), in comparison with the stationary PDFs obtained by Monte Carlo simulations of the S-ODE and S-DDE of the intermittent control model. In each panel, values of probability $\mathbf{p}$ are color-coded. (a) Stationary PDF of the Markov chain model, obtained by calculating kernel of the matrix $\mathbf{K}^{\dagger}$ in Eq.~(\ref{eq:FEEq_2D_hybrid_SS_3}) and Eq.~(\ref{eq:SteadyState_withoutDuplication}). (b) and (c) Stationary PDFs of the S-ODE and S-DDE at $t=50$ s of Monte Carlo simulations, which are identical with the bottom row of Figs.~\ref{fig:Fig_03.eps}(b) and (c).}
  \label{fig:Fig_04.eps}
\end{figure}

Figure \ref{fig:Fig_04.eps} shows the stationary PDFs for the Markov chain model (FEM simulation of the hybrid FP equations), in comparison with the stationary PDFs obtained through time evolutions of PDFs by Monte Carlo simulations of the S-ODE and S-DDE of the intermittent control model. Figure \ref{fig:Fig_04.eps} looks exactly the same as the bottom row of Fig.~\ref{fig:Fig_03.eps}, but the stationary PDF for the Markov chain model (Fig.~\ref{fig:Fig_04.eps}(a)) was obtained differently. That is, it was obtained by calculating kernel of the matrix $\mathbf{K}^{\dagger}$ in Eq.~(\ref{eq:FEEq_2D_hybrid_SS_3}) and Eq.~(\ref{eq:SteadyState_withoutDuplication}), instead of iterating Eq.~(\ref{eq:Markov}).

\subsection{Parameter-dependent changes in the stationary PDF}
\begin{figure}[htbp]
  \includegraphics[keepaspectratio=true,width=140mm]{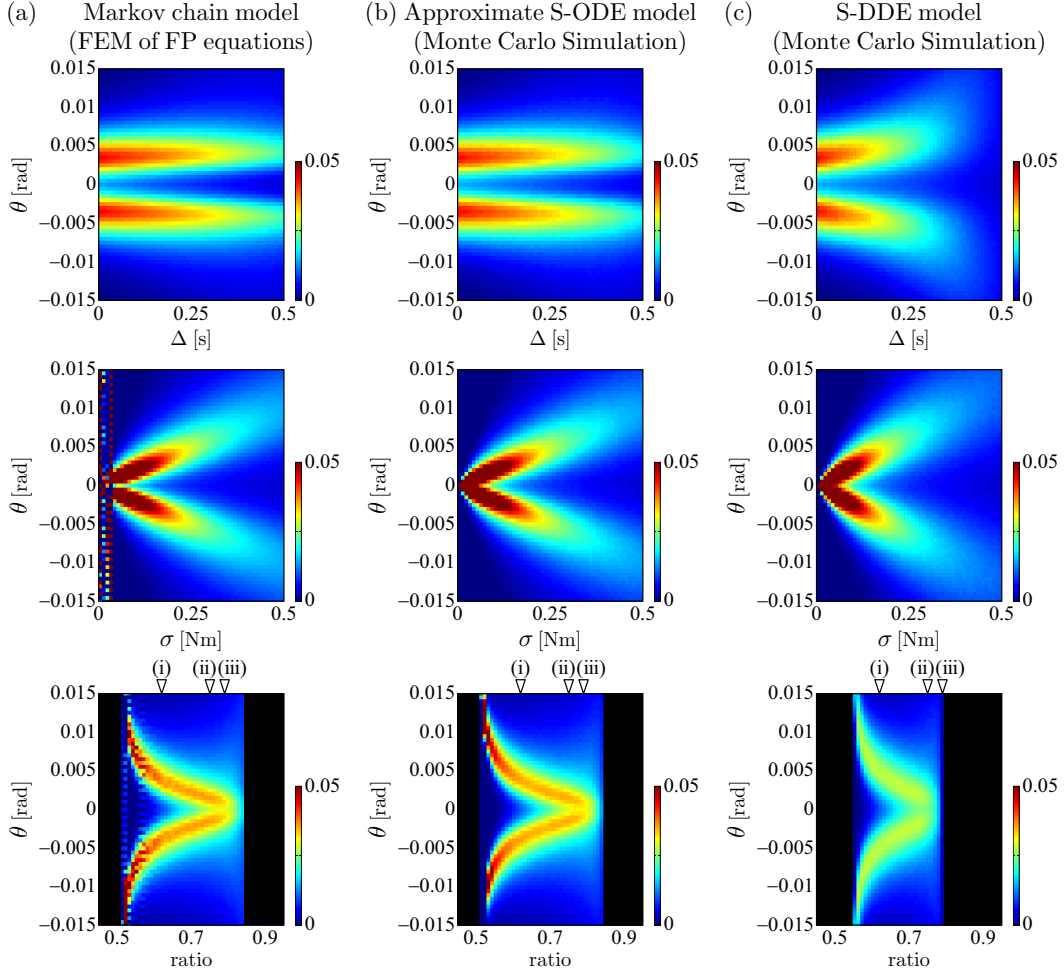}
  \caption{Parameter-dependent stationary distribution of $\theta$ as functions of $\Delta$, $\sigma$, and ratio between the area of $\text{S}_{\text{on}}$ and the whole area of $\text{S}_{\text{on}}\cup\text{S}_{\text{off}}$. In each panel, values of probability $\mathbf{p}$ are color-coded. Black color used for the bottom row means instability of the models. Column (a) Parameter-dependence of stationary distribution for the Markov chain model, which are computed using Eqs.~(\ref{eq:FEEq_2D_hybrid_SS_3}) and (\ref{eq:SteadyState_withoutDuplication}). Columns (b) and (c) Parameter-dependence of stationary distribution for the S-ODE and S-DDE models, respectively, obtained by Monte Carlo simulations. Arrows numbered by (i), (ii) and (iii) indicate the cases with ratio$=0.62$, ratio$=0.75$ and ratio$=0.79$, respectively, which correspond the slope values of $a=-0.4$, $-1.0$ and $-1.3$. These values are used for calculating PSDs in Fig.~\ref{fig:Fig_06.eps}.}
  \label{fig:Fig_05.eps}
\end{figure}

We investigate parameter-dependent changes in the stationary PDFs of model-simulated postural sway to obtain stochastic bifurcation diagrams, where the stationary PDFs for varied parameter values of the model are obtained by calculating kernel of the matrix $\mathbf{K}^{\dagger}$ in Eq.~(\ref{eq:FEEq_2D_hybrid_SS_3}) for the Markov chain model, and by Monte Carlo simulations for the S-ODE and S-DDE models. Specifically, as shown in Fig.~\ref{fig:Fig_05.eps}, we consider the following parameters to change: the feedback delay time $\Delta$, the noise intensity $\sigma$, and the parameter $a$ representing the slope of the switching boundary $\omega=a\theta$. Since $a$ goes to $-\infty$ as the switching boundary becomes close to the $\omega$-axis, we define the alternative parameter that represents the ratio between the area of $\text{S}_{\text{on}}$ and the whole area of $\text{S}_{\text{on}}\cup\text{S}_{\text{off}}$, referred to as the {\it ratio}, which is defined by $\text{ratio}=1/2-\arctan(a)/\pi$ as a function of the slope parameter $a$. That is, ratio$=0.5$ when $a=0$ (the $\theta$-axis is the switching boundary), and ratio$=1.0$ when $a=-\infty$ (the $\omega$-axis is the switching boundary), for which there is no OFF region and the model becomes the continuous control model. We calculated stationary PDFs of the Markov chain model, the S-ODE and S-DDE models, for varied parameter values, with non-varied parameters are fixed as in Table \ref{tab:ModelParameters}. Stationary PDFs are calculated using Eqs.~(\ref{eq:FEEq_2D_hybrid_SS_3}) and (\ref{eq:SteadyState_withoutDuplication}) for the Markov chain model, and using Monte Carlo simulations for the S-ODE and S-DDE models.

Three panels in the top row of Fig.~\ref{fig:Fig_05.eps} are the stationary distributions of $\theta$ as a function of the delay time $\Delta\in[0,0.5]$ s. The S-DDE model could maintain upright posture even for relatively large delay, which can be expressed by the S-ODE and the corresponding Markov chain model. For all models, stationary distributions of $\theta$ do not change significantly as $\Delta$ increases, i.e., two peaks on the left and right half planes as shown in Fig.~\ref{fig:Fig_04.eps}, although kurtosis of those peaks decreases as $\Delta$ increases.

Panels in the middle row of Fig.~\ref{fig:Fig_05.eps} are the stationary distributions of $\theta$ as a function of the noise intensity $\sigma\in[0,0.5]$ s. Diagrams for the S-ODE and the corresponding Markov chain models shows a good agreement with the one for the S-DDE. In all models, two peaks in the left and right half planes become more separated in the left and right directions as $\sigma$ increase. The Markov chain model becomes unreliable for small noise intensity about $\sigma<0.05$, as can be seen from the strange behavior in the left panel of the middle row of Fig.~\ref{fig:Fig_05.eps}. This is due to the fact that the switched FP equations cannot describe deterministic dynamics of the intermittent control model.

Panels in the bottom row of Fig.~\ref{fig:Fig_05.eps} are the stationary distributions of $\theta$ as a function of the $\text{ratio}\in[0.45,0.95]$. Black regions in each diagram indicate the instability of stationary distribution (instability of the upright posture) of the models. That is, $\text{ratio}<0.5$ and $\text{ratio}>0.85$ for the S-ODE model, $\text{ratio}<0.54$ and $\text{ratio}>0.79$ for the S-DDE model. See \cite{Nomura_2013} for detailed stochastic bifurcation analysis of the S-DDE model. For stable parameter regime, as the ratio increases, i.e., as the area of $\text{S}_{\text{off}}$ becomes smaller, $\theta$ becomes more densely distributed around the origin. As mentioned above, the upright posture becomes unstable as the off-region of the intermittent control model is diminished, i.e., as the intermittent control model becomes close to the traditional continuous control model.

\subsection{PSD of simulated postural sway}
\begin{figure}[htbp]
  \includegraphics[keepaspectratio=true,width=100mm]{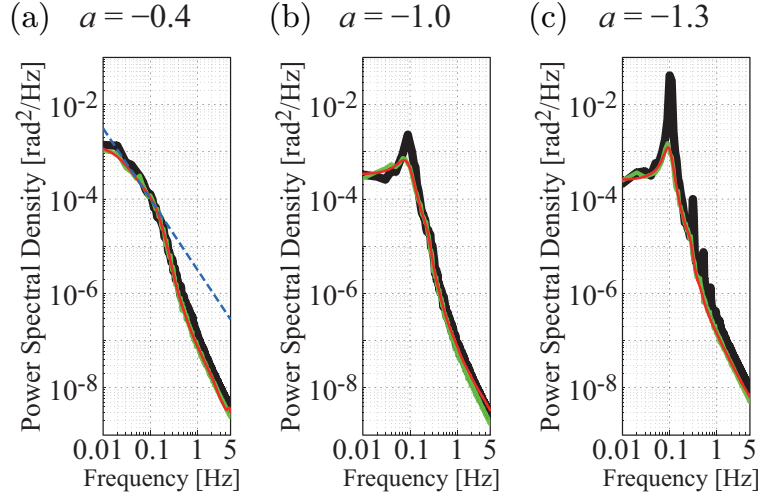}
  \caption{PSDs of model-simulated postural sway $\theta$ for intermittent control model obtained by finite element analysis and Monte Carlo simulations. In each panel, black thick curve is PSD of non-approximated (DDE) intermittent control model calculated from result of Monte Carlo simulations, green curve is that of ODE-approximated intermittent control model, and red curve is PSD of ODE-approximated intermittent control model obtained by finite element analysis. Three slope parameter values of $a$ (and thus the corresponding on-off ratios) are considered for computing the corresponding PSDs. (a) $a=-0.4$ ($\text{ratio}=0.62$). (b) $-1.0$ ($0.75$). (c) $-1.3$ ($0.79$). These three parameters are indicated by the indices of (i), (ii) and (iii) at the bottom row of Fig.~\ref{fig:Fig_05.eps}. In (a), a blue dashed line with the slope of $-3/2$ is depicted as a reference to be compared with the spectrum at the low frequency regime.}
  \label{fig:Fig_06.eps}
\end{figure}

Figure \ref{fig:Fig_06.eps} shows PSDs of model-simulated postural sway $\theta$ for the Markov chain model (FEM simulation of the switched FP equations) in comparison with those obtained by Monte Carlo simulations of the S-ODE and S-DDE of the intermittent control model. For each model, three slope parameter values of $a$ are considered, i.e., $a=-0.4$, $-1.0$, and $-1.3$, which corresponds to $\text{ratio}=0.62$, $0.75$, and $0.79$ that are indicated by the indices of (i), (ii) and (iii) at the bottom row of Fig.~\ref{fig:Fig_05.eps}. In each panel, the PSD with red curve is obtained by Fast Fourier Transform (FFT) of the autocorrelation function in Eq.~(\ref{eq:AutoCorr}). PSDs with green and black curves are for the S-ODE and S-DDE models, for which ensemble averages of FFT of 50 stationary sample paths for 120 s of Monte Carlo simulations are computed. PSDs obtained by Eq.~(\ref{eq:AutoCorr}) for the Markov chain model matches well with those for Monte Carlo simulations of the S-ODE and S-DDE model. Particularly, the PSDs exhibit $f^{-3/2}$-like scaling behavior for the parameter value of $a=-0.4$, a typical value for the intermittent control model as used in the previous study \cite{Asai_2009}.

\section{Concluding Remarks}
We developed a comprehensive numerical recipe to represent and simulate switched-type hybrid Fokker-Planck (FP) equations, which govern time evolutions of probability density functions (PDFs) of two stochastic ordinary differential equations (S-ODEs) that are switched with each other, as finite Markov chain models. To this end, the finite element method was utilized. The major achievement of this study is the derivation of Eq.~(\ref{eq:Markov}) that formulates the Markov chain approximation of switched-type hybrid S-ODEs, and Eqs.~(\ref{eq:FEEq_2D_hybrid_SS_3}) and (\ref{eq:SteadyState_withoutDuplication}) for solving stationary PDFs of the switched-type hybrid S-ODEs. Those formulations can be utilized for a wide range of switched-type S-ODEs expressed by Eqs.~(\ref{eq:SDE_2D_hybrid_A}) and (\ref{eq:SDE_2D_hybrid_B}), as far as solutions of the deterministic version of the switched-type ODE (with no noise) are $C^{0}$-continuous at the switching boundary. For example, typical formulations of model of the sliding-mode control, if it is operated with additive noise, may be in this category. As can be expected from the step-by-step procedure detailed in this paper, the basic idea used for deriving Eq.~(\ref{eq:Markov}) can be extended to more general stochastic hybrid systems exhibiting discontinuous jumps by certain modifications of ways to evaluate the values of $\bm{\upbeta}_{k}(\mathbf{x},t)$ at jump-type-switching boundaries. We then applied the recipe to the intermittent postural control model \cite{Asai_2009}, in which dynamics of the model are determined by state-dependent switching between the on-subsystem as a stochastic delay differential equation (S-DDE) and the off-subsystem as a S-ODE. In this study, the S-DDE for the on-subsystem was approximated by using a small delay expansion of the ODE \cite{Stepan_2000}, by which the S-DDE was approximated by the S-ODE. The hybrid FP equations, composed of two S-ODEs were approximated by a finite state Markov chain model under certain assumptions and by using the finite element method (FEM). Then stochastic on-off switching dynamics of the Markov chain model, including time evolutions of PDFs, stationary PDFs, and power spectral density functions (PSD) of model-simulated postural sway were analyzed. We also investigated how the stationary PDF alters as values of important parameters of the model change. Dynamics of the Markov chain model were compared with those of the original intermittent control model based on Monte Carlo simulations, by which the developed numerical recipe and the resultant Markov chain model were validated. We are planning to associate the Markov chain model obtained in this study with a Markov decision process and a reinforcement learning of the postural control strategy in our future study.

\begin{acknowledgments}
This work was supported in part by JSPS/MEXT KAKENHI 20K11989 to YS, 19H04181 and 22H036620 and 22H047750 to TN, and 21J13652 to AN.
\end{acknowledgments}

\vspace{1cm}

\noindent
Correspondence: suzuki@bpe.es.osaka-u.ac.jp or taishin@bpe.es.osaka-u.ac.jp.

\bibliography{Suzuki_2022_RevTex}

\begin{thebibliography}{37}%
\makeatletter
\providecommand \@ifxundefined [1]{%
 \@ifx{#1\undefined}
}%
\providecommand \@ifnum [1]{%
 \ifnum #1\expandafter \@firstoftwo
 \else \expandafter \@secondoftwo
 \fi
}%
\providecommand \@ifx [1]{%
 \ifx #1\expandafter \@firstoftwo
 \else \expandafter \@secondoftwo
 \fi
}%
\providecommand \natexlab [1]{#1}%
\providecommand \enquote  [1]{``#1''}%
\providecommand \bibnamefont  [1]{#1}%
\providecommand \bibfnamefont [1]{#1}%
\providecommand \citenamefont [1]{#1}%
\providecommand \href@noop [0]{\@secondoftwo}%
\providecommand \href [0]{\begingroup \@sanitize@url \@href}%
\providecommand \@href[1]{\@@startlink{#1}\@@href}%
\providecommand \@@href[1]{\endgroup#1\@@endlink}%
\providecommand \@sanitize@url [0]{\catcode `\\12\catcode `\$12\catcode
  `\&12\catcode `\#12\catcode `\^12\catcode `\_12\catcode `\%12\relax}%
\providecommand \@@startlink[1]{}%
\providecommand \@@endlink[0]{}%
\providecommand \url  [0]{\begingroup\@sanitize@url \@url }%
\providecommand \@url [1]{\endgroup\@href {#1}{\urlprefix }}%
\providecommand \urlprefix  [0]{URL }%
\providecommand \Eprint [0]{\href }%
\providecommand \doibase [0]{https://doi.org/}%
\providecommand \selectlanguage [0]{\@gobble}%
\providecommand \bibinfo  [0]{\@secondoftwo}%
\providecommand \bibfield  [0]{\@secondoftwo}%
\providecommand \translation [1]{[#1]}%
\providecommand \BibitemOpen [0]{}%
\providecommand \bibitemStop [0]{}%
\providecommand \bibitemNoStop [0]{.\EOS\space}%
\providecommand \EOS [0]{\spacefactor3000\relax}%
\providecommand \BibitemShut  [1]{\csname bibitem#1\endcsname}%
\let\auto@bib@innerbib\@empty
\bibitem [{\citenamefont {Billman}(2020)}]{Billman_2020}%
  \BibitemOpen
  \bibfield  {author} {\bibinfo {author} {\bibfnamefont {G.~E.}\ \bibnamefont
  {Billman}},\ }\href {https://doi.org/10.3389/fphys.2020.00200} {\bibfield
  {journal} {\bibinfo  {journal} {Frontiers in Physiology}\ }\textbf {\bibinfo
  {volume} {11}},\ \bibinfo {pages} {200} (\bibinfo {year} {2020})}\BibitemShut
  {NoStop}%
\bibitem [{\citenamefont {Bernhardt}\ \emph {et~al.}(2020)\citenamefont
  {Bernhardt}, \citenamefont {O'Connor}, \citenamefont {Sunday},\ and\
  \citenamefont {Gonzalez}}]{Bernhardt_2020}%
  \BibitemOpen
  \bibfield  {author} {\bibinfo {author} {\bibfnamefont {J.~R.}\ \bibnamefont
  {Bernhardt}}, \bibinfo {author} {\bibfnamefont {M.~I.}\ \bibnamefont
  {O'Connor}}, \bibinfo {author} {\bibfnamefont {J.~M.}\ \bibnamefont
  {Sunday}},\ and\ \bibinfo {author} {\bibfnamefont {A.}~\bibnamefont
  {Gonzalez}},\ }\href {https://doi.org/10.1098/rstb.2019.0454} {\bibfield
  {journal} {\bibinfo  {journal} {Philosophical Transactions of the Royal
  Society B: Biological Sciences}\ }\textbf {\bibinfo {volume} {375}},\
  \bibinfo {pages} {20190454} (\bibinfo {year} {2020})},\ \Eprint
  {https://arxiv.org/abs/https://royalsocietypublishing.org/doi/pdf/10.1098/rstb.2019.0454}
  {https://royalsocietypublishing.org/doi/pdf/10.1098/rstb.2019.0454}
  \BibitemShut {NoStop}%
\bibitem [{\citenamefont {Fang}\ \emph {et~al.}(2019)\citenamefont {Fang},
  \citenamefont {Kruse}, \citenamefont {Lu},\ and\ \citenamefont
  {Wang}}]{Fang_2019}%
  \BibitemOpen
  \bibfield  {author} {\bibinfo {author} {\bibfnamefont {X.}~\bibnamefont
  {Fang}}, \bibinfo {author} {\bibfnamefont {K.}~\bibnamefont {Kruse}},
  \bibinfo {author} {\bibfnamefont {T.}~\bibnamefont {Lu}},\ and\ \bibinfo
  {author} {\bibfnamefont {J.}~\bibnamefont {Wang}},\ }\href
  {https://doi.org/10.1103/RevModPhys.91.045004} {\bibfield  {journal}
  {\bibinfo  {journal} {Rev. Mod. Phys.}\ }\textbf {\bibinfo {volume} {91}},\
  \bibinfo {pages} {045004} (\bibinfo {year} {2019})}\BibitemShut {NoStop}%
\bibitem [{\citenamefont {Seydnejad}\ and\ \citenamefont
  {Kitney}(2001)}]{Seydnejad_2001}%
  \BibitemOpen
  \bibfield  {author} {\bibinfo {author} {\bibfnamefont {S.}~\bibnamefont
  {Seydnejad}}\ and\ \bibinfo {author} {\bibfnamefont {R.}~\bibnamefont
  {Kitney}},\ }\href {https://doi.org/10.1109/51.917729} {\bibfield  {journal}
  {\bibinfo  {journal} {IEEE Engineering in Medicine and Biology Magazine}\
  }\textbf {\bibinfo {volume} {20}},\ \bibinfo {pages} {92} (\bibinfo {year}
  {2001})}\BibitemShut {NoStop}%
\bibitem [{\citenamefont {Ivanov}\ \emph {et~al.}(1998)\citenamefont {Ivanov},
  \citenamefont {Amaral}, \citenamefont {Goldberger},\ and\ \citenamefont
  {Stanley}}]{Ivanov_1998}%
  \BibitemOpen
  \bibfield  {author} {\bibinfo {author} {\bibfnamefont {P.~C.}\ \bibnamefont
  {Ivanov}}, \bibinfo {author} {\bibfnamefont {L.~A.~N.}\ \bibnamefont
  {Amaral}}, \bibinfo {author} {\bibfnamefont {A.~L.}\ \bibnamefont
  {Goldberger}},\ and\ \bibinfo {author} {\bibfnamefont {H.~E.}\ \bibnamefont
  {Stanley}},\ }\href {https://doi.org/10.1209/epl/i1998-00366-3} {\bibfield
  {journal} {\bibinfo  {journal} {Europhysics Letters ({EPL})}\ }\textbf
  {\bibinfo {volume} {43}},\ \bibinfo {pages} {363} (\bibinfo {year}
  {1998})}\BibitemShut {NoStop}%
\bibitem [{\citenamefont {Milton}\ \emph {et~al.}(1989)\citenamefont {Milton},
  \citenamefont {Longtin}, \citenamefont {Beuter}, \citenamefont {Mackey},\
  and\ \citenamefont {Glass}}]{Milton_1989}%
  \BibitemOpen
  \bibfield  {author} {\bibinfo {author} {\bibfnamefont {J.~G.}\ \bibnamefont
  {Milton}}, \bibinfo {author} {\bibfnamefont {A.}~\bibnamefont {Longtin}},
  \bibinfo {author} {\bibfnamefont {A.}~\bibnamefont {Beuter}}, \bibinfo
  {author} {\bibfnamefont {M.~C.}\ \bibnamefont {Mackey}},\ and\ \bibinfo
  {author} {\bibfnamefont {L.}~\bibnamefont {Glass}},\ }\href
  {https://doi.org/https://doi.org/10.1016/S0022-5193(89)80135-3} {\bibfield
  {journal} {\bibinfo  {journal} {Journal of Theoretical Biology}\ }\textbf
  {\bibinfo {volume} {138}},\ \bibinfo {pages} {129} (\bibinfo {year}
  {1989})}\BibitemShut {NoStop}%
\bibitem [{\citenamefont {Glass}(2015)}]{Glass_2015}%
  \BibitemOpen
  \bibfield  {author} {\bibinfo {author} {\bibfnamefont {L.}~\bibnamefont
  {Glass}},\ }\href {https://doi.org/10.1063/1.4915529} {\bibfield  {journal}
  {\bibinfo  {journal} {Chaos: An Interdisciplinary Journal of Nonlinear
  Science}\ }\textbf {\bibinfo {volume} {25}},\ \bibinfo {pages} {097603}
  (\bibinfo {year} {2015})},\ \Eprint
  {https://arxiv.org/abs/https://doi.org/10.1063/1.4915529}
  {https://doi.org/10.1063/1.4915529} \BibitemShut {NoStop}%
\bibitem [{\citenamefont {Belair}\ \emph {et~al.}(2021)\citenamefont {Belair},
  \citenamefont {Nekka},\ and\ \citenamefont {Milton}}]{Belair_2021}%
  \BibitemOpen
  \bibfield  {author} {\bibinfo {author} {\bibfnamefont {J.}~\bibnamefont
  {Belair}}, \bibinfo {author} {\bibfnamefont {F.}~\bibnamefont {Nekka}},\ and\
  \bibinfo {author} {\bibfnamefont {J.~G.}\ \bibnamefont {Milton}},\ }\href
  {https://doi.org/10.1063/5.0058345} {\bibfield  {journal} {\bibinfo
  {journal} {Chaos: An Interdisciplinary Journal of Nonlinear Science}\
  }\textbf {\bibinfo {volume} {31}},\ \bibinfo {pages} {060401} (\bibinfo
  {year} {2021})},\ \Eprint
  {https://arxiv.org/abs/https://doi.org/10.1063/5.0058345}
  {https://doi.org/10.1063/5.0058345} \BibitemShut {NoStop}%
\bibitem [{\citenamefont {Maurer}\ and\ \citenamefont
  {Peterka}(2005)}]{Maurer_2004}%
  \BibitemOpen
  \bibfield  {author} {\bibinfo {author} {\bibfnamefont {C.}~\bibnamefont
  {Maurer}}\ and\ \bibinfo {author} {\bibfnamefont {R.~J.}\ \bibnamefont
  {Peterka}},\ }\href {https://doi.org/10.1152/jn.00221.2004} {\bibfield
  {journal} {\bibinfo  {journal} {Journal of Neurophysiology}\ }\textbf
  {\bibinfo {volume} {93}},\ \bibinfo {pages} {189} (\bibinfo {year} {2005})},\
  \bibinfo {note} {pMID: 15331614},\ \Eprint
  {https://arxiv.org/abs/https://doi.org/10.1152/jn.00221.2004}
  {https://doi.org/10.1152/jn.00221.2004} \BibitemShut {NoStop}%
\bibitem [{\citenamefont {Kiemel}\ \emph {et~al.}(2011)\citenamefont {Kiemel},
  \citenamefont {Zhang},\ and\ \citenamefont {Jeka}}]{Kiemel_2011}%
  \BibitemOpen
  \bibfield  {author} {\bibinfo {author} {\bibfnamefont {T.}~\bibnamefont
  {Kiemel}}, \bibinfo {author} {\bibfnamefont {Y.}~\bibnamefont {Zhang}},\ and\
  \bibinfo {author} {\bibfnamefont {J.~J.}\ \bibnamefont {Jeka}},\ }\href
  {https://doi.org/10.1523/JNEUROSCI.1013-11.2011} {\bibfield  {journal}
  {\bibinfo  {journal} {Journal of Neuroscience}\ }\textbf {\bibinfo {volume}
  {31}},\ \bibinfo {pages} {15144} (\bibinfo {year} {2011})},\ \Eprint
  {https://arxiv.org/abs/https://www.jneurosci.org/content/31/42/15144.full.pdf}
  {https://www.jneurosci.org/content/31/42/15144.full.pdf} \BibitemShut
  {NoStop}%
\bibitem [{\citenamefont {Cabrera}\ and\ \citenamefont
  {Milton}(2002)}]{Cabrera_2002}%
  \BibitemOpen
  \bibfield  {author} {\bibinfo {author} {\bibfnamefont {J.~L.}\ \bibnamefont
  {Cabrera}}\ and\ \bibinfo {author} {\bibfnamefont {J.~G.}\ \bibnamefont
  {Milton}},\ }\href {https://doi.org/10.1103/PhysRevLett.89.158702} {\bibfield
   {journal} {\bibinfo  {journal} {Phys. Rev. Lett.}\ }\textbf {\bibinfo
  {volume} {89}},\ \bibinfo {pages} {158702} (\bibinfo {year}
  {2002})}\BibitemShut {NoStop}%
\bibitem [{\citenamefont {Collins}\ and\ \citenamefont
  {De~Luca}(1994)}]{Collins_1994}%
  \BibitemOpen
  \bibfield  {author} {\bibinfo {author} {\bibfnamefont {J.~J.}\ \bibnamefont
  {Collins}}\ and\ \bibinfo {author} {\bibfnamefont {C.~J.}\ \bibnamefont
  {De~Luca}},\ }\href {https://doi.org/10.1103/PhysRevLett.73.764} {\bibfield
  {journal} {\bibinfo  {journal} {Phys. Rev. Lett.}\ }\textbf {\bibinfo
  {volume} {73}},\ \bibinfo {pages} {764} (\bibinfo {year} {1994})}\BibitemShut
  {NoStop}%
\bibitem [{\citenamefont {Yamamoto}\ \emph {et~al.}(2015)\citenamefont
  {Yamamoto}, \citenamefont {Smith}, \citenamefont {Suzuki}, \citenamefont
  {Kiyono}, \citenamefont {Tanahashi}, \citenamefont {Sakoda}, \citenamefont
  {Morasso},\ and\ \citenamefont {Nomura}}]{Yamamoto_2015}%
  \BibitemOpen
  \bibfield  {author} {\bibinfo {author} {\bibfnamefont {T.}~\bibnamefont
  {Yamamoto}}, \bibinfo {author} {\bibfnamefont {C.~E.}\ \bibnamefont {Smith}},
  \bibinfo {author} {\bibfnamefont {Y.}~\bibnamefont {Suzuki}}, \bibinfo
  {author} {\bibfnamefont {K.}~\bibnamefont {Kiyono}}, \bibinfo {author}
  {\bibfnamefont {T.}~\bibnamefont {Tanahashi}}, \bibinfo {author}
  {\bibfnamefont {S.}~\bibnamefont {Sakoda}}, \bibinfo {author} {\bibfnamefont
  {P.}~\bibnamefont {Morasso}},\ and\ \bibinfo {author} {\bibfnamefont
  {T.}~\bibnamefont {Nomura}},\ }\href
  {https://doi.org/https://doi.org/10.14814/phy2.12329} {\bibfield  {journal}
  {\bibinfo  {journal} {Physiological Reports}\ }\textbf {\bibinfo {volume}
  {3}},\ \bibinfo {pages} {e12329} (\bibinfo {year} {2015})},\ \Eprint
  {https://arxiv.org/abs/https://physoc.onlinelibrary.wiley.com/doi/pdf/10.14814/phy2.12329}
  {https://physoc.onlinelibrary.wiley.com/doi/pdf/10.14814/phy2.12329}
  \BibitemShut {NoStop}%
\bibitem [{\citenamefont {Santos}\ \emph {et~al.}(2015)\citenamefont {Santos},
  \citenamefont {Picoli}, \citenamefont {Depr{\'{a}}},\ and\ \citenamefont
  {Mendes}}]{Santos_2015}%
  \BibitemOpen
  \bibfield  {author} {\bibinfo {author} {\bibfnamefont {E.~S.~D.}\
  \bibnamefont {Santos}}, \bibinfo {author} {\bibfnamefont {S.}~\bibnamefont
  {Picoli}}, \bibinfo {author} {\bibfnamefont {P.~P.}\ \bibnamefont
  {Depr{\'{a}}}},\ and\ \bibinfo {author} {\bibfnamefont {R.~S.}\ \bibnamefont
  {Mendes}},\ }\href {https://doi.org/10.1209/0295-5075/109/48001} {\bibfield
  {journal} {\bibinfo  {journal} {{EPL} (Europhysics Letters)}\ }\textbf
  {\bibinfo {volume} {109}},\ \bibinfo {pages} {48001} (\bibinfo {year}
  {2015})}\BibitemShut {NoStop}%
\bibitem [{\citenamefont {Suzuki}\ \emph {et~al.}(2020)\citenamefont {Suzuki},
  \citenamefont {Nakamura}, \citenamefont {Milosevic}, \citenamefont {Nomura},
  \citenamefont {Tanahashi}, \citenamefont {Endo}, \citenamefont {Sakoda},
  \citenamefont {Morasso},\ and\ \citenamefont {Nomura}}]{Suzuki_2020}%
  \BibitemOpen
  \bibfield  {author} {\bibinfo {author} {\bibfnamefont {Y.}~\bibnamefont
  {Suzuki}}, \bibinfo {author} {\bibfnamefont {A.}~\bibnamefont {Nakamura}},
  \bibinfo {author} {\bibfnamefont {M.}~\bibnamefont {Milosevic}}, \bibinfo
  {author} {\bibfnamefont {K.}~\bibnamefont {Nomura}}, \bibinfo {author}
  {\bibfnamefont {T.}~\bibnamefont {Tanahashi}}, \bibinfo {author}
  {\bibfnamefont {T.}~\bibnamefont {Endo}}, \bibinfo {author} {\bibfnamefont
  {S.}~\bibnamefont {Sakoda}}, \bibinfo {author} {\bibfnamefont
  {P.}~\bibnamefont {Morasso}},\ and\ \bibinfo {author} {\bibfnamefont
  {T.}~\bibnamefont {Nomura}},\ }\href {https://doi.org/10.1063/5.0022319}
  {\bibfield  {journal} {\bibinfo  {journal} {Chaos: An Interdisciplinary
  Journal of Nonlinear Science}\ }\textbf {\bibinfo {volume} {30}},\ \bibinfo
  {pages} {113140} (\bibinfo {year} {2020})},\ \Eprint
  {https://arxiv.org/abs/https://doi.org/10.1063/5.0022319}
  {https://doi.org/10.1063/5.0022319} \BibitemShut {NoStop}%
\bibitem [{\citenamefont {Eurich}\ and\ \citenamefont
  {Milton}(1996)}]{Eurich_1996}%
  \BibitemOpen
  \bibfield  {author} {\bibinfo {author} {\bibfnamefont {C.~W.}\ \bibnamefont
  {Eurich}}\ and\ \bibinfo {author} {\bibfnamefont {J.~G.}\ \bibnamefont
  {Milton}},\ }\href {https://doi.org/10.1103/PhysRevE.54.6681} {\bibfield
  {journal} {\bibinfo  {journal} {Phys. Rev. E}\ }\textbf {\bibinfo {volume}
  {54}},\ \bibinfo {pages} {6681} (\bibinfo {year} {1996})}\BibitemShut
  {NoStop}%
\bibitem [{\citenamefont {Bottaro}\ \emph {et~al.}(2008)\citenamefont
  {Bottaro}, \citenamefont {Yasutake}, \citenamefont {Nomura}, \citenamefont
  {Casadio},\ and\ \citenamefont {Morasso}}]{Bottaro_2008}%
  \BibitemOpen
  \bibfield  {author} {\bibinfo {author} {\bibfnamefont {A.}~\bibnamefont
  {Bottaro}}, \bibinfo {author} {\bibfnamefont {Y.}~\bibnamefont {Yasutake}},
  \bibinfo {author} {\bibfnamefont {T.}~\bibnamefont {Nomura}}, \bibinfo
  {author} {\bibfnamefont {M.}~\bibnamefont {Casadio}},\ and\ \bibinfo {author}
  {\bibfnamefont {P.}~\bibnamefont {Morasso}},\ }\href
  {https://doi.org/https://doi.org/10.1016/j.humov.2007.11.005} {\bibfield
  {journal} {\bibinfo  {journal} {Human Movement Science}\ }\textbf {\bibinfo
  {volume} {27}},\ \bibinfo {pages} {473} (\bibinfo {year} {2008})}\BibitemShut
  {NoStop}%
\bibitem [{\citenamefont {Asai}\ \emph {et~al.}(2009)\citenamefont {Asai},
  \citenamefont {Tasaka}, \citenamefont {Nomura}, \citenamefont {Nomura},
  \citenamefont {Casadio},\ and\ \citenamefont {Morasso}}]{Asai_2009}%
  \BibitemOpen
  \bibfield  {author} {\bibinfo {author} {\bibfnamefont {Y.}~\bibnamefont
  {Asai}}, \bibinfo {author} {\bibfnamefont {Y.}~\bibnamefont {Tasaka}},
  \bibinfo {author} {\bibfnamefont {K.}~\bibnamefont {Nomura}}, \bibinfo
  {author} {\bibfnamefont {T.}~\bibnamefont {Nomura}}, \bibinfo {author}
  {\bibfnamefont {M.}~\bibnamefont {Casadio}},\ and\ \bibinfo {author}
  {\bibfnamefont {P.}~\bibnamefont {Morasso}},\ }\href
  {https://doi.org/10.1371/journal.pone.0006169} {\bibfield  {journal}
  {\bibinfo  {journal} {PLOS ONE}\ }\textbf {\bibinfo {volume} {4}},\ \bibinfo
  {pages} {e6169} (\bibinfo {year} {2009})}\BibitemShut {NoStop}%
\bibitem [{\citenamefont {Nomura}\ \emph {et~al.}(2013)\citenamefont {Nomura},
  \citenamefont {Oshikawa}, \citenamefont {Suzuki}, \citenamefont {Kiyono},\
  and\ \citenamefont {Morasso}}]{Nomura_2013}%
  \BibitemOpen
  \bibfield  {author} {\bibinfo {author} {\bibfnamefont {T.}~\bibnamefont
  {Nomura}}, \bibinfo {author} {\bibfnamefont {S.}~\bibnamefont {Oshikawa}},
  \bibinfo {author} {\bibfnamefont {Y.}~\bibnamefont {Suzuki}}, \bibinfo
  {author} {\bibfnamefont {K.}~\bibnamefont {Kiyono}},\ and\ \bibinfo {author}
  {\bibfnamefont {P.}~\bibnamefont {Morasso}},\ }\href
  {https://doi.org/https://doi.org/10.1016/j.mbs.2013.02.002} {\bibfield
  {journal} {\bibinfo  {journal} {Mathematical Biosciences}\ }\textbf {\bibinfo
  {volume} {245}},\ \bibinfo {pages} {86} (\bibinfo {year} {2013})}\BibitemShut
  {NoStop}%
\bibitem [{\citenamefont {Nomura}\ \emph {et~al.}(2020)\citenamefont {Nomura},
  \citenamefont {Suzuki},\ and\ \citenamefont {Morasso}}]{Nomura_2020}%
  \BibitemOpen
  \bibfield  {author} {\bibinfo {author} {\bibfnamefont {T.}~\bibnamefont
  {Nomura}}, \bibinfo {author} {\bibfnamefont {Y.}~\bibnamefont {Suzuki}},\
  and\ \bibinfo {author} {\bibfnamefont {P.~G.}\ \bibnamefont {Morasso}},\
  }\bibinfo {title} {A model of the intermittent control strategy for
  stabilizing human quiet stance},\ in\ \href
  {https://doi.org/10.1007/978-1-4614-7320-6\_100698-1} {\emph {\bibinfo
  {booktitle} {Encyclopedia of Computational Neuroscience}}},\ \bibinfo
  {editor} {edited by\ \bibinfo {editor} {\bibfnamefont {D.}~\bibnamefont
  {Jaeger}}\ and\ \bibinfo {editor} {\bibfnamefont {R.}~\bibnamefont {Jung}}}\
  (\bibinfo  {publisher} {Springer New York},\ \bibinfo {address} {New York,
  NY},\ \bibinfo {year} {2020})\ pp.\ \bibinfo {pages} {1--10}\BibitemShut
  {NoStop}%
\bibitem [{\citenamefont {Michimoto}\ \emph {et~al.}(2016)\citenamefont
  {Michimoto}, \citenamefont {Suzuki}, \citenamefont {Kiyono}, \citenamefont
  {Kobayashi}, \citenamefont {Morasso},\ and\ \citenamefont
  {Nomura}}]{Michimoto_2016}%
  \BibitemOpen
  \bibfield  {author} {\bibinfo {author} {\bibfnamefont {K.}~\bibnamefont
  {Michimoto}}, \bibinfo {author} {\bibfnamefont {Y.}~\bibnamefont {Suzuki}},
  \bibinfo {author} {\bibfnamefont {K.}~\bibnamefont {Kiyono}}, \bibinfo
  {author} {\bibfnamefont {Y.}~\bibnamefont {Kobayashi}}, \bibinfo {author}
  {\bibfnamefont {P.}~\bibnamefont {Morasso}},\ and\ \bibinfo {author}
  {\bibfnamefont {T.}~\bibnamefont {Nomura}},\ }in\ \href
  {https://doi.org/10.1109/EMBC.2016.7590634} {\emph {\bibinfo {booktitle}
  {2016 38th Annual International Conference of the IEEE Engineering in
  Medicine and Biology Society (EMBC)}}}\ (\bibinfo {year} {2016})\ pp.\
  \bibinfo {pages} {37--40}\BibitemShut {NoStop}%
\bibitem [{\citenamefont {Richmond}\ \emph {et~al.}(2021)\citenamefont
  {Richmond}, \citenamefont {Fling}, \citenamefont {Lee},\ and\ \citenamefont
  {Peterson}}]{Richmond_2021}%
  \BibitemOpen
  \bibfield  {author} {\bibinfo {author} {\bibfnamefont {S.~B.}\ \bibnamefont
  {Richmond}}, \bibinfo {author} {\bibfnamefont {B.~W.}\ \bibnamefont {Fling}},
  \bibinfo {author} {\bibfnamefont {H.}~\bibnamefont {Lee}},\ and\ \bibinfo
  {author} {\bibfnamefont {D.~S.}\ \bibnamefont {Peterson}},\ }\href
  {https://doi.org/https://doi.org/10.1016/j.jbiomech.2021.110485} {\bibfield
  {journal} {\bibinfo  {journal} {Journal of Biomechanics}\ }\textbf {\bibinfo
  {volume} {123}},\ \bibinfo {pages} {110485} (\bibinfo {year}
  {2021})}\BibitemShut {NoStop}%
\bibitem [{\citenamefont {Xiang}\ \emph {et~al.}(2018)\citenamefont {Xiang},
  \citenamefont {Glasauer},\ and\ \citenamefont {Seemungal}}]{Xiang_2018}%
  \BibitemOpen
  \bibfield  {author} {\bibinfo {author} {\bibfnamefont {M.}~\bibnamefont
  {Xiang}}, \bibinfo {author} {\bibfnamefont {S.}~\bibnamefont {Glasauer}},\
  and\ \bibinfo {author} {\bibfnamefont {B.~M.}\ \bibnamefont {Seemungal}},\
  }\href {https://doi.org/10.1093/brain/awy250} {\bibfield  {journal} {\bibinfo
   {journal} {Brain}\ }\textbf {\bibinfo {volume} {141}},\ \bibinfo {pages}
  {2824} (\bibinfo {year} {2018})}\BibitemShut {NoStop}%
\bibitem [{\citenamefont {Tigrini}\ \emph {et~al.}(2022)\citenamefont
  {Tigrini}, \citenamefont {Verdini}, \citenamefont {Fioretti},\ and\
  \citenamefont {Mengarelli}}]{Tigrini_2022}%
  \BibitemOpen
  \bibfield  {author} {\bibinfo {author} {\bibfnamefont {A.}~\bibnamefont
  {Tigrini}}, \bibinfo {author} {\bibfnamefont {F.}~\bibnamefont {Verdini}},
  \bibinfo {author} {\bibfnamefont {S.}~\bibnamefont {Fioretti}},\ and\
  \bibinfo {author} {\bibfnamefont {A.}~\bibnamefont {Mengarelli}},\ }\href
  {https://doi.org/https://doi.org/10.1016/j.cnsns.2021.106198} {\bibfield
  {journal} {\bibinfo  {journal} {Communications in Nonlinear Science and
  Numerical Simulation}\ }\textbf {\bibinfo {volume} {108}},\ \bibinfo {pages}
  {106198} (\bibinfo {year} {2022})}\BibitemShut {NoStop}%
\bibitem [{\citenamefont {Loram}\ and\ \citenamefont
  {Lakie}(2002)}]{Loram_2002}%
  \BibitemOpen
  \bibfield  {author} {\bibinfo {author} {\bibfnamefont {I.~D.}\ \bibnamefont
  {Loram}}\ and\ \bibinfo {author} {\bibfnamefont {M.}~\bibnamefont {Lakie}},\
  }\href {https://doi.org/https://doi.org/10.1113/jphysiol.2002.025049}
  {\bibfield  {journal} {\bibinfo  {journal} {The Journal of Physiology}\
  }\textbf {\bibinfo {volume} {545}},\ \bibinfo {pages} {1041} (\bibinfo {year}
  {2002})},\ \Eprint
  {https://arxiv.org/abs/https://physoc.onlinelibrary.wiley.com/doi/pdf/10.1113/jphysiol.2002.025049}
  {https://physoc.onlinelibrary.wiley.com/doi/pdf/10.1113/jphysiol.2002.025049}
  \BibitemShut {NoStop}%
\bibitem [{\citenamefont {Ott}\ \emph {et~al.}(1990)\citenamefont {Ott},
  \citenamefont {Grebogi},\ and\ \citenamefont {Yorke}}]{Ott_1990}%
  \BibitemOpen
  \bibfield  {author} {\bibinfo {author} {\bibfnamefont {E.}~\bibnamefont
  {Ott}}, \bibinfo {author} {\bibfnamefont {C.}~\bibnamefont {Grebogi}},\ and\
  \bibinfo {author} {\bibfnamefont {J.~A.}\ \bibnamefont {Yorke}},\ }\href
  {https://doi.org/10.1103/PhysRevLett.64.1196} {\bibfield  {journal} {\bibinfo
   {journal} {Phys. Rev. Lett.}\ }\textbf {\bibinfo {volume} {64}},\ \bibinfo
  {pages} {1196} (\bibinfo {year} {1990})}\BibitemShut {NoStop}%
\bibitem [{\citenamefont {Guillouzic}\ \emph {et~al.}(1999)\citenamefont
  {Guillouzic}, \citenamefont {L'Heureux},\ and\ \citenamefont
  {Longtin}}]{Longtin_1999}%
  \BibitemOpen
  \bibfield  {author} {\bibinfo {author} {\bibfnamefont {S.}~\bibnamefont
  {Guillouzic}}, \bibinfo {author} {\bibfnamefont {I.}~\bibnamefont
  {L'Heureux}},\ and\ \bibinfo {author} {\bibfnamefont {A.}~\bibnamefont
  {Longtin}},\ }\href {https://doi.org/10.1103/PhysRevE.59.3970} {\bibfield
  {journal} {\bibinfo  {journal} {Phys. Rev. E}\ }\textbf {\bibinfo {volume}
  {59}},\ \bibinfo {pages} {3970} (\bibinfo {year} {1999})}\BibitemShut
  {NoStop}%
\bibitem [{\citenamefont {Bect}\ \emph {et~al.}(2006)\citenamefont {Bect},
  \citenamefont {Phulpin}, \citenamefont {Baili},\ and\ \citenamefont
  {Fleury}}]{Bect_2006}%
  \BibitemOpen
  \bibfield  {author} {\bibinfo {author} {\bibfnamefont {J.}~\bibnamefont
  {Bect}}, \bibinfo {author} {\bibfnamefont {Y.}~\bibnamefont {Phulpin}},
  \bibinfo {author} {\bibfnamefont {H.}~\bibnamefont {Baili}},\ and\ \bibinfo
  {author} {\bibfnamefont {G.}~\bibnamefont {Fleury}},\ }in\ \href
  {https://doi.org/10.1109/PMAPS.2006.360298} {\emph {\bibinfo {booktitle}
  {2006 International Conference on Probabilistic Methods Applied to Power
  Systems}}}\ (\bibinfo {year} {2006})\ pp.\ \bibinfo {pages}
  {1--6}\BibitemShut {NoStop}%
\bibitem [{\citenamefont {Bect}(2008)}]{Bect_2008}%
  \BibitemOpen
  \bibfield  {author} {\bibinfo {author} {\bibfnamefont {J.}~\bibnamefont
  {Bect}},\ }\href
  {https://doi.org/https://doi.org/10.3182/20080706-5-KR-1001.02618} {\bibfield
   {journal} {\bibinfo  {journal} {IFAC Proceedings Volumes}\ }\textbf
  {\bibinfo {volume} {41}},\ \bibinfo {pages} {15480} (\bibinfo {year}
  {2008})},\ \bibinfo {note} {17th IFAC World Congress}\BibitemShut {NoStop}%
\bibitem [{\citenamefont {Kumar}\ \emph {et~al.}(2007)\citenamefont {Kumar},
  \citenamefont {Chakravorty},\ and\ \citenamefont {Junkins}}]{Kumar_2007}%
  \BibitemOpen
  \bibfield  {author} {\bibinfo {author} {\bibfnamefont {M.}~\bibnamefont
  {Kumar}}, \bibinfo {author} {\bibfnamefont {S.}~\bibnamefont {Chakravorty}},\
  and\ \bibinfo {author} {\bibfnamefont {J.~L.}\ \bibnamefont {Junkins}},\ }in\
  \href {https://doi.org/10.1109/CDC.2007.4434816} {\emph {\bibinfo {booktitle}
  {2007 46th IEEE Conference on Decision and Control}}}\ (\bibinfo {year}
  {2007})\ pp.\ \bibinfo {pages} {3078--3083}\BibitemShut {NoStop}%
\bibitem [{\citenamefont {Wang}\ and\ \citenamefont {Lee}(2020)}]{Wang_2020}%
  \BibitemOpen
  \bibfield  {author} {\bibinfo {author} {\bibfnamefont {W.}~\bibnamefont
  {Wang}}\ and\ \bibinfo {author} {\bibfnamefont {T.}~\bibnamefont {Lee}},\
  }in\ \href {https://doi.org/10.23919/ACC45564.2020.9147910} {\emph {\bibinfo
  {booktitle} {2020 American Control Conference (ACC)}}}\ (\bibinfo {year}
  {2020})\ pp.\ \bibinfo {pages} {1803--1808}\BibitemShut {NoStop}%
\bibitem [{\citenamefont {Stepan}\ and\ \citenamefont
  {Kollar}(2000)}]{Stepan_2000}%
  \BibitemOpen
  \bibfield  {author} {\bibinfo {author} {\bibfnamefont {G.}~\bibnamefont
  {Stepan}}\ and\ \bibinfo {author} {\bibfnamefont {L.}~\bibnamefont
  {Kollar}},\ }\href
  {https://doi.org/https://doi.org/10.1016/S0895-7177(00)00039-X} {\bibfield
  {journal} {\bibinfo  {journal} {Mathematical and Computer Modelling}\
  }\textbf {\bibinfo {volume} {31}},\ \bibinfo {pages} {199} (\bibinfo {year}
  {2000})},\ \bibinfo {note} {proceedings of the Conference on Dynamical
  Systems in Biology and Medicine}\BibitemShut {NoStop}%
\bibitem [{\citenamefont {Winter}\ \emph {et~al.}(1998)\citenamefont {Winter},
  \citenamefont {Patla}, \citenamefont {Prince}, \citenamefont {Ishac},\ and\
  \citenamefont {Gielo-Perczak}}]{Winter_1998}%
  \BibitemOpen
  \bibfield  {author} {\bibinfo {author} {\bibfnamefont {D.~A.}\ \bibnamefont
  {Winter}}, \bibinfo {author} {\bibfnamefont {A.~E.}\ \bibnamefont {Patla}},
  \bibinfo {author} {\bibfnamefont {F.}~\bibnamefont {Prince}}, \bibinfo
  {author} {\bibfnamefont {M.}~\bibnamefont {Ishac}},\ and\ \bibinfo {author}
  {\bibfnamefont {K.}~\bibnamefont {Gielo-Perczak}},\ }\href
  {https://doi.org/10.1152/jn.1998.80.3.1211} {\bibfield  {journal} {\bibinfo
  {journal} {Journal of Neurophysiology}\ }\textbf {\bibinfo {volume} {80}},\
  \bibinfo {pages} {1211} (\bibinfo {year} {1998})},\ \bibinfo {note} {pMID:
  9744933},\ \Eprint
  {https://arxiv.org/abs/https://doi.org/10.1152/jn.1998.80.3.1211}
  {https://doi.org/10.1152/jn.1998.80.3.1211} \BibitemShut {NoStop}%
\bibitem [{\citenamefont {Morasso}\ and\ \citenamefont
  {Sanguineti}(2002)}]{Morasso_2002}%
  \BibitemOpen
  \bibfield  {author} {\bibinfo {author} {\bibfnamefont {P.~G.}\ \bibnamefont
  {Morasso}}\ and\ \bibinfo {author} {\bibfnamefont {V.}~\bibnamefont
  {Sanguineti}},\ }\href {https://doi.org/10.1152/jn.2002.88.4.2157} {\bibfield
   {journal} {\bibinfo  {journal} {Journal of Neurophysiology}\ }\textbf
  {\bibinfo {volume} {88}},\ \bibinfo {pages} {2157} (\bibinfo {year}
  {2002})},\ \bibinfo {note} {pMID: 12364538},\ \Eprint
  {https://arxiv.org/abs/https://doi.org/10.1152/jn.2002.88.4.2157}
  {https://doi.org/10.1152/jn.2002.88.4.2157} \BibitemShut {NoStop}%
\bibitem [{\citenamefont {Burkhardt}\ \emph {et~al.}(2000)\citenamefont
  {Burkhardt}, \citenamefont {Franklin},\ and\ \citenamefont
  {Gawronski}}]{Burkhardt_2000}%
  \BibitemOpen
  \bibfield  {author} {\bibinfo {author} {\bibfnamefont {T.~W.}\ \bibnamefont
  {Burkhardt}}, \bibinfo {author} {\bibfnamefont {J.}~\bibnamefont
  {Franklin}},\ and\ \bibinfo {author} {\bibfnamefont {R.~R.}\ \bibnamefont
  {Gawronski}},\ }\href {https://doi.org/10.1103/PhysRevE.61.2376} {\bibfield
  {journal} {\bibinfo  {journal} {Phys. Rev. E}\ }\textbf {\bibinfo {volume}
  {61}},\ \bibinfo {pages} {2376} (\bibinfo {year} {2000})}\BibitemShut
  {NoStop}%
\bibitem [{\citenamefont {Burkhardt}(2007)}]{Burkhardt_2007}%
  \BibitemOpen
  \bibfield  {author} {\bibinfo {author} {\bibfnamefont {T.~W.}\ \bibnamefont
  {Burkhardt}},\ }\href {https://doi.org/10.1088/1742-5468/2007/07/p07004}
  {\bibfield  {journal} {\bibinfo  {journal} {Journal of Statistical Mechanics:
  Theory and Experiment}\ }\textbf {\bibinfo {volume} {2007}},\ \bibinfo
  {pages} {P07004} (\bibinfo {year} {2007})}\BibitemShut {NoStop}%
\bibitem [{\citenamefont {Kloeden}\ and\ \citenamefont
  {Platen}(2011)}]{Kloeden_2011}%
  \BibitemOpen
  \bibfield  {author} {\bibinfo {author} {\bibfnamefont {P.}~\bibnamefont
  {Kloeden}}\ and\ \bibinfo {author} {\bibfnamefont {E.}~\bibnamefont
  {Platen}},\ }\href {https://books.google.co.jp/books?id=BCvtssom1CMC} {\emph
  {\bibinfo {title} {Numerical Solution of Stochastic Differential
  Equations}}},\ Stochastic Modelling and Applied Probability\ (\bibinfo
  {publisher} {Springer Berlin Heidelberg},\ \bibinfo {year}
  {2011})\BibitemShut {NoStop}%
\end{thebibliography}%

\end{document}